\documentclass[12pt,oneside]{article}
\pdfoutput=1
\usepackage{amsthm}
\usepackage{enumitem}
\usepackage{jheppub}
\usepackage[skip=2pt,font=footnotesize]{caption}
\usepackage[utf8]{inputenc}
\usepackage{verbatim}
\usepackage{amsmath}
\usepackage{amssymb}
\usepackage{amsthm}
\usepackage{slashed}
\usepackage{amsfonts}
\usepackage{makecell}
\usepackage{tabularx}
\usepackage{comment}
\usepackage{graphicx,subfigure}
\usepackage{float}
\usepackage{color}
\usepackage{psfrag}
\usepackage{color}
\definecolor{darkblue}{rgb}{0.1,0.1,.7}
\usepackage{hyperref} 

\makeatletter
\newcommand\xleftrightarrow[2][]{%
  \ext@arrow 9999{\longleftrightarrowfill@}{#1}{#2}}
\newcommand\longleftrightarrowfill@{%
  \arrowfill@\leftarrow\relbar\rightarrow}
\makeatother

\hypersetup{breaklinks}
\usepackage{comment}

\def\1{{\rm 1-loop}}
\newcommand{\dDisc}{\text{dDisc}}

\newcommand\numberthis{\addtocounter{equation}{1}\tag{\theequation}}
\usepackage{braket}

\def\<{\langle}
\def\>{\rangle}

\newcommand \s {\sigma}
\newcommand   \f  {\phi}

\newcommand{\bea}{\begin{eqnarray}}
\newcommand{\eea}{\end{eqnarray}}

\def\z{\zeta}

\def\ll{\<\!\<}
\def\rr{\>\!\>}
\def\disc{\text{disc}}

\def\a{\alpha}

\def\O{{\cal O}}

\def\a{\alpha}

\newcommand {\be} {\begin {equation}}
\newcommand {\ee} {\end {equation}}

\newcommand {\bes} {\begin {equation*}}
\newcommand {\ees} {\end {equation*}}

\renewcommand{\Re}{{\text{Re}\hspace{.05cm}}}
\renewcommand{\Im}{{\text{Im}\hspace{.05cm}}}

\newcommand{\beq}{\begin{equation}}
\newcommand{\eeq}{\end{equation}}

\def\be{ \begin{equation} }
\def\ee{ \end{equation} }

 \newmuskip\pFqmuskip

\usepackage{enumitem}

\newlist{primenumerate}{enumerate}{1}
\setlist[primenumerate,1]{label={3$'$.}}

\newcommand*\pFq[6][8]{%
  \begingroup 
  \pFqmuskip=#1mu\relax
  \mathcode`\,=\string"8000
  \begingroup\lccode`\~=`\,
  \lowercase{\endgroup\let~}\pFqcomma
  {}_{#2}F_{#3}{\left[\genfrac..{0pt}{}{#4}{#5};#6\right]}%
  \endgroup
}
\newcommand{\pFqcomma}{\mskip\pFqmuskip}

\makeatletter
\renewcommand{\@maketitle}{
\newpage
 \begin{center}%
  {\large\bfseries \@title \par}%
 \end{center}%
 \par} \makeatother

\numberwithin{equation}{section}

\begin{document}
\hfill \hbox{{CALT-TH-2021-012}}

\title{\Large Dispersion Formulas in QFTs, CFTs, and Holography}
\author{David Meltzer}

\affiliation{Walter Burke Institute for Theoretical Physics, California Institute of Technology,\\ Pasadena, California, 91125}

\emailAdd{dmeltzer@caltech.edu}

\abstract{ 
We study momentum space dispersion formulas in general QFTs and their applications for CFT correlation functions.
We show, using two independent methods, that QFT dispersion formulas can be written in terms of causal commutators.
The first derivation uses analyticity properties of retarded correlators in momentum space.
The second derivation uses the largest time equation and the defining properties of the time-ordered product.
At four points we show that the momentum space QFT dispersion formula depends on the same causal double-commutators as the CFT dispersion formula.
At $n$-points, the QFT dispersion formula depends on a sum of nested advanced commutators.
For CFT four-point functions, we show that the momentum space dispersion formula is equivalent to the CFT dispersion formula, up to possible semi-local terms.
We also show that the Polyakov-Regge expansions associated to the momentum space and CFT dispersion formulas are related by a Fourier transform.
In the process, we prove that the momentum space conformal blocks of the causal double-commutator are equal to cut Witten diagrams.
Finally, by combining the momentum space dispersion formulas with the AdS Cutkosky rules, we find a complete, bulk unitarity method for AdS/CFT correlators in momentum space.
}

\maketitle

\section{Introduction} 

In the past decade, there has been a resurgence of interest in studying quantum field theory (QFT) and quantum gravity with bootstrap methods. 
The idea of the bootstrap is that by leveraging physical principles and symmetries, we can map out the space of consistent theories.
As an example, the conformal bootstrap uses unitarity, causality, and locality to derive rigorous bounds on the space of conformal field theories (CFTs) \cite{Simmons-Duffin:2016gjk,Poland:2018epd}.
Lorentzian methods have played a central role in the analytic conformal bootstrap, including the discovery of the large-spin expansion \cite{Fitzpatrick:2012yx,Komargodski:2012ek}, the Lorentzian inversion formula \cite{Caron-Huot:2017vep}, and the conformal dispersion formula \cite{Carmi:2019cub}.
The goal of this work is to understand how the momentum space dispersion formula, which is valid for general QFTs, is related to the conformal dispersion formula.
In the process, we will demonstrate how methods from the conformal and S-matrix bootstraps \cite{Eden:1966dnq} are related.

Dispersion formulas are one of the central ingredients in the S-matrix bootstrap \cite{Eden:1966dnq}.
The useful feature of dispersion formulas is that they express an amplitude, $M(s,t)$, in terms of its imaginary part, $\Im M(s,t)$.
Unitarity implies that the imaginary part of an amplitude factorizes into a product of sub-amplitudes  \cite{Cutkosky:1960sp,Veltman:1963th}. Therefore, dispersion formulas can be used to compute one-loop amplitudes by recycling the tree-level amplitudes.\footnote{We are assuming the amplitude is 1PI and that there are no anomalous thresholds.}
The imaginary part of the amplitude in the forward limit is also positive, which can be used to derive sign constraints on effective field theory (EFT) couplings \cite{Adams:2006sv}.\footnote{For related recent work see \cite{Bellazzini:2016xrt,deRham:2017zjm,deRham:2017avq} and references therein.} 
Recently, stronger bounds on EFTs have been derived by leveraging crossing symmetry \cite{Bellazzini:2020cot,Tolley:2020gtv,Caron-Huot:2020cmc,Arkani-Hamed:2020blm,Sinha:2020win,Caron-Huot:2021rmr}. 
These new results show that EFT couplings have to scale as one would expect from dimensional analysis.
Despite their long history, dispersion formulas are still being used to discover new properties of scattering amplitudes.

The application of dispersive methods to conformal correlation functions is a comparatively newer subject.
The discovery of the Lorentzian inversion formula \cite{Caron-Huot:2017vep}, and the following derivation of the CFT dispersion formula \cite{Carmi:2019cub}, prove that a Euclidean CFT four-point function $\<\f\f\f\f\>_{\text{E}}$ can be expressed as an integral over a double-commutator $\<[\f,\f][\f,\f]\>$, up to possible low-spin ambiguities.
In this sense, the double-commutator of a four-point function is the CFT generalization of $\Im M(s,t)$. 
Moreover, in non-perturbative CFTs the double-commutator can often be well-approximated by the contribution of a few low-dimension operators \cite{dsdi,Caron-Huot:2017vep,Albayrak:2019gnz,Carmi:2019cub,Liu:2020tpf,Caron-Huot:2020ouj}. 
Finally, the positivity properties of the CFT double-commutator can be used to generalize bounds on flat-space EFTs \cite{Adams:2006sv} to corresponding bounds on AdS EFTs  \cite{Hartman:2015lfa}. 

One of the motivations for this work is to further develop dispersive methods for AdS/CFT correlation functions \cite{Fitzpatrick:2011dm,Aharony:2016dwx,Caron-Huot:2017vep,Ponomarev:2019ofr,Meltzer:2019nbs,Meltzer:2020qbr}.
In \cite{Caron-Huot:2017vep} it was shown that the double-commutator of a one-loop Witten diagram is completely fixed in terms of tree-level CFT data.
In other words, the double-commutator factorizes one-loop Witten diagrams into a product of tree-level diagrams.
The proof that the double-commutator factorizes Witten diagrams was later rederived from a bulk, Lorentzian perspective by generalizing the flat space Cutkosky rules \cite{Cutkosky:1960sp,Veltman:1963th}  to AdS momentum space \cite{Meltzer:2020qbr}.\footnote{See \cite{Ponomarev:2019ofr,Meltzer:2019nbs} for an alternative bulk method where cuts are performed by replacing an AdS propagator with a harmonic function and then projecting onto physical states.}
Once we have the CFT double-commutator in momentum space, we then need to Fourier transform it to position space before we can use the CFT dispersion formula.
While these Fourier transforms are in principle computable, this transition between momentum and position space for CFT correlators can often be difficult in practice.
In this work we will bypass this complication by directly studying momentum space dispersion formulas for CFT correlators.
Combined with the results of \cite{Meltzer:2020qbr}, this gives a bulk unitarity method for AdS/CFT correlators in momentum space.

AdS/CFT unitarity methods explain how bulk locality and unitarity constrain the analytic form of the boundary CFT correlators.
Similar questions have also been studied in the cosmological bootstrap \cite{Arkani-Hamed:2015bza,Arkani-Hamed:2018kmz,Baumann:2019oyu,Baumann:2020dch,Sleight:2019mgd,Sleight:2019hfp,Hillman:2019wgh,Goodhew:2020hob}, where inflationary correlators are fixed using consistency conditions such as factorization, the absence of folded singularities, and conformal symmetry.
To relate on-shell methods in AdS/CFT and dS, we recall that in perturbation theory the AdS partition function is related by analytic continuation to the dS wavefunction \cite{Maldacena:2002vr,Harlow:2011ke,Isono:2020qew}.
Based on insights from the modern amplitudes program \cite{Elvang:2013cua}, it is natural to expect that on-shell methods for AdS and dS correlators will reveal new symmetries and structures that are not manifest using a Lagrangian approach.
For example, there has been progress in the study of on-shell recursion relations \cite{Raju:2010by,Raju:2011mp,Raju:2012zr,Raju:2012zs}, positive geometries \cite{Arkani-Hamed:2017fdk}, and the ambitwistor string \cite{Roehrig:2020kck,Eberhardt:2020ewh} for (A)dS correlators.
Requiring that AdS correlators factorize correctly when we put propagators on shell also gives a useful consistency check for new conjectures, including on the possible generalization of double-copy \cite{Bern:2019prr} to AdS/CFT correlators \cite{Farrow:2018yni,Lipstein:2019mpu,Armstrong:2020woi,Albayrak:2020fyp}.

Taking a step back from holography, another motivation for this work is to understand: how is the CFT dispersion formula related to dispersion formulas in general QFTs? 
The CFT dispersion formula reconstructs a Euclidean correlator from its double-commutator, which can be computed by taking a double-discontinuity in cross-ratio space \cite{Caron-Huot:2017vep}.
By contrast, a QFT momentum space dispersion formula depends on a single discontinuity with respect to a Mandelstam invariant.
One hint that the two dispersion formulas are related is that the absorptive part of a scattering amplitude can be computed by applying the LSZ reduction procedure \cite{LSZ_I,LSZ_II} to a retarded double-commutator \cite{Lehmann:1958ita} (see \cite{Sommer:1970mr} for a review).
Similarly, the momentum space discontinuity of a CFT four-point function is equal to a causal double-commutator \cite{Polyakov:1974gs}.
We can use the latter identity to show that the momentum space and CFT dispersion formulas rely on the same physical input, the $t$- and $u$-channel causal double-commutators.
Furthermore, we will show that at $n$-points the QFT dispersion formula can be written in terms of sums of nested causal commutators.
The appearance of commutators in the CFT dispersion formula is not an accident of conformal symmetry, but is rather a universal feature of QFT dispersion formulas in both massive and massless theories. 

\subsection{Background}
To set the stage for the rest of the paper, we will first briefly review the S-matrix and CFT dispersion formulas.
S-matrix dispersion formulas are derived by leveraging the analyticity properties of scattering amplitudes \cite{Eden:1966dnq,Sommer:1970mr,Itzykson:1980rh,Bogolyubov:1990kw}. 
The simplest amplitude to study is a $2\rightarrow 2$ scattering amplitude, $M(s,t)$, for identical scalars of mass $m>0$. 
Here $s$ and $t$ are the Mandelstam invariants, $s=-(p_1+p_2)^2$ and $t=-(p_2+p_3)^2$. The third invariant $u=-(p_1+p_3)^2$ is fixed by $s+t+u=4m^2$.

\begin{figure}
\begin{center}
\includegraphics[scale=.45]{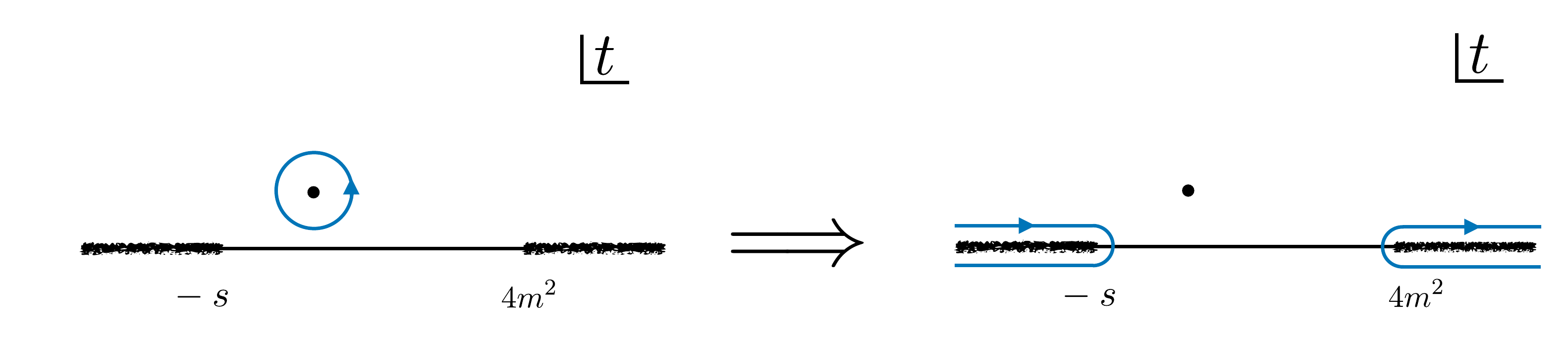}
\end{center}
\caption{The amplitude is first written as a contour integral and then this contour is deformed to wrap the branch cuts on the real line. We have assumed that the arcs at infinity vanish.}
\label{fig:Contour_Deformation}
\end{figure}

Under favorable circumstances, the amplitude is analytic in the entire cut $t$-plane for $s<0$ and fixed.
To derive a dispersion formula, we first express the amplitude in terms of a contour integral,
\begin{align}
M(s,t)=\frac{1}{2\pi i}\oint \frac{dt'}{t'-t}M(s,t'),\label{eq:M_contour}
\end{align}
where $t'$ encircles $t$ counter-clockwise.
We will assume the amplitude decays in the large $t$ Regge limit,
\begin{align}
\lim\limits_{|t|\rightarrow\infty}M(s,t)=0 \quad \text{for } s<0, \label{eq:vanishing_Regge}
\end{align}
where we take $t$ large while keeping its phase fixed. 
If \eqref{eq:vanishing_Regge} is true, then we can deform the contour, drop the arcs at infinity, and express the amplitude in terms of its discontinuities,
\begin{align}
M(s,t)=\frac{1}{2\pi i}\int\limits_{4m^2}^{\infty}\frac{dt'}{t'-t}\disc_{t'}M(s,t')+\text{($u$-channel)}.\label{eq:dispamp}
\end{align}
Here we assumed that there are no poles due to single-particle exchange and that the cut starts at the two particle threshold, $4m^2$.
By crossing symmetry, there is also a $u$-channel cut for $t\in(-\infty,-s]$. This procedure is shown schematically in figure \ref{fig:Contour_Deformation}.

In order to write down \eqref{eq:dispamp} we had to make several assumptions about the analytic structure of the amplitude. 
For example, we assumed that $M(s,t)$ only has non-analyticities which come from physical states being put on shell.
In general, amplitudes can also have anomalous thresholds.
These are non-analyticities in the complex $t$-plane that cannot be associated to the exchange of a physical state \cite{Eden:1966dnq,Itzykson:1980rh,Zwicky:2016lka}.
Anomalous thresholds can potentially obstruct crossing symmetry, or the existence of a crossing path between the physical $t$- and $u$-channel processes.
Proving crossing symmetry for the S-matrix is a difficult problem and has only been shown for $2\rightarrow 2$ \cite{Bros:1964iho,Bros:1965kbd} and $2\rightarrow 3$ \cite{BEG23,BROS1986325} scattering of stable, external particles.
Proving crossing symmetry for general higher-point amplitudes remains an open problem, see \cite{Mizera:2021ujs} for a recent discussion.

The CFT dispersion formula \cite{Carmi:2019cub} has a similar structure to the S-matrix dispersion formula,
\begin{align}
\<\f(x_1)\ldots \f(x_4)\>_{\text{E}}=\int d^{d}y_1\ldots d^{d}y_4 K(x_i,y_i) \<[\f(y_1),\f(y_4)]_A[\f(y_2),\f(y_3)]_A\> +\text{($u$-channel)}. \label{eq:introCFTdispersion}
\end{align}
On the left hand side of \eqref{eq:introCFTdispersion} we included the ``E" subscript to emphasize that this is a Euclidean correlation function, and not a Wightman function. The kernel $K$ in \eqref{eq:introCFTdispersion} is fixed by conformal symmetry and is given in \cite{Carmi:2019cub}. On the right hand side we introduced the advanced commutator,\footnote{The CFT inversion and dispersion formulas are typically written with the causal restrictions put into the integration region. For the inversion formula, see for example equation 1.7 of \cite{ssw}. The CFT dispersion formula is derived from the inversion formula and therefore also depends on causal commutators.
\label{footnote:causal}}
\[
    [\f(x_1),\f(x_2)]_{A}= 
\begin{cases}
    [\f(x_1),\f(x_2)],& \text{ if } x_2^0>x_1^0\\
    0,&  \text{otherwise}.\numberthis \label{eq:advanced_comm}
\end{cases} 
\]
We will not review the derivation given in \cite{Carmi:2019cub} here, but will instead only emphasize that this dispersion formula has been derived for all CFT four-point functions by leveraging conformal symmetry, locality, and unitarity. 
At most, the CFT dispersion formula requires a finite number of subtractions if the correlator is not sufficiently bounded in the Regge limit \cite{Carmi:2019cub}.
The CFT dispersion formula also manifestly cannot contain anomalous thresholds. 
Given the double-commutator in \eqref{eq:introCFTdispersion}, we can insert a complete set of states,
\begin{align}
 \<[\f(x_1),\f(x_4)]_A[\f(x_2),\f(x_3)]_A\>=\sum\limits_{\Psi} \frac{\<[\f(x_1),\f(x_4)]_A|\Psi\>\<\Psi|[\f(x_2),\f(x_3)]_A\>}{\<\Psi|\Psi\>},
\end{align}
and this gives a convergent Lorentzian OPE \cite{Gillioz:2016jnn,Gillioz:2018kwh}. 
In other words, the integrand of the CFT dispersion formula is fixed in terms of physical states and how they couple to the external operators. For CFTs with a graphical expansion, e.g., Feynman diagrams for weakly coupled theories or Witten diagrams for holographic theories, this means we only need to consider unitarity cuts of the diagrams.

The two dispersion formulas, \eqref{eq:dispamp} and \eqref{eq:introCFTdispersion}, cannot be used in the same theory as CFTs do not have an S-matrix.
However, we can ask: how is the CFT dispersion formula related to the four-point dispersion formula for QFT correlation functions?
By studying this question, we will see how ideas developed for massive QFTs, with the original aim of studying the S-matrix, can be used in the CFT bootstrap.

\subsection{Summary and Outline}
We will now give a summary and outline for the rest of the paper.
In Section \ref{sec:OperatorOrderings} we introduce conventions and identities which will be used throughout this work.
We start by defining the time-ordered, retarded, and advanced products.
We also review coincidence relations between these three products in momentum space.

In sections \ref{sec:disp_from_analyticity} and \ref{sec:disp_from_largest_time} we explain how to derive the following momentum space dispersion formula for time-ordered, scalar, four-point functions in a generic QFT,
\begin{align}
\<T[\f_1\f_2\f_3\f_4]\>=& \ \frac{1}{i \pi }\int\limits_{0}^{\infty}\frac{dt'}{t'-t-i\epsilon}\<[\f_2,\f_3]_{A}[\f_1,\f_4]_{A}\>
+\text{($u$-channel)}. \label{eq:disp_summary}
\end{align}
To be general, we have taken the $t$ integral to run from $0$ to $\infty$, but in a massive theory the integrand will only be non-zero when $t$ is above some threshold value $t_0$.
The $u$-channel contribution can be found by taking $1\leftrightarrow 2$.
We assume the QFT is defined in $d\geq3$ spacetime dimensions. 
We also generalize \eqref{eq:disp_summary} to $n$-point functions, where each advanced commutator is replaced with the advanced product defined in Section \ref{sec:OperatorOrderings}.

The formula \eqref{eq:disp_summary} is schematic and we will give more precise formulas in Sections \ref{sec:disp_from_analyticity} and \ref{sec:disp_from_largest_time}. The left hand side of  \eqref{eq:disp_summary} is a function of the Mandelstam invariants $s$, $t$, and $u$ as well as the ``norms" $\z_i=-p_i^2$. Similarly, on the right hand side we integrate over the internal invariant $t'$ with $\z_i$ and $s$ held fixed. Finally, \eqref{eq:disp_summary} is derived in the configuration where $\z_i$, $s<0$ such that $\sum_{i}\z_i-s<0$. 
These kinematics are chosen so that the correlator is analytic in the complex $t$-plane with non-overlapping cuts on the real line.
Throughout this work, we will always assume $\z_i$ and $s$ are chosen such that the above inequalities hold. 

In Section \ref{sec:disp_from_analyticity} we review how to derive dispersion formulas for correlation functions using analyticity arguments.
We recall that time-ordered correlators with real momenta $p_i$ are boundary values of the retarded correlator $r_{n}(k_1,\ldots,k_{n-1})$ with complex momenta $k_i$. 
One can show that $r_{n}(k_1,\ldots,k_{n-1})$ is analytic in the following domain \cite{Steinmann_1,Steinmann_2,Ruelle_1,Araki_burgoyne,Araki_2,Epstein:1966yea},\footnote{In massive QFTs we can relax the second condition to $\{\Im(k_I)=0, \ k_I^{2}>-m_{I,\text{thr}}^{2}\}$ where $m_{I,\text{thr}}$ is the threshold mass for the $I^{\text{th}}$ channel.}
\begin{align}
\mathcal{D}=\{\Im(k_{I})\neq 0, \ \Im(k_I)^2<0\} \cup \{\Im(k_I)=0, \ k_I^{2}>0\}, \label{eq:intro_regionAnalyticity}
\end{align}
where $I$ is a subset of $\{1,\ldots,n\}$ and we use the notation
\begin{align}
k_{I}=\sum\limits_{i\in I}k_{i}.\label{eq:defofkI}
\end{align}
The region \eqref{eq:intro_regionAnalyticity} is known as the primitive domain of analyticity.
To derive \eqref{eq:intro_regionAnalyticity} one needs to impose microcausality,
\begin{align}
[\f(x),\f(y)]=0 \quad \text{if } \ (x-y)^2>0,
\end{align}
or that operators commute at spacelike separation.
We also need to assume that all the physical states in our Hilbert space, $|\Psi(p)\>\in\mathcal{H}$, have positive energy.
The retarded correlator $r_{n}(k_1,\ldots,k_{n-1})$ obeys a dispersion formula inside the primitive domain $\mathcal{D}$, which can then be used to derive a dispersion formula for the time-ordered correlator with real momenta.
We also explain how to rewrite the integrand of this dispersion formula in terms of causal commutators.
In this derivation, the number of subtractions needed to write down a dispersion formula will only depend on how the correlator scales in the Regge limit.

The results of Section \ref{sec:disp_from_analyticity} are based on methods used to derive S-matrix dispersion formulas.
The primitive domain of analyticity \eqref{eq:intro_regionAnalyticity} does not intersect the complex mass-shell, $k_i^2=-m_i^2$, and cannot be used to derive dispersion formulas for scattering amplitudes.
Instead, \eqref{eq:intro_regionAnalyticity} implies that there are fixed-$s$ dispersion formulas for off-shell values of $k_{i}^{2}$.
To derive a dispersion formula for the S-matrix, one needs to amputate the external lines and analytically continue $k_i^2$ to the mass-shell.
In the process of analytically continuing, anomalous thresholds can move onto the physical sheet and the dispersion integral may be deformed into the complex plane \cite{Sommer:1970mr,Bogolyubov:1990kw}.
In this work we are interested in correlation functions and will study dispersion formulas inside the primitive domain $\mathcal{D}$, where we have analyticity in the entire cut $t$-plane.
Section \ref{sec:disp_from_analyticity} will then largely be a review on how these off-shell dispersion formulas are derived.
To our knowledge, the main new results in this section is the expression for the five-point dispersion formula in terms of a double advanced product\footnote{This formula should be equivalent to the one found in \cite{AlvarezEstrada:1973yy} after using LSZ reduction and the positive spectrum condition. We thank Julio Parra-Martinez for bringing this paper to our attention.} and the method used to relate momentum space discontinuities to causal commutators, at general $n$-points, using the axiomatic definitions of the time-ordered and causal products \cite{steinmann1968,Bogolyubov:1990kw}.

In Section \ref{sec:disp_from_largest_time} we give an alternative derivation of \eqref{eq:disp_summary} using the largest time equation \cite{Veltman:1963th,tHooft:1973wag}. 
In \cite{Remiddi:1981hn}, the largest time equation was used to derive Lorentz invariant dispersion formulas for Feynman diagrams with up to four external points. 
We will build on \cite{Remiddi:1981hn} by generalizing their method of deriving dispersion formulas to $n$-point correlation functions in a generic QFT. 
We also use the positive spectrum condition to rewrite the $n$-point dispersion formula in terms of a double advanced product.
For this method, the number of subtractions needed will depend on how the correlator scales in high-energy limits that go beyond the usual Regge limit.
While the methods used in this section will require more assumptions than the ones used in Section \ref{sec:disp_from_analyticity}, one virtue of this approach is it is simpler to generalize to $n$-points.
The reader interested in applications to CFT correlators can safely skip this section.

In Sections \ref{sec:equivalence} and \ref{sec:app_to_CFT}, we will specialize to CFT correlation functions. 
In Section \ref{sec:equivalence} we show that the momentum space and conformal dispersion formulas are equivalent, modulo possible semi-local terms.
We then study the Polyakov-Regge expansions generated by each dispersion formula and show that the corresponding Polyakov-Regge blocks are related by a Fourier transform.
We also give conditions on when Polyakov-Regge blocks are superbounded in position and momentum space.
In Section \ref{sec:app_to_CFT} we will test the momentum space dispersion formula on different CFT correlators and show that the dispersion formula works as expected. 
We also study the dispersion formula for spinning external operators.

The appendices contain various technical details.
In Appendix \ref{app:Domain_Analyticity} we study the primitive domain of analyticity for four- and five-point functions. 
In Appendix \ref{app:AdSUnitarity} we review the AdS Cutkosky rules.
In Appendix \ref{app:holoblocks} we prove that the $t$-channel, momentum space conformal blocks of the double-commutator $\<[\f,\f]_A[\f,\f]_A\>$ are equal to cut Witten diagrams.
In Appendix \ref{app:Regge_Blocks} we study the Regge limit for $t$-channel conformal blocks in momentum space.
In Appendix \ref{app:details} we go into more detail on the assumptions used to derive the dispersion formula via the largest time equation.

\section{Operator Orderings}
\label{sec:OperatorOrderings}
In order to make this work self-contained, in this section we will review how to define operator orderings in general QFTs. The definitions introduced here will be used in both Sections \ref{sec:disp_from_analyticity} and \ref{sec:disp_from_largest_time}.

In free-field theories the time-ordered product is defined by,
\begin{align}
T[\f(x_1)\ldots \f(x_n)]=\sum\limits_{\pi}\theta(x^{0}_{\pi_1}-x^{0}_{\pi_2})\ldots\theta(x^{0}_{\pi_{n-1}}-x^{0}_{\pi_n})\f(x_{\pi_1})\ldots\f(x_{\pi_n}), \label{eq:TOrdering}
\end{align}
where we sum over all permutations, $\pi$, of the set $\{1,\ldots,n\}$. 
We will also refer to \eqref{eq:TOrdering} as the $T$-product.
The definition \eqref{eq:TOrdering} cannot be used for non-perturbative QFTs as it involves the product of two distributions, the discontinuous $\theta$-function and the operator-valued fields $\f(x)$. This product of distributions is not well-defined in general and we need an alternative definition.
We will discuss subtleties on the product of $\theta$-functions and correlation functions in more detail in Section \ref{sec:disp_from_largest_time}.

One possible alternative to \eqref{eq:TOrdering} is to start with Euclidean correlation functions and then define Lorentzian correlators by analytic continuation \cite{Streater:1989vi,Haag:1992hx} (see \cite{Hartman:2015lfa} for a review). 
Through different analytic continuations, or different $i\epsilon$ prescriptions, one can generate all Lorentzian correlators. 
The analytic continuation from Euclidean to Lorentzian signature defines a healthy Lorentzian QFT if the Osterwalder-Schrader reconstruction theorems hold \cite{Osterwalder:1973dx,Osterwalder:1974tc}.

We will use a different approach and, following  \cite{steinmann1968,Bogolyubov:1990kw}, define the time-ordered product using a set of physically well-motivated conditions. 
For simplicity, we restrict to real, scalar operators. 
We will assume the $T$-product exists and has the following properties:
\begin{enumerate}
\item The time-ordered product $T[\f(x_1)\ldots\f(x_n)]$ is permutation invariant,
\begin{align}
T[\f(x_{\pi_1})\ldots\f(x_{\pi_{n}})]=T[\f(x_1)\ldots\f(x_n)],
\end{align}
for all permutations $\pi$ of $\{1,\ldots,n\}$. \label{item_1}
\item The $T$-product is Poincar\'e covariant:
\begin{align}
U(a,\Lambda)T[\f(x_1)\ldots \f(x_n)]U^{-1}(a,\Lambda) = T[\f(\Lambda x_1+a)\ldots \f(\Lambda x_n+a)],
\end{align}
where $\Lambda$ is a Lorentz transformation and $a$ is a constant vector.
\item Given two sets of points $X=\{x_1,\ldots, x_n\}$ and $Y=\{y_{1},\ldots,y_m\}$ such that $x^0_i\geq y^0_j$ for all $x_i\in X$ and $y_j \in Y$, the $T$-product factorizes as:
\begin{align}
T[\f(x_{1})\ldots\f(x_{n})\f(y_{1})\ldots\f(y_{m})]=T[\f(x_{1})\ldots\f(x_{n})]T[\f(y_{1})\ldots\f(y_{m})].\label{eq:Tfactorization}
\end{align}
This is of course the condition that $T$ orders the operators based on their time.
\item For all $n$, the following unitarity condition holds:
\begin{align}
\sum\limits_{r=0}^{n}(-1)^{r}\sum\limits_{\alpha\in\Pi_r(n)}\<\overline{T}[\f(x_{\alpha_1})\ldots \f(x_{\alpha_r})]T[\f(x_{\alpha_{r+1}})\ldots \f(x_{\alpha_n})]\>=0, \label{eq:QFT_Optical}
\end{align}
where $\Pi_{r}(n)$ is set of partitions of $\{1,\ldots,n\}$ into two sets of size $r$ and $n-r$. Here the anti-time-ordered product $\overline{T}$ is defined by,
\begin{align}
T[\f(x_1)\ldots\f(x_n)]^{\dagger}=\overline{T}[\f(x_1)\ldots\f(x_n)].\label{eq:anti-time-ordering}
\end{align} \label{item_4}
\end{enumerate}
When the original definition \eqref{eq:TOrdering} is valid, one can verify these identities using the definition of the $\theta$-function.
In practice, to compute out-of-time ordered correlation functions of the form $\<\overline{T}[\f\ldots \f]T[\f\ldots \f]\>$ one can use the Schwinger-Keldysh rules \cite{Schwinger:1960qe,Keldysh:1964ud}.
For more general out-of-time ordered correlators, such as those involving retarded and advanced commutators, one can use the rules given in \cite{Haehl:2017qfl}.

To see why the identity \eqref{eq:QFT_Optical} corresponds to a unitarity condition, we can pull out the $r=0$ and $r=n$ terms in the sum and take the vacuum expectation value:
\begin{align}
\<T[\f(x_1)\ldots \f(x_n)]\>+&(-1)^n\<\overline{T}[\f(x_1)\ldots\f(x_n)]\>
\nonumber \\&=\sum\limits_{r=1}^{n-1}(-1)^{r+1}\sum\limits_{\alpha\in\Pi_r(n)}\<\overline{T}[\f(x_{\alpha_1})\ldots \f(x_{\alpha_r})]T[\f(x_{\alpha_{r+1}})\ldots \f(x_{\alpha_n})]\>.\label{eq:QFTopticalV2}
\end{align}
This relation shows that the real (imaginary) part of an even (odd) point, time-ordered correlator can be expressed as a sum of partially time-ordered correlators $\<\overline{T}[\f\ldots \f]T[\f\ldots\f]\>$. The terms on the right-hand side of \eqref{eq:QFTopticalV2} can be factorized by inserting a complete set of states between the pairs of ordered operators. 
For theories with an S-matrix, one can use the LSZ prescription to go from \eqref{eq:QFTopticalV2} to the standard S-matrix unitarity condition $S^{\dagger}S=1$ \cite{Schweber:1961zz,Bogolyubov:1990kw}. In appendix \ref{app:AdSUnitarity}, we review how to use \eqref{eq:QFTopticalV2} to derive the AdS cutting rules.

Another natural object to consider is the retarded product of operators. 
In free-field theories the $R$-product can be defined in terms of commutators,
\begin{align}
R[\f(x);\f(x_1)...\f(x_{n})]=(-i)^n\sum\limits_{\pi}&\theta(x^0-x^0_{\pi_1})\theta(x^0_{\pi_1}-x^0_{\pi_2})...\theta(x^0_{\pi_{n-1}}-x^0_{\pi_n})
\nonumber \\
&[...[[\f(x),\f(x_{\pi_1})],\f(x_{\pi_2})]...,\f(x_{\pi_n})].
\label{eq:retarded_product}
\end{align}
The overall power of $i$ is chosen so that the $R$-product is Hermitian.
On the right hand side of \eqref{eq:retarded_product} we sum over all permutations of the operators $\{\f(x_1),\ldots,\f(x_n)\}$, but leave $\f(x)$ fixed. 
The $R$-product is then symmetric in the $\f(x_i)$ and is only non-zero if $\f(x)$ is in the future lightcone of all the other operators.
This causal restriction implies that correlation functions involving the $R$-product will be analytic in certain cones of complex momentum space. 
We discuss this in more detail in Section \ref{sec:disp_from_analyticity}.
For $n=1$, the $R$-product reduces to the retarded commutator,
\begin{align}
R[\f(x);\f(x_1)]=-i\theta(x^0-x_1^0)[\f(x),\f(x_1)].
\end{align}

As with the original definition of the $T$-product, the definition \eqref{eq:retarded_product} may not be well-defined for general theories.
To define the $R$-product in general, we will first assume the $T$-product exists and has the properties (\ref{item_1})-(\ref{item_4}) listed above. If this holds, we can define the $R$-product as \cite{Bogolyubov:1990kw},\footnote{In the literature, the roles are often reversed: the $R$-product is assumed to exist and is then used to define the $T$-product \cite{steinmann1968}.}
\begin{align}
R[\f(x);\f(x_1)...\f(x_n)]&=i^n\sum\limits_{r=0}^{n}(-1)^{n-r}\sum\limits_{\alpha \in \Pi_{r}(n)}\overline{T}[\f(x_{\a_{1}})...\f(x_{\a_r})]T[\f(x_{\a_{r+1}})...\f(x_{\a_n})\f(x)] 
\nonumber \\
&=(-i)^n\sum\limits_{r=0}^{n}(-1)^{r}\sum\limits_{\a\in\Pi_{r}(n)}\overline{T}[\f(x)\f(x_{\a_{1}})...\f(x_{\a_r})]T[\f(x_{\a_{r+1}})...\f(x_{\a_{n}})] \label{eq:retarded_productV2}.
\end{align} 
If we assume both \eqref{eq:TOrdering} and \eqref{eq:retarded_product} are valid, we can prove \eqref{eq:retarded_productV2} by expanding each product in terms of $\theta$-functions. 

It will also be convenient to introduce the advanced product,
\begin{align}
A[\f(x);\f(x_1)...\f(x_n)]=(-i)^n\sum\limits_{\pi}&\theta(x^0_{\pi_1}-x^0)\theta(x^0_{\pi_2}-x^0_{\pi_1})...\theta(x^0_{\pi_{n}}-x^0_{\pi_{n-1}})
\nonumber \\
&[...[[\f(x),\f(x_{\pi_1})],\f(x_{\pi_2})]...,\f(x_{\pi_n})].
\label{eq:advanced_product}
\end{align}
For $n=1$, the advanced product reduces to the advanced commutator defined in \eqref{eq:advanced_comm},
\begin{align}
A[\f(x);\f(x_1)]=-i\theta(x_1^0-x)[\f(x),\f(x_1)].
\end{align}
To define the $A$-product in general theories, we again take a sum of (anti-)time-ordered products,
\begin{align}
A[\f(x);\f(x_1)...\f(x_n)]&=(-i)^n\sum\limits_{r=0}^{n}(-1)^r\sum\limits_{\a\in\Pi_{r}(n)}T[\f(x_{\a_1})...\f(x_{\a_r})\f(x)] \overline{T}[\f(x_{\a_{r+1}})...\f(x_{\a_n})]
\nonumber \\
&=i^n\sum\limits_{r=0}^{n}(-1)^{n-r}\sum\limits_{\a\in\Pi_{r}(n)}T[\f(x_{\a_1})...\f(x_{\a_{r}})]\overline{T}[\f(x)\f(x_{\a_{r+1}})...\f(x_{\a_n})] \label{eq:advanced_productV2}.
\end{align} 

In the following sections, it will be important that the time-ordered, retarded, and advanced products have identical actions on the vacuum when the external momenta, and their sums, are spacelike. When going to momentum space, we will use the following conventions for the Fourier transform,
\begin{align}
f(p)&=\int d^{d}xe^{ip \cdot x}f(x),
\\
f(x)&=\int \frac{d^{d}p}{(2\pi)^d}e^{-ip\cdot x}f(p).
\end{align}
To give the necessary conditions for the various products to agree, we need to introduce the open lightcones,
\begin{align}
V_{\pm}=\{p\ |\ p^{2}<0, \ \pm p^0>0\}.
\end{align}
Here we are using the mostly plus, flat space metric $\eta_{\mu\nu}$.
The corresponding closed lightcones, $\overline{V}_{\pm}$, are found by using non-strict inequalities.

To relate the time-ordered, retarded and advanced products, we will use the positive spectrum condition \cite{Haag:1992hx}. This says that all the physical states in our Hilbert space, $|\Psi(p)\>\in\mathcal{H}$, have momentum lying in the closed, forward lightcone, $p\in\overline{V}_+$. This implies,
\begin{align}
O[\f(p_1)\ldots \f(p_n)]|0\>=0  \ \ \text{if} \ \ \sum\limits_{i=1}^{n}p_{i}\notin \overline{V}_{+}, \label{eq:pos_energy}
\end{align} 
where $O$ can be any ordering operation, i.e. the $T$-, $R$-, or $A$-products. Similarly, a string of operators acting on the left vacuum $\<0|$ vanishes if the sum of their momenta lies outside $\overline{V}_{-}$.
Together, \eqref{eq:retarded_productV2} and \eqref{eq:pos_energy} imply:
\begin{align}
R[\f(p);\f(p_1)\ldots\f(p_n)]|0\>&=i^n \overline{T}[\f(p)\f(p_1)\ldots \f(p_n)]|0\> \ \ \ \ \ \ \ \  \hspace{.65cm} \text{if} \ \ p_{I}\notin \overline{V}_{+},
\vspace{1cm} \label{eq:RtoT}
\\[3pt]
\<0|R[\f(p);\f(p_1)\ldots\f(p_n)]&=(-i)^n\<0|T[\f(p)\f(p_1)\ldots \f(p_n)]  \ \ \ \ \ \ \ \ \text{if}\ \ p_{I}\notin \overline{V}_{-},
\label{eq:RtoTb}
\end{align}
for all $I\subseteq\{1,\ldots,n\}$ and where the notation $p_I$ was defined in \eqref{eq:defofkI}.
By using \eqref{eq:advanced_productV2} one can derive similar relations for the advanced product,
\begin{align}
A[\f(p);\f(p_1)\ldots\f(p_n)]|0\>&=i^nT[\f(p)\f(p_1)\ldots \f(p_n)]|0\> \ \ \ \ \ \ \ \  \hspace{.65cm} \text{if}\ \ p_{I}\notin \overline{V}_{+} ,\label{eq:AtoT}
\\[3pt]
\<0|A[\f(p);\f(p_1)\ldots\f(p_n)]&=(-i)^n\<0|\overline{T}[\f(p)\f(p_1)\ldots \f(p_n)] \ \ \ \ \ \ \ \  \text{if} \ \ p_{I}\notin \overline{V}_{-}, \label{eq:AtoTb}
\end{align}
for all $I\subseteq\{1,\ldots,n\}$. 
These relations show that partially time-ordered correlators $\<\overline{T}[\f\ldots \f]T[\f\ldots\f]\>$ are equal to the double advanced product $\<A[\f\ldots \f]A[\f\ldots\f]\>$ if enough external momenta and their sums are spacelike. We will use the identities \eqref{eq:AtoT} and \eqref{eq:AtoTb} repeatedly to write dispersion formulas in terms of the $A$-product.

Finally, we will introduce some useful notation for $n$-point correlators in momentum space.
We define correlators such as $\<T[\f(p_1)\ldots \f(p_n)]\>$ in terms of their Fourier transform,
\begin{align}
\<T[\f(p_1)\ldots \f(p_n)]\>=\int d^{d}x_1\ldots d^{d}x_n  \<T[\f(x_1)\ldots \f(x_n)]\>e^{i(p_1\cdot x_1+\ldots +p_n\cdot x_n)}.
\end{align}
It will also be convenient to factor out the overall momentum conserving $\delta$-function using a double-bracket notation,
\begin{align}
\<T[\f(p_1)\ldots \f(p_n)]\>=(2\pi)^d\delta(p_1+\ldots+p_n)\ll T[\f(p_1)\ldots \f(p_n)]\rr. \label{eq:double_bracket}
\end{align}
The double-bracket correlators can be computed by fixing one point to the origin and Fourier transforming in the remaining $n-1$ positions.
To make formulas more compact, we also introduce the following notation,
\begin{align}
\tau_{n}(p_1,\ldots,p_{n-1})&=\ll T[\f(p_1) \ldots \f(p_{n})]\rr,
\\
r_{n}(p_1,\ldots,p_{n-1})&=\ll R[\f(p);\f(p_1)\ldots\f(p_{n-1})]\rr,
\\
a_{n}(p_1,\ldots,p_{n-1})&=\ll A[\f(p);\f(p_1)\ldots\f(p_{n-1})]\rr.
\end{align}
There is an asymmetry between our definitions for the time-ordered correlator $\tau_{n}$ and the causal correlators $r_{n}$ and $a_{n}$. This is because the $T$-product is symmetric in all $n$ positions while the $R$- and $A$-products distinguish the first position and are symmetric in the remaining $n-1$ positions. In order to use the relations \eqref{eq:RtoT}-\eqref{eq:AtoTb}, it is simplest to study the causal correlators as functions of the $n-1$ momenta $p_i$ and fix the first argument, $p$, using momentum conservation.

From the coincidence relations \eqref{eq:RtoT}-\eqref{eq:AtoTb}, we find,
\begin{align}
r_n(p_1,\ldots,p_{n-1})=(-1)^na_n(p_1,\ldots,p_{n-1})=& \ (-i)^{n-1}\tau_n(p_1,\ldots,p_{n-1})
\nonumber\\ =& \ i^{n-1}\tau^*_n(p_1,\ldots,p_{n-1}),
 \label{eq:coinc_vev}
\end{align}
if the momenta $p_i$ and their sums are all spacelike.

Depending on the problem, it can be simpler to think of the correlators as functions of the momenta $p_i$ or as functions of the Lorentz invariants $p_i\cdot p_j$. 
The former will be convenient when we need to use operator identities, such as the coincidence relations \eqref{eq:RtoT}-\eqref{eq:AtoTb}, or study how partially ordered correlators $\<\overline{T}[\f\ldots\f]T[\f\ldots\f]\>$ can be factorized. 
To write down dispersion formulas, it will be convenient to consider the correlators as functions of the Lorentz invariants, 
\begin{align}
\z_i&=-p_i^2,
\\
s_{i_1\ldots i_n}&=-\left(p_{i_1}+\ldots+p_{i_n}\right)^2. \label{eq:multi_mand}
\end{align}
We have distinguished the single-index Lorentz invariant $\z_i=s_i$ to be consistent with the notation used in the literature.
We will also abuse notation and use $\tau_{n}(p_i)$ and $\tau_{n}(\z_i,s_{ij})$ to refer to the time-ordered correlator as a function of the momenta $p_i$ and the Lorentz invariants $\{\z_i,s_{ij}\}$, respectively.
Finally, throughout this work we will drop disconnected contributions to the correlators.
At four- and five-points the disconnected terms vanish automatically if we assume all the external operators are distinct.

\section{Derivation 1: Dispersion from Analyticity}
\label{sec:disp_from_analyticity}
In this section we review how dispersion formulas for the retarded and time-ordered correlator can be derived using analyticity arguments.
We will also use unitarity and the positive spectrum condition to explain how these dispersion formulas can be rewritten in terms of advanced commutators.

\subsection{Analyticity in $k$-Space}
\label{sec:analytick} 

To find the domain of analyticity for the retarded correlator in momentum space, we use its definition in terms of the $R$-product,
\begin{align}
r_{n}(p_1,\ldots,p_{n-1})=\int d^{d}x_1\ldots d^{d}x_{n-1}\<R[\f(0);\f(x_1)\ldots\f(x_{n-1})]\>e^{i(p_1\cdot x_1+\ldots+p_{n-1}\cdot x_{n-1})}. \label{eq:rp_definition}
\end{align} 
The correlator $\<R[\f(0);\f(x_1)\ldots\f(x_{n-1})]\>$ only has support for $x_i\in \overline{V}_{-}$. 
Therefore, if we make the replacement
\begin{align}
p_i\rightarrow k_i=p_i+iq_i \quad \text{with } \  q\in V_+,
\end{align}
then the integral \eqref{eq:rp_definition} becomes exponentially damped,
\begin{align}
r_{n}(k_1,\ldots,k_{n-1})=\int d^{d}x_1\ldots d^{d}x_{n-1}\<R[\f(0);\f(x_1)\ldots\f(x_{n-1})]\>&e^{i(p_1\cdot x_1+\ldots+p_{n-1}\cdot x_{n-1})}
\nonumber
\\
&e^{-(q_1\cdot x_1+\ldots+q_{n-1}\cdot x_{n-1})}. \label{eq:rk_definition}
\end{align}
The right hand side of \eqref{eq:rk_definition} is a Laplace transform of the tempered distribution $\<R[\f(0);\f(x_1)\ldots\f(x_{n-1})]\>$, which implies $r_{n}(k_i)$ is analytic in $k_i$ for $\Im(k_i)\in V_+$ \cite{Streater:1989vi,Haag:1992hx}.
To fix the notation, in the remainder of this section we use $k_i=p_i+iq_i$ to denote complex momentum and use $p_i$ and $q_i$ to denote real momenta.

Through the same arguments, one can show that the advanced correlator, $a_{n}(k_i)$, is analytic in the opposite cone, $\Im(k_i)\in V_-$. 
Next, the identity \eqref{eq:coinc_vev} says that the two analytic functions $r_{n}(k_i)$ and $a_{n}(k_i)$ agree if the momenta $k_i$ and their sums are all real and spacelike.
According to the edge of the wedge theorem \cite{Streater:1989vi,Haag:1992hx}, this implies the retarded and advanced correlators, $r_{n}(k_i)$ and $(-1)^{n-1}a_{n}(k_i)$, are related by analytic continuation. 

More generally, we can study a larger class of correlators known as the generalized retarded functions \cite{Ruelle_1,Araki_burgoyne,Araki_2,Epstein:1966yea}.
Using the same arguments as before, namely coincidence relations in momentum space and the edge of the wedge theorem, one can also prove that the generalized retarded correlators are related by analytic continuation to $r_{n}(k_i)$. This is used to prove that $r_{n}(k_i)$ is analytic in the primitive domain of analyticity written earlier \eqref{eq:intro_regionAnalyticity},
\begin{align}
\mathcal{D}=\{\Im(k_{I})\neq 0, \ \Im(k_I)^2<0\} \cup \{\Im(k_I)=0, \ k_I^{2}>0\}. \label{eq:regionAnalyticity}
\end{align}

One correlator which appears as the boundary value of $r_{n}(k_i)$ is the time-ordered correlation function $\tau_{n}(p_i)$ \cite{Ruelle_1,Araki_burgoyne,Araki_2,Epstein:1966yea}.
To write down the appropriate limit, we introduce the cone function,
\[
    C(p)= 
\begin{cases}
    V_{\pm},& \text{ if } p\in \overline{V}_{\pm}\\
    V_{+}\cup V_-,&  \text{ if } p^{2}>0.\numberthis
\end{cases}
\]
If $p$ is timelike or null, then $C(p)$ gives the corresponding open lightcone. If $p$ is spacelike, then $C(p)$ is the union of the forwards and backwards lightcone. The time-ordered correlator is then the following boundary value of the retarded correlator,
\begin{align}
\tau_{n}(p_1,\ldots,p_{n-1})=i^{n-1}\lim_{\substack{q_I\rightarrow 0 \\ q_I \in C(p_I)}}r_{n}(k_1,\ldots,k_{n-1}),\label{eq:ret_corr_to_time}
\end{align}
where $k_i=p_i+iq_i$ and $I\subseteq\{1,\ldots,n-1\}$. That is, we take the momenta $k_I$ to be real such that $\Im(k_I)$ lies in the same cone as its real part.
This is the standard $i\epsilon$ prescription written in a slightly different way. To recover the $i\epsilon$ prescription, we can assume the real part of one momentum is timelike, $p_i^{2}<0$, and take the small $q_i$ limit,
\begin{align}
k_i^2 \approx p_i^2+2ip_i\cdot q_i,
\end{align}
where we dropped the $q_i^{2}$ term. By assumption $p_i$ and $q_i$ are in the same cone, which implies $p_i\cdot q_i <0$. If we relabel $2p_i\cdot q_i\rightarrow -\epsilon$ with $\epsilon>0$, then we recover the $i\epsilon$ prescription for time-ordered correlators,
\begin{align}
k_i^{2}= p_i^{2}-i\epsilon.
\end{align}
If $p_i$ is spacelike, then \eqref{eq:ret_corr_to_time} says $k_i^2$ can be taken real from above or below in the complex plane, i.e., the correlation function is single-valued around this point. 

We will also need the anti-time-ordered correlator with real momenta $\tau^*_{n}(p_i)$.
This is found by taking a similar limit of the retarded correlator, but now with $q_I$ lying in the opposite cone of $p_I$,
\begin{align}
\tau^*_{n}(p_1,\ldots,p_{n-1})=(-i)^{n-1}\lim_{\substack{q_I\rightarrow 0 \\ -q_I \in C(p_I)}}r_{n}(k_1,\ldots,k_{n-1}). \label{eq:ret_corr_to_anti_time}
\end{align}

To keep the notation consistent, we will always refer to the analytic, $n$-point momentum space correlator as the retarded correlator, $r_{n}(k_i)$.
However, one should keep in mind that the retarded correlator with real momenta, $r_{n}(p_i)$, is just one of its boundary values and we can equally think of $r_{n}(k_i)$ as the analytically continued time-ordered or advanced correlator.

\subsection{Two- and Three-Point Functions}
\label{sec:2_3_ptanalyticity}
As a warm-up for the four- and five-point functions, we can use the analyticity properties reviewed in the previous section to write down dispersion formulas for two- and three-point functions of scalars. 
This will let us see how the unitarity condition \eqref{eq:QFTopticalV2} appears when studying discontinuities in momentum space.
We will use $r_{2}(\zeta)$ and $r_{3}(\zeta_1,\zeta_2,\z_3)$ to denote the retarded two- and three-point correlators as functions of $\z_i$.

From \eqref{eq:regionAnalyticity}, we see the two-point function $r_{2}(\zeta)$ is analytic away from the positive real $\z$ axis. For complex $\z$, we can write down a dispersion formula,
\begin{align}
r_{2}(\zeta)=\frac{1}{2\pi i}\oint \frac{d\zeta'}{\zeta'-\zeta}r_{2}(\zeta'), \label{eq:2ptanalyticPt1}
\end{align} 
where the $\zeta'$ contour encircles $\zeta$ counter-clockwise.
If we assume the correlator decays at high energies,
\begin{align}
\lim\limits_{|\zeta|\rightarrow\infty}r_{2}(\zeta)=0,
\end{align}
where we take $\zeta$ large and complex, then we can deform the contour in \eqref{eq:2ptanalyticPt1} and pick up non-analyticities on the real line,\footnote{Depending on the behavior of $r_{2}(\z)$ near $\z=0$, we may need to use a keyhole contour around $\z=0$ before taking the contour to wrap the branch cut.}
\begin{align}
r_{2}(\zeta)=\frac{1}{2\pi i}\int\limits_{0}^{\infty} \frac{d\zeta'}{\zeta'-\zeta}\disc_{\zeta'}r_{2}(\zeta').\label{eq:2pt_r_disp}
\end{align}
To find a dispersion formula for the time-ordered correlator, we use \eqref{eq:ret_corr_to_time}:
\begin{align}
\tau_2(\zeta)=i r_2(\zeta+i\epsilon)\quad \text{ for } \ \zeta\in\mathbb{R},
\end{align}
where we leave the $\epsilon\rightarrow 0$ limit implicit. If we take $\z$ to be real from above, then the left hand side of \eqref{eq:2pt_r_disp} becomes a time-ordered correlator.
To rewrite the integrand in terms of $\tau_2$, we use that by definition:
\begin{align}
\disc_{\zeta}f(\zeta)=f(\zeta+i\epsilon)-f(\zeta-i\epsilon).
\end{align}
Then the integrand of \eqref{eq:2pt_r_disp} becomes,
\begin{align}
\disc_{\zeta}r_2(\zeta)&=i\disc_\zeta\tau_{2}(\zeta)
\nonumber
\\
&=i(\tau_{2}(\zeta)+\tau^{*}_2(\zeta))
\nonumber\\
&=2i\Re \tau_{2}(\zeta), \label{eq:chain_2pt}
\end{align}
where we used \eqref{eq:ret_corr_to_anti_time} to arrive at the second line. The first line of \eqref{eq:chain_2pt} gives the expected K\"all\'en-Lehmann spectral representation:
\begin{align}
\tau_{2}(\zeta)=\frac{1}{2\pi i}\int\limits_{0}^{\infty} \frac{d\zeta'}{\zeta'-\zeta-i\epsilon}\disc_{\zeta'}\tau_{2}(\zeta').\label{eq:2ptdispfromanalyticity}
\end{align}
To rewrite \eqref{eq:2ptdispfromanalyticity} in terms of Wightman functions, we use $\disc_\zeta\tau_{2}(\zeta)=2\Re \tau_{2}(\zeta)$ and the identity \eqref{eq:QFTopticalV2}  for $n=2$,
\begin{align}
2\Re\<T[\f(x)\f(0)]\>=\<\f(x)\f(0)\>+\<\f(0)\f(x)\>.
\end{align}

The two-point dispersion formula illustrates the main points that will be important for deriving the four- and five-point dispersion formulas. 
We wrote down a contour integral representation for the retarded correlator, $r_2(\z)$, in its primitive domain of analyticity \eqref{eq:2ptanalyticPt1}, deformed the contour to derive a dispersion formula for $r_{2}(\zeta)$ \eqref{eq:2pt_r_disp}, and finally used \eqref{eq:ret_corr_to_time} and \eqref{eq:ret_corr_to_anti_time} to rewrite the dispersion formula in terms of the time-ordered correlator $\tau_{2}(\z)$ \eqref{eq:2ptdispfromanalyticity}. 
The one feature of higher-point dispersion formulas that is not captured by the two-point example is the appearance of commutators. 
To see how commutators emerge, we will study the three-point function.

For the three-point function, $r_{3}(\zeta_i)$, we will fix $\zeta_{1,2}<0$ and disperse in $\z_3$. 
The retarded three-point function is analytic in the $\zeta_3$ plane away from the positive real axis.
If we assume the retarded correlator vanishes at large $\z_3$,
\begin{align}
\lim\limits_{|\z_3|\rightarrow \infty}r_3(\z_1,\z_2,\z_3)=0 \quad \text{if } \ \z_{1,2}<0,
\end{align}
then we can write down the following dispersion formula for complex $\z_3$,
\begin{align}
r_{3}(\zeta_1,\zeta_2,\zeta_3)&=\frac{1}{2\pi i} \int\limits_{0}^{\infty}\frac{d\z'_3}{\z'_3-\z_3}\disc_{\z_3'}r_3(\z_1,\z_2,\z'_3),\label{eq:r3pt_disp_pt1}
\end{align}  

One difference in comparison to the two-point dispersion formula is that the discontinuity with respect to $\zeta_3$ now gives the imaginary part of the time-ordered correlator,
\begin{align}
\disc_{\zeta_3}r_3(\z_1,\z_2,\z_3)&=-\disc_{\z_3}\tau_3(\z_1,\z_2,\z_3)
\nonumber
\\
&=-\left(\tau_3(\z_1,\z_2,\z_3)-\tau^*_3(\z_1,\z_2,\z_3)\right)
\nonumber
\\
&=-2i\Im\tau_3(\z_1,\z_2,\z_3). \label{eq:disc3pt}
\end{align}
The relative factors between \eqref{eq:ret_corr_to_time} and \eqref{eq:ret_corr_to_anti_time} ensure that the discontinuity of the $n$-point retarded correlator, for $n$ even or odd, gives the real or imaginary part of the time-ordered correlator, respectively. 

By taking $\z_3$ real from above in \eqref{eq:r3pt_disp_pt1} we find,
\begin{align}
\tau_{3}(\zeta_1,\zeta_2,\zeta_3)=\frac{1}{2\pi i}\int\limits_{0}^{\infty} \frac{d\z'_3}{\z'_3-\z_3-i\epsilon}\disc_{\z'_3}\tau_3(\z_1,\z_2,\z'_3),\label{eq:dispersion3ptanalytic}
\end{align}  
where we used the first line of \eqref{eq:disc3pt}.
Using the identity $\disc_{\z_3}\tau_3(\z_i)=2i\Im\tau_3(\z_i)$, we can rewrite the integrand of \eqref{eq:dispersion3ptanalytic} using advanced commutators. To prove this, we take the unitarity condition \eqref{eq:QFTopticalV2} and set $n=3$,
\begin{align}
2i\Im\<T[\f(p_1)\f(p_2)\f(p_3)]\>=& \ \<\f(p_1)T[\f(p_2)\f(p_3)]\>+\<\f(p_2)T[\f(p_1)\f(p_3)]\>
\nonumber \\ &+\<\f(p_3)T[\f(p_1)\f(p_2)]\> - \ldots,
\label{eq:unitarity_3pt}
\end{align}
where we suppressed terms where two operators are in the anti-time-ordered product.
In the integrand of \eqref{eq:dispersion3ptanalytic} we have $\z_{1,2}<0$, or that $p_{1}$ and $p_{2}$ are spacelike. Using the positive spectrum condition \eqref{eq:pos_energy}, this implies the only non-zero terms in \eqref{eq:unitarity_3pt} have $\f(p_3)$ acting directly on the left or right vacuum,
\begin{align}
2i\Im\<T[\f(p_1)\f(p_2)\f(p_3)]\>=\<\f(p_3)T[\f(p_1)\f(p_2)]\>-\<\overline{T}[\f(p_1)\f(p_2)]\f(p_3)\>.\label{eq:ImtoTTb}
\end{align}
Since $p_{1,2}$ are spacelike, we can also use \eqref{eq:AtoT} and \eqref{eq:AtoTb} to rewrite the right hand side of \eqref{eq:ImtoTTb} in terms of the advanced product,
\begin{align}
2\Im\<T[\f(p_1)\f(p_2)\f(p_3)]\>=\<\f(p_3)A[\f(p_1);\f(p_2)]\>+\<A[\f(p_1);\f(p_2)]\f(p_3)\>.\label{eq:im3pttoComm}
\end{align}
Finally, by combining \eqref{eq:disc3pt} and \eqref{eq:im3pttoComm}, we find that the discontinuity with respect to $\z_3$ computes a three-point function involving advanced commutators,
\begin{align}
\disc_{\zeta_3}\<T[\f(p_1)\f(p_2)\f(p_3)]\>=i\big(\<A[\f(p_1);\f(p_2)]\f(p_3)\>+\<\f(p_3)A[\f(p_1);\f(p_2)]\>\big).\label{eq:disc3pt_to_comm}
\end{align}
This proves that in a generic QFT, either massive or massless, the integrand of the $\z_3$ dispersion formula can be rewritten in terms of advanced commutators.
Finally, we can use that causal commutators are analytic in the momenta to analytically continue \eqref{eq:disc3pt_to_comm} in $\z_{1,2}$.

\subsection{Four-Point Function}
\label{ssec:4ptanalyticity}
In this section we will study a single-variable dispersion formula for the four-point function. With four momenta we can construct the following Lorentz invariants:
\begin{align}
\zeta_i &= -k_i^{2},
\\
s&=-(k_1+k_2)^{2},
\\
t&=-(k_2+k_3)^{2},
\\
u&=-(k_1+k_3)^{2}.
\end{align}
These variables obey the relation,
\begin{align}
s+t+u=\zeta_1+\zeta_2+\zeta_3+\zeta_4. \label{eq:4pt_relation}
\end{align}
We will take our independent invariants to be $\z_i$, $s$ and $t$. 

At four points, there are multiple dispersion formulas one can write down depending on which variables are analytically continued.
To make a connection with the S-matrix and CFT dispersion formulas, we will disperse in $t$ for fixed $\z_i$ and $s$.
In Appendix \ref{app:Domain_Analyticity} we will review the argument that the retarded correlator $r_{4}(\z_i,s,t,u)$ is analytic in the cut $t$-plane when $\z_i,$ $s<0$. 
Using this analyticity property, we can write down the following contour integral,
\begin{align}
r_4(\zeta_i,s,t,u)=\frac{1}{2\pi i}\oint \frac{dt'}{t'-t}r_4(\zeta_i,s,t',u),\label{eq:dispersion_4pt_Pt1}
\end{align}
where $t$ is complex.

When deforming the contour in $t'$, we can pick up two possible branch cuts. 
One is the $t$-channel cut, which runs from $t=0$ to $t=\infty$.
From \eqref{eq:4pt_relation} there is also a $u$-channel cut which starts at $u=0$ or
\begin{align}
t=\zeta_1+\zeta_2+\zeta_3+\zeta_4-s,
\end{align}
and runs to $t=-\infty$.
To ensure that the $t$- and $u$-channel cuts do not overlap, we take $\sum_{i}\z_i-s<0$. 
If we were studying a dispersion formula for a $2\rightarrow 2$ scattering amplitude $M(s,t)$, then we would have $\z_i>0$ and $s<0$ and the two cuts would necessarily overlap in gapless theories. 
Here we are able to separate the cuts because we are studying correlation functions and can take the external momenta to be spacelike.

If we assume the retarded correlator vanishes in the large $t$ Regge limit
\begin{align}
\lim\limits_{|t|\rightarrow\infty}r_{4}(\zeta_i,s,t,u)=0 \quad \text{ for } \ s, \ \z_i <0, \label{eq:decay4ptRegge}
\end{align}
where $t$ is taken large away from the real line, then we can blow up the contour in \eqref{eq:dispersion_4pt_Pt1} and drop the arcs at infinity. 
Assuming \eqref{eq:decay4ptRegge} holds we find,
\begin{align}
r_4(\zeta_i,s,t,u)=\frac{1}{2\pi i} \bigg(&\int\limits_{0}^{\infty}\frac{dt'}{t'-t}\disc_{t'}r_4(\zeta_i,s,t',u)
+\int\limits_{0}^{\infty}\frac{du'}{u'-u}\disc_{u'}r_4(\zeta_i,s,t,u')
\bigg).
\label{eq:dispersion_4pt_pt2}
\end{align}
This procedure is exactly the same as the one shown in figure \ref{fig:Contour_Deformation}.

Next, we take $t$ to be real from above. In this limit the left hand side of \eqref{eq:dispersion_4pt_pt2} becomes a time-ordered correlator. To also rewrite the integrand in terms of the time-ordered correlator, we use \eqref{eq:ret_corr_to_time} and \eqref{eq:ret_corr_to_anti_time} to find,
\begin{align}
\disc_{t}r_4(\zeta_i,s,t,u)&=-i\disc_{t}\tau_{4}(\zeta_i,s,t,u)
\nonumber
\\
&=-i(\tau_4(\zeta_i,s,t,u)+\tau^*_4(\zeta_i,s,t,u))
\nonumber
\\
&=-2i\Re\tau_4(\zeta_i,s,t,u). \label{eq:disc_and_re}
\end{align}
The discontinuity in $t$ is taken with $\zeta_i,s,u<0$.
This gives a dispersion formula purely in terms of the time-ordered correlator with real momenta,
\begin{align}
\tau_4(\zeta_i,s,t,u)=\frac{1}{2\pi i} \left(\int\limits_{0}^{\infty}\frac{dt'}{t'-t-i\epsilon}\disc_{t'} \tau_4(\zeta_i,s,t',u)
+\int\limits_{0}^{\infty}\frac{du'}{u'-u-i\epsilon}\disc_{u'} \tau_4(\zeta_i,s,t,u')
\right).
\label{eq:dispersion_4pt_pt3}
\end{align}

If the original correlator $r_4$ does not vanish in the limit $|t|\rightarrow \infty$, then we can apply the dispersion argument to a rescaled correlator,
\begin{align}
r_{4}(\zeta_i,s,t,u)\rightarrow \frac{r_{4}(\zeta_i,s,t,u)}{(t-t_1)\ldots (t-t_N)},
\end{align}
where the $t_i$ are some arbitrary subtraction points. We can take $t_i$ to have a small positive imaginary part, in which case the subtracted dispersion formula for the retarded correlator becomes a subtracted dispersion formula for the time-ordered correlator with real momenta,  
\begin{align}
\frac{\tau_{4}(\zeta_i,s,t,u)}{(t-t_1)\ldots (t-t_N)}=\frac{1}{2\pi i}\bigg(\int\limits_{0}^{\infty}\frac{dt'}{t'-t-i\epsilon}\disc_{t'}\frac{\tau_{4}(\zeta_i,s,t',u)}{(t'-t_1)\ldots (t'-t_N)}+\text{($u$-channel)}\bigg).
\end{align}

Finally, we will explain how the integrand of the unsubtracted dispersion formula \eqref{eq:dispersion_4pt_pt3} is related to the integrand of the CFT dispersion formula \cite{Carmi:2019cub}. 
We will leave the arguments that the two dispersion formulas are equivalent to Section \ref{sec:equivalence}. 
Here we will show that for spacelike external momenta, the right hand side of \eqref{eq:dispersion_4pt_pt3} depends on the same causal double-commutators as the CFT dispersion formula \eqref{eq:introCFTdispersion}.

To prove this, we will follow the same procedure used to study the three-point function. 
First, the identity \eqref{eq:disc_and_re} says that taking a single $t$-channel discontinuity of the time-ordered correlator gives $\Re \tau_4(p_i)$. 
To relate the real part of the time-ordered correlator to a double-commutator, we will take the identity \eqref{eq:QFTopticalV2} and set $n=4$:
\begin{align}
2\Re \<T[\f(p_1)\ldots \f(p_4)]\>= & \ \<\f(p_1)T[\f(p_2)\f(p_3)\f(p_4)]\>+\<\overline{T}[\f(p_2)\f(p_3)\f(p_4)]\f(p_1)\>
\nonumber \\ & \ -\left[\<\overline{T}[\f(p_2)\f(p_3)]T[\f(p_1)\f(p_4)]\>+(p_2,p_3)\leftrightarrow (p_1,p_4)\right]
\nonumber \\ & \ +\text{(partitions)},
\label{eq:QFToptical4pt}
\end{align}
where we suppressed the other partitions of the external operators.

The dispersion formula \eqref{eq:dispersion_4pt_pt3} was derived under the assumption $\z_i$, $s<0$. In addition, in the $t$-channel term of \eqref{eq:dispersion_4pt_pt3} we have $u<0$. This means the momenta $p_i$, $p_1+p_2$, and $p_1+p_3$ are spacelike in the $t$-channel term of the dispersion formula.
Under these assumptions, the positive spectrum condition \eqref{eq:pos_energy} implies that the right hand side of \eqref{eq:QFToptical4pt} only contains two non-zero terms:
\begin{align}
2\Re \<T[\f(p_1)\ldots \f(p_4)]\>=-\big(\<\overline{T}[\f(p_2)\f(p_3)]T[\f(p_1)\f(p_4)]\>+(p_2,p_3)\leftrightarrow (p_1,p_4)\big).
\label{eq:4pttoTbT}
\end{align}
We can now use \eqref{eq:AtoT} and \eqref{eq:AtoTb} to rewrite the (anti-)time-ordered products in \eqref{eq:4pttoTbT} in terms of the advanced product,
\begin{align}
2\Re \<T[\f(p_1)\ldots \f(p_4)]\>=& \ -\big(\<A[\f(p_2);\f(p_3)]A[\f(p_1);\f(p_4)]\>+(p_2,p_3)\leftrightarrow (p_1,p_4)\big).  \label{eq:ReTtoDC}
\end{align}
Combining \eqref{eq:disc_and_re} and \eqref{eq:ReTtoDC} yields,
\begin{align}
\disc_{t} \<T[\f(p_1)\ldots \f(p_4)]\>=& \ -\big(\<A[\f(p_2);\f(p_3)]A[\f(p_1);\f(p_4)]\> +(p_2,p_3)\leftrightarrow (p_1,p_4)\big). \label{eq:disctToDC} 
\end{align}
One can repeat an identical analysis for the $u$-channel discontinuity by making the replacement $p_1\leftrightarrow p_2$. 
This proves that the four-point QFT dispersion formula \eqref{eq:dispersion_4pt_pt3} can be rewritten in terms of advanced double-commutators.

The momentum space dispersion formula \eqref{eq:dispersion_4pt_pt3} and the CFT dispersion formula \eqref{eq:introCFTdispersion} both express the full correlator as an integral over $t$- and $u$-channel, causal double-commutators.
One na\"ive difference between these formulas is that the momentum space dispersion formula appears to depend on more double-commutators than the CFT dispersion formula.
From \eqref{eq:disctToDC} we see that the $t$- and $u$-channel terms in the momentum space formula  \eqref{eq:dispersion_4pt_pt3} each depend on a sum of double-commutators.
On the other hand, in the CFT dispersion formula \eqref{eq:introCFTdispersion} the $t$- and $u$-channel terms each contain a single double-commutator. 
However, this is only a difference in presentation.
The kernel $K$ in \eqref{eq:introCFTdispersion} is only nonzero when the pairs of positions $(y_1,y_4)$ and $(y_2,y_3)$ are spacelike separated from each other, see the discussion below equation 1.7 of \cite{ssw}.\footnote{The CFT dispersion formula was derived in \cite{Carmi:2019cub} by plugging the Lorentzian inversion formula into the Euclidean partial wave expansion. Therefore, the kernel $K$ of the CFT dispersion formula also only has support when the pairs $(y_1,y_4)$ and $(y_2,y_3)$ are spacelike separated.}
We are then free to make the following replacement in \eqref{eq:introCFTdispersion}:
\begin{align}
\<[\f(y_1),\f(y_4)]_A&[\f(y_2),\f(y_3)]_A\>
\nonumber
\\
&\Longrightarrow \frac{1}{2}\big(\<[\f(y_1),\f(y_4)]_A[\f(y_2),\f(y_3)]_A\>+\<[\f(y_2),\f(y_3)]_A[\f(y_1),\f(y_4)]_A\>\big),
\end{align}
which matches the integrand of \eqref{eq:dispersion_4pt_pt3} after using \eqref{eq:disctToDC}.
 
This is not a proof that the momentum space dispersion formula, applied to a CFT correlation function, is equivalent to the conformal dispersion formula. 
In order to show that, we need to understand on what space of functions each dispersion formula is valid, which we will consider in Section \ref{sec:equivalence}.
It does prove however that both dispersion formulas depend on the same physical input.
It is also important to emphasize that the momentum space formula \eqref{eq:dispersion_4pt_pt3} is valid for massive QFTs as well.

\subsection{Five-Point Function}
\label{ssec:five-point}
Finally, in this section we will study single-variable dispersion formulas for the five-point, time-ordered correlator. 
We will follow the same procedure used for the four-point dispersion formula.

We can take our independent invariants to be $\z_i$, for $i=1,\ldots,5$, and the Mandelstams $\{s_{13},s_{14},s_{15},s_{23},s_{35}\}$. The five remaining Lorentz invariants are given by the linear relations,
\begin{align}
s_{12}&=2 \z_1+\z_2+\z_3+\z_4+\z_5-s_{13}-s_{14}-s_{15},\label{eq:s5pt_relations1}
\\
s_{24}&=-\z_1-\z_3-\z_5+s_{13}+s_{15}+s_{35},\label{eq:s5pt_relations2}
\\
s_{25}&=\z_2+\z_3+\z_5+s_{14}-s_{23}-s_{35},\label{eq:s5pt_relations3}
\\
s_{34}&=\z_1+\z_2+2 \z_3+\z_4+\z_5-s_{13}-s_{23}-s_{35},\label{eq:s5pt_relations4}
\\
s_{45}&=\z_1+\z_4+\z_5-s_{14}-s_{15}+s_{23}.\label{eq:s5pt_relations5}
\end{align}
In Appendix \ref{app:Domain_Analyticity} we will show that the five-point retarded correlator is analytic in the cut $s_{35}$ plane if we assume the other independent Lorentz invariants are negative, or below threshold.

We will study the following contour integral representation of the retarded correlator,
\begin{align}
r_{5}(\z_i,s_{35},\ldots)=\frac{1}{2\pi i}\oint \frac{ds_{35}'}{s_{35}'-s_{35}}r_5(\z_i,s'_{35},\ldots).\label{eq:five_pt_r_contour}
\end{align}
We have suppressed the independent Mandelstam invariants $\{s_{13},s_{14},s_{15},s_{23}\}$ that are held fixed at their external values.

There are four branch cuts in the complex $s_{35}$ plane:
\begin{align}
s_{35}&\in [0,\infty), \label{eq:branch_cut_1}
\\
s_{35}&\in [\z_1+\z_3+\z_5-s_{13}-s_{15},\infty), \label{eq:branch_cut_2}
\\
s_{35}&\in (-\infty,\z_2+\z_3+\z_5+s_{14}-s_{23}], \label{eq:branch_cut_3}
\\
s_{35}&\in (-\infty,\z_1+\z_2+2 \z_3+\z_4+\z_5-s_{13}-s_{23}]. \label{eq:branch_cut_4}
\end{align}
The first cut \eqref{eq:branch_cut_1} corresponds to states being created in the $s_{35}$ channel. The other three branch cuts \eqref{eq:branch_cut_2}-\eqref{eq:branch_cut_4} correspond to the $s_{24}$, $s_{25}$ and $s_{34}$ channels, respectively, and are found using \eqref{eq:s5pt_relations2}-\eqref{eq:s5pt_relations4}. We will assume the independent invariants are chosen such that the two cuts which run to positive infinity, \eqref{eq:branch_cut_1} and \eqref{eq:branch_cut_2}, do not overlap with the cuts that run to negative infinity, \eqref{eq:branch_cut_3} and \eqref{eq:branch_cut_4}.

If we assume the five-point correlator vanishes in the large $s_{35}$ limit,
\begin{align}
\lim\limits_{|s_{35}|\rightarrow\infty}r_{5}(\z_i,s_{35},\ldots)=0,
\end{align}
with the other independent invariants held negative and fixed, then we can write down an unsubtracted dispersion formula for the retarded five-point function. Taking $s_{35}$ to be real from above, this gives an unsubtracted dispersion formula for the time-ordered correlator with real momenta,
\begin{align}
\tau_{5}(\z_i,s_{35},\ldots)=\frac{1}{2\pi i}\bigg( \ &\int\limits_{s^{+}_{35}}^{\infty} \frac{ds'_{35}}{s'_{35}-s_{35}-i\epsilon} \disc_{s'_{35}}\tau_{5}(\z_i,s'_{35},\ldots)
\nonumber
\\&+\int\limits_{-\infty}^{s^{-}_{35}} \frac{ds'_{35}}{s'_{35}-s_{35}-i\epsilon} \disc_{s'_{35}}\tau_{5}(\z_i,s'_{35},\ldots)\bigg).\label{eq:5pt_dispersion}
\end{align}
The $s^{\pm}_{35}$ are defined by,
\begin{align}
s^{+}_{35}&=\min(0,\z_1+\z_3+\z_5-s_{13}-s_{15}),
\\
s^{-}_{35}&=\max(\z_2+\z_3+\z_5+s_{14}-s_{23},\z_1+\z_2+2 \z_3+\z_4+\z_5-s_{13}-s_{23}).
\end{align}
To rewrite \eqref{eq:5pt_dispersion} in terms of commutators, we use \eqref{eq:ret_corr_to_time} and \eqref{eq:ret_corr_to_anti_time} to find,
\begin{align}
\disc_{s_{35}}r_{5}(\z_i,s_{ij})&=\disc_{s_{35}}\tau_{5}(\z_i,s_{ij})
\nonumber \\
&=\tau_{5}(\z_i,s_{ij})-\tau^{*}_5(\z_i,s_{ij})
\nonumber \\
&=2i\Im\tau_{5}(\z_i,s_{ij}).\label{eq:disc_to_Im}
\end{align}
We will now relate the imaginary part of the time-ordered correlator to a double advanced product. For simplicity, we will focus on the first term in \eqref{eq:5pt_dispersion}. In this configuration we have $s_{35}\geq 0$ and/or $s_{24}\geq 0$, while all the other Lorentz invariants are negative. Using the positive spectrum condition \eqref{eq:pos_energy}, this implies the unitarity condition  \eqref{eq:QFTopticalV2} becomes,
\begin{align}
2i\Im\<T[\f(p_1)\ldots \f(p_5)]\>=& \ \<\overline{T}[\f(p_1)\f(p_2)\f(p_4)]T [\f(p_3)\f(p_5)]\>
\nonumber \\ &-\<\overline{T}[\f(p_2)\f(p_4)]T [\f(p_1)\f(p_3)\f(p_5)]\> + (p_2,p_4)\leftrightarrow(p_3,p_5).\label{eq:im5pttoTTb}
\end{align}
Using \eqref{eq:AtoT} and \eqref{eq:AtoTb} we can rewrite the right hand side of \eqref{eq:im5pttoTTb} using advanced commutators. Combined with the relation \eqref{eq:disc_to_Im} this gives,
\begin{align}
\disc_{s_{35}}\<T[\f(p_1)\ldots \f(p_5)]\>=& \ i\bigg(\<A[\f(p_2);\f(p_1)\f(p_4)]A[\f(p_3);\f(p_5)]\>
\nonumber
\\
&+\<A[\f(p_2);\f(p_4)]A[\f(p_3);\f(p_1)\f(p_5)]\>\bigg)+ (p_2,p_4)\leftrightarrow(p_3,p_5), \label{eq:disc5pt_to_Adv}
\end{align}
in the configuration where $s_{35}\geq0$ and/or $s_{24}\geq0$, but the other invariants are below threshold.
It is straightforward to follow the same procedure for the second term of \eqref{eq:5pt_dispersion}.
Finally, we have found that the integrand of the five-point dispersion formula \eqref{eq:5pt_dispersion} depends on a double advanced product.
It would be interesting to compare this result against the five-point CFT dispersion formula, which is not currently known.


\section{Derivation 2: Dispersion from Largest Time}
\label{sec:disp_from_largest_time}
In this section we will derive dispersion formulas using the largest time equation \cite{Veltman:1963th,tHooft:1973wag}. 
In \cite{Remiddi:1981hn} the largest time equation and an infinite momentum limit was used to derive Lorentz invariant dispersion formulas for Feynman diagrams.
In this section we will show how the same dispersion formulas can be derived for $n$-point correlation functions.
We will not need to assume that the theory is weakly coupled or has a dual AdS description.\footnote{For weakly coupled theories or theories with a weakly coupled AdS dual, we can go between the correlation function and graphical derivations using the cutting rules.}
We also show how the $n$-point dispersion formula can be rewritten in terms of a double advanced product.

\subsection{Largest Time Equations}

In Section \ref{sec:OperatorOrderings} we used an axiomatic definition of the time-ordered product that does not rely on $\theta$-functions \cite{Bogolyubov:1990kw,steinmann1968}. One of the axioms for the time-ordered product was the unitarity condition \eqref{eq:QFTopticalV2}. This identity is one half of a full-fledged unitarity method: it explains how to compute the real or imaginary part of a correlator in terms of ``cut", partially time-ordered correlators, but it does not explain how to compute the time-ordered correlator itself.

Can we also use the axiomatic properties of the time-ordered product to write down a dispersion formula? 
The answer is yes but, unlike the unitarity condition \eqref{eq:QFTopticalV2}, the resulting dispersion formula will be sensitive to the UV properties of our theory. 
In Section \ref{sec:disp_from_analyticity} the UV sensitivity came from deforming a contour in the complex $t$-plane and requiring that the arcs at infinity vanish.
In this section it will instead come from properly defining products of $\theta$-functions and correlation functions.

Before writing down a dispersion formula using the definition of the time-ordered product, it is useful to motivate why the unitarity condition \eqref{eq:QFT_Optical} holds. We will show \eqref{eq:QFT_Optical} follows from the factorization condition \eqref{eq:Tfactorization} when no two points are coincident. For coincident point terms, \eqref{eq:QFT_Optical} has to be taken as an independent axiom. Without loss of generality, we assume one operator has the largest time, for example $x_{n}^0\geq x_{i}^0$ for all $i$. We then split the sum \eqref{eq:QFT_Optical} into two terms, depending on whether $\f(x_n)$ appears in the time or anti-time-ordered product:
\begin{align}
&\sum\limits_{r=0}^{n}(-1)^{r}\sum\limits_{\alpha\in\Pi_r(n)}\<\overline{T}[\f(x_{\alpha_1})\ldots \f(x_{\alpha_r})]T[\f(x_{\alpha_{r+1}})\ldots \f(x_{\alpha_n})]\>
\nonumber
\\
=&\sum\limits_{r=0}^{n-1}(-1)^{r}\sum\limits_{\alpha\in\Pi_r(n-1)}\bigg(\<\overline{T}[\f(x_{\alpha_1})\ldots \f(x_{\alpha_r})]T[\f(x_{\alpha_{r+1}})\ldots \f(x_{\alpha_{n-1}})\f(x_n)]\>
\nonumber \\
&\hspace{1.3in} -\<\overline{T}[\f(x_n)\f(x_{\alpha_1})\ldots \f(x_{\alpha_r})]T[\f(x_{\alpha_{r+1}})\ldots \f(x_{\alpha_{n-1}})]\>\bigg).
\label{eq:provingunitarity}
\end{align}
As a reminder, $\Pi_{r}(n)$ is the set of partitions of $\{1,\ldots,n\}$ into two sets of size $r$ and $n-r$.
Since $\f(x_n)$ has the largest time, we can write,
\begin{align}
T[\f(x_{\alpha_{r+1}})\ldots \f(x_{\alpha_{n-1}})\f(x_n)]&=\f(x_n)T[\f(x_{\alpha_{r+1}})\ldots \f(x_{\alpha_{n-1}})],
\\[3pt]
\overline{T}[\f(x_n)\f(x_{\alpha_1})\ldots \f(x_{\alpha_r})]&=\overline{T}[\f(x_{\alpha_1})\ldots \f(x_{\alpha_r})]\f(x_n).
\end{align}
Using these identities, we see the two terms in \eqref{eq:provingunitarity} cancel against each other. Our assumption that $\f(x_n)$ has the largest time was arbitrary and we can equally assume any other operator has the largest time. This proves the original unitarity condition at non-coincident points: 
\begin{align}
\sum\limits_{r=0}^{n}(-1)^{r}\sum\limits_{\alpha\in\Pi_r(n)}\<\overline{T}[\f(x_{\alpha_1})\ldots \f(x_{\alpha_r})]T[\f(x_{\alpha_{r+1}})\ldots \f(x_{\alpha_n})]\>=0.\label{eq:QFT_OpticalRestated}
\end{align}
For this reason, \eqref{eq:QFT_OpticalRestated} is often referred to as the largest-time equation \cite{Veltman:1963th,tHooft:1973wag}.

We can use similar manipulations to write down dispersion formulas.
We start by making the assumption that $\f(x_n)$ always has a smaller time component than another operator.
For example, we can assume $x_{n}^{0}<x_{n-1}^{0}$.
In this case, we do not have to sum over all possible partitions of the external operators as we did in \eqref{eq:QFT_OpticalRestated}. 
Instead, we can sum over all partitions where $\f(x_n)$ is always in the time-ordered product:
\begin{align}
\theta(x_{n-1}^{0}-x_{n}^{0})\sum\limits_{r=0}^{n-1}(-1)^r\sum\limits_{\alpha\in\Pi_{r}(n-1 )}\<\overline{T}[\f(x_{\alpha_1})...\f(x_{\alpha_{r}})]T[\f(x_{\alpha_{r+1}})...\f(x_{\alpha_{n-1}})\f(x_{n})]\>=0.\label{eq:dispersionsetupPt1}
\end{align}
We multiplied by $\theta(x_{n-1}^{0}-x_{n}^{0})$ to enforce the condition that $x_{n}^{0}<x_{n-1}^{0}$. 
To prove \eqref{eq:dispersionsetupPt1}, we can assume any other operator, say $\f(x_1)$, has the largest time. 
We can then split \eqref{eq:dispersionsetupPt1} into two terms as we did in \eqref{eq:provingunitarity} and use the defining properties of $T$ and $\overline{T}$ to show the resulting sum cancels. 
The choice of which operator has the largest time was arbitrary, and therefore \eqref{eq:dispersionsetupPt1} holds in general. 
However, in comparison to \eqref{eq:QFT_OpticalRestated}, we have to assume that products of $\theta$-functions and correlation functions are well-defined in order for \eqref{eq:dispersionsetupPt1} to hold.

There is an identical equality with the same form of \eqref{eq:dispersionsetupPt1} but with $x_{n}\leftrightarrow x_{n-1}$. The sum of these two identities yields:
\begin{align}
0=&\sum\limits_{r=0}^{n-2}(-1)^{r+1}\sum\limits_{\alpha \in \Pi_{r}(n-2)}\bigg[  \<\overline{T}[\f(x_{\alpha_1})...\f(x_{\alpha_{r}})]T[\f(x_{\alpha_{r+1}})...\f(x_{\alpha_{n-2}})\f(x_{n-1})\f(x_n)]\>
\nonumber 
\\   &\hspace{.36in}-\bigg( \theta(x_{n}^0-x_{n-1}^0)\<\overline{T}[\f(x_n)\f(x_{\alpha_1})...\f(x_{\alpha_{r}})]T[\f(x_{\alpha_{r+1}})...\f(x_{\alpha_{n-2}})\f(x_{n-1})]\>
\nonumber 
\\
  &\hspace{4in}+ (x_{n-1}\leftrightarrow x_{n})\bigg)\bigg].
\label{eq:dispersion_Largesttime}
\end{align}
This formula is similar in structure to \eqref{eq:QFT_OpticalRestated}, but with two important differences. The first is that the fully anti-time-ordered correlator does not appear in \eqref{eq:dispersion_Largesttime}. Each term in the above identity contains at least one operator in the time-ordered product. 
On the other hand, the time-ordered correlator does appear in the first line of \eqref{eq:dispersion_Largesttime} when we set $r=0$. In this sense, \eqref{eq:dispersion_Largesttime} is already a position-space dispersion formula: it expresses the time-ordered correlator in terms of the factorized, partially time-ordered correlators. The second important difference is that \eqref{eq:dispersion_Largesttime} is not manifestly Lorentz invariant due to the explicit time dependence of the $\theta$-functions.

To eventually derive a more standard, momentum space dispersion formula, we need to use the Fourier space representation of the $\theta$-function:
\begin{align}
\theta(x^0)=\int\limits_{-\infty}^{\infty} \frac{dk^0}{2\pi}\frac{i}{k^0+i\epsilon}e^{-ik^0x^0}=\int \frac{d^{d}k}{2\pi}\frac{i}{k^0+i\epsilon}\delta(\vec{k})e^{ik\cdot x}, \label{eq:theta_k_space}
\end{align}
where we use $\vec{k}$ to indicate momenta along the $d-1$ spatial directions. We introduced the trivial spatial integrals to simplify later expressions.

We can then take the Fourier transform of \eqref{eq:dispersion_Largesttime} to find, 
\begin{align}
0=&\sum\limits_{r=0}^{n-2}(-1)^{r+1}\sum\limits_{\alpha \in \Pi_{r}(n-2)}\bigg[\ll \overline{T}[\f(p_{\alpha_1})...\f(p_{\alpha_{r}})]T[\f(p_{\alpha_{r+1}})...\f(p_{\alpha_{n-2}})\f(p_{n-1})\f(p_n)]\rr
\nonumber 
\\
-&\int\frac{d^{d}k}{2\pi}\frac{i}{k^0+i\epsilon}\delta(\vec{k})\bigg(\ll\overline{T}[\f(p_n+k)\f(p_{\alpha_1})...\f(p_{\alpha_{r}})]T[\f(p_{\alpha_{r+1}})...\f(p_{\alpha_{n-2}})\f(p_{n-1}-k)]\rr
\nonumber 
\\
&\hspace{4.3in}+(p_{n-1}\leftrightarrow p_n)\bigg)\bigg], \label{eq:dispersion_Largesttime_kspace}
\end{align}
where we used the double-bracket notation defined in \eqref{eq:double_bracket}.
The fact we broke Lorentz invariance in \eqref{eq:dispersion_Largesttime} by choosing a time coordinate is now manifested in \eqref{eq:dispersion_Largesttime_kspace} by the fact our momentum integral runs solely over $k^0$. 

We can simplify \eqref{eq:dispersion_Largesttime_kspace} if we assume the external momenta $p_i$, for $i=1,\ldots,n-2$, and their sums are spacelike. For this choice of momenta, the positive spectrum condition \eqref{eq:pos_energy} implies all terms with $r\geq1$ in the first line of \eqref{eq:dispersion_Largesttime_kspace} vanish. This gives,
\begin{align}
\ll T[\f(p_1)\ldots\f(p_n)]\rr=&\sum\limits_{r=0}^{n-2}(-1)^{r}\sum\limits_{\alpha \in \Pi_{r}(n-2)}\bigg[\int\frac{d^{d}k}{2\pi}\frac{i}{k^0+i\epsilon}\delta(\vec{k})
\nonumber
\\
&\bigg(\ll\overline{T}[\f(p_n+k)\f(p_{\alpha_1})...\f(p_{\alpha_{r}})]T[\f(p_{\alpha_{r+1}})...\f(p_{\alpha_{n-2}})\f(p_{n-1}-k)]\rr
\nonumber 
\\
&\hspace{2.6in}+(p_{n-1}\leftrightarrow p_n)\bigg)\bigg], \label{eq:dispersion_Largesttime_kspaceV2}
\end{align}
under the condition that $\sum\limits_{i\in I}p_{i}$ is spacelike for all $I\subseteq\{1,\ldots,n-2\}$.
In the following sections we will use exactly these kinematics to derive Lorentz invariant relations from \eqref{eq:dispersion_Largesttime_kspaceV2} \cite{Remiddi:1981hn}. 
Another helpful feature of these kinematics is they let us rewrite \eqref{eq:dispersion_Largesttime_kspaceV2} in terms of the advanced product. Using the operator identities \eqref{eq:AtoT} and \eqref{eq:AtoTb}, we find that each term on the right hand side of \eqref{eq:dispersion_Largesttime_kspaceV2} is equal to a double advanced product,
\begin{align}
\<\overline{T}[\f(p_n+k)\f(p_{\alpha_1})&...\f(p_{\alpha_{r}})]T[\f(p_{\alpha_{r+1}})...\f(p_{\alpha_{n-2}})\f(p_{n-1}-k)]\>
\nonumber \\
&=\<A[\f(p_n+k);\f(p_{\alpha_1})...\f(p_{\alpha_{r}})]A[\f(p_{n-1}-k);\f(p_{\alpha_{r+1}})...\f(p_{\alpha_{n-2}})]\>.\label{eq:TbTtoAA}
\end{align}
In the following sections we will continue to use \eqref{eq:dispersion_Largesttime_kspaceV2} and express all formulas in terms of the (anti-)time-ordered products. One should keep in mind however that \eqref{eq:TbTtoAA} implies the dispersion formulas can be rewritten using advanced commutators.

In order to arrive at \eqref{eq:dispersion_Largesttime_kspaceV2}, we used $\theta$-function identities which may not be valid when the $\theta$-functions multiply distributions. 
We can define the product of distributions in position space by taking the Fourier transform of their convolution in momentum space  \cite{doi:10.1063/1.1703706}. For example, $\theta(x)A(x)$ is defined by:
\begin{align}
\theta(x)A(x)&=\int\limits_{-\infty}^{\infty}\frac{dp}{2\pi}e^{-ip\cdot x}\int\limits_{-\infty}^{\infty}\frac{dk}{2\pi}\tilde{\theta}(k)\tilde{A}(p-k)
\nonumber \\
&=i\int\limits_{-\infty}^{\infty}\frac{dp}{2\pi}e^{-ip\cdot x}\int\limits_{-\infty}^{\infty}\frac{dk}{2\pi}\frac{\tilde{A}(p-k)}{k+i\epsilon}.\label{eq:1dispersion}
\end{align}
We can recognize that the $k$ integral is a $1d$ dispersion integral. In order for the $k$ integral to converge we need 
\begin{align}
\lim\limits_{p\rightarrow\infty}\tilde{A}(p)=0.
\end{align}
That is, we need $\tilde{A}(p)$ to have soft UV behavior in order for $\theta(x)A(x)$ to be well-defined. 
For example, if $A(x)$ is a $\delta$-function, or derivatives thereof, then $\tilde{A}(p)$ is a polynomial and the $k$ integral in \eqref{eq:1dispersion} diverges. 
Since the identity \eqref{eq:dispersion_Largesttime} involves the product of $\theta(x)$ with various correlation functions, \eqref{eq:dispersion_Largesttime} is only valid if our correlators have sufficiently soft UV behavior.

In general, correlation functions do not decay at large energies, but they are assumed to be tempered distributions, or polynomially bounded \cite{Streater:1989vi}.
If $\tilde{A}(p)$ in \eqref{eq:1dispersion} is polynomially bounded at large $p$, then one can define a subtracted distribution, $\tilde{A}_{\text{sub}}(p)$, which does decay at large $p$. 
For the subtracted distribution, it does make sense to consider the product $\theta(x)A_{\text{sub}}(x)$.
This explains how the dispersion formula \eqref{eq:dispersion_Largesttime} is dependent on the UV properties of the theory.

The derivation of the four-point dispersion formula given in the following section will depend on different high-energy limits in comparison to the one given in Section \ref{sec:disp_from_analyticity}.
In Section \ref{sec:disp_from_analyticity} we saw that the unsubtracted dispersion formula \eqref{eq:dispersion_4pt_pt3} is valid if the correlator decays in the Regge limit, or when we take $t$ large with $\z_i$ and $s$ held fixed.
By contrast, requiring that the $k^0$ integral in \eqref{eq:dispersion_Largesttime_kspaceV2} converges puts bounds on how fast the correlator can scale when we take $\z_{n-1}$, $\z_n$, $s_{1n}$ and $s_{2n}$ to be large at the same rate.
Therefore, the number of subtractions needed in order for \eqref{eq:dispersion_Largesttime_kspaceV2} to hold will not depend on how the correlator scales in the Regge limit.
For the remainder of this section we will assume that the $k$-integral in \eqref{eq:dispersion_Largesttime_kspaceV2} converges, although it would be interesting to understand the number of subtractions needed in general theories.\footnote{In perturbation theory, the $k^0$ integral diverges when a loop amplitude is UV divergent and needs to be regularized \cite{tHooft:1973wag}.}

\subsection{Four-Point Function}
\label{sec:FourPtTimeOrdering}
In this section we will use \eqref{eq:dispersion_Largesttime_kspaceV2} to find a Lorentz invariant, four-point dispersion formula. We will skip the analysis of two- and three-point functions as these are covered in detail in previous works. For example, at $n=2$ the position space formula \eqref{eq:dispersion_Largesttime} reduces to the original definition of the time-ordered product \eqref{eq:TOrdering},
\begin{align}
\<T[\f(x_1)\f(x_2)]\>=\theta(x_{12}^0)\<\f(x_1)\f(x_2)\>+(1\leftrightarrow 2). \label{eq:Timeordered_twopt}
\end{align}
The relation between \eqref{eq:Timeordered_twopt} and the two-point dispersion formula, or K\"all\'en-Lehmann spectral representation, \eqref{eq:2ptdispfromanalyticity} is well-known \cite{tHooft:1973wag,Veltman:1994wz,Remiddi:1981hn}.\footnote{For applications of the K\"all\'en-Lehmann spectral representation to the QFT bootstrap see \cite{Karateev:2020axc}.} At three-points, the derivation of the Lorentz invariant dispersion formula from the non-Lorentz invariant one was carried out in \cite{Remiddi:1981hn}. We will focus on four-point correlators as certain details at four-points were left implicit in \cite{Remiddi:1981hn} and also because the generalization to $n>4$-point correlators will be manifest.

At four-points we will use the same notation as before,
\begin{align}
\z_i&=-p_i^2,
\\
s&=-(p_1+p_2)^2,
\\
t&=-(p_2+p_3)^2,
\\
u&=-(p_1+p_3)^2.
\end{align}
We will work in the following kinematics,
\begin{align}
p_1&=(P,P,p_1^\perp),\label{eq:4ptKinematics1}
\\
p_2&=(-P,-P,p_2^\perp),
\\
p_3&=(0,0,p_3^\perp),
\\
p_4&=(0,0,p_4^\perp). \label{eq:4ptKinematics4}
\end{align}
Here $p_i^{\perp}$ are the components of the momenta in the $d-2$ transverse, spatial directions.\footnote{In this section we assume $d\geq4$ so that we can assume the momenta are linearly independent.}
We will leave momentum conservation along these transverse spatial dimensions implicit.
Later we will take the limit $P\rightarrow \infty$.

In the above configuration, all the external Lorentz invariants are below threshold, $\z_{i}<0$ and $s,t,u<0$. We can then use \eqref{eq:dispersion_Largesttime_kspaceV2} for $n=4$: 
\begin{align}
\ll T[\f(p_1)&\f(p_2)\f(p_3)\f(p_4)]\rr
=\int \frac{d^{d}k}{(2\pi)}\frac{i}{k^0+i\epsilon}\delta(\vec{k})
\bigg[
\ll\f(p_4+k)T[\f(p_1)\f(p_2)\f(p_3-k)]\rr
\nonumber
\\[3pt]
&+\ll \overline{T}[\f(p_1)\f(p_2)\f(p_4+k)]\f(p_3-k)\rr
- \ll \overline{T}[\f(p_2)\f(p_3+k)]T[\f(p_1)\f(p_4-k)]\rr
\nonumber
\\[3pt]
&-  \ll \overline{T}[\f(p_2)\f(p_4+k)]T[\f(p_1)\f(p_3-k)]\rr  + (p_3\leftrightarrow p_4) \bigg].
\label{eq:4ptDispLargTimeV1}
\end{align}

To recover a Lorentz invariant relation, we will follow \cite{Remiddi:1981hn} and study the infinite $P$ limit of \eqref{eq:4ptDispLargTimeV1} under the integral. Given our choice of kinematics, we will find a dispersion formula in $t$ at fixed $\z_i$, $s<0$.\footnote{By choosing other kinematics, one can also find dispersion relations in $\z_i$ \cite{Remiddi:1981hn}.}
In this section we focus on the third and fourth correlators of \eqref{eq:4ptDispLargTimeV1}, which in the large $P$ limit will be non-zero and yield the expected Lorentz invariant dispersion formula \eqref{eq:dispersion_4pt_pt3}.
A sufficient set of conditions to set the remaining $6$ correlators to zero is given in Appendix \ref{app:details}.

To find a Lorentz invariant dispersion formula, it will be convenient to factor out the explicit energy dependence of the partially time-ordered correlators:
\begin{align}
\ll T[\f(p_1)\f(p_2)\f(p_3)\f(p_4)]\rr&=\tau_{4}(\zeta_i,s,t,u), \label{eq:4pttaulargest}
\\[3pt]
\vspace{.3cm}
\ll \overline{T}[\f(p_2)\f(p_3)]T[\f(p_1)\f(p_4)]\rr&=-\theta(p_1^0+p_4^0)\disc_{t}\tau_{4}(\zeta_i,s,t,u),\label{eq:largest_time_disct}
\\[3pt]
\ll \overline{T}[\f(p_2)\f(p_4)]T[\f(p_1)\f(p_3)]\rr&=-\theta(p_1^0+p_3^0)\disc_{u}\tau_{4}(\zeta_i,s,t,u).
\label{eq:largest_time_discu}
\end{align}
These identities follow from \eqref{eq:QFTopticalV2} and \eqref{eq:disc_and_re}.\footnote{Strictly speaking, in this section it is not necessary to introduce discontinuities as all the external momenta are real. However, using discontinuities will allow us to match onto the results of Section \ref{sec:disp_from_analyticity}.}

To rewrite the contribution of $\ll \overline{T}[\f(p_2)\f(p_3+k)]T[\f(p_1)\f(p_4-k)]\rr$ to the dispersion formula \eqref{eq:4ptDispLargTimeV1} in a Lorentz invariant way, we need to remove the explicit energy dependence in \eqref{eq:largest_time_disct}. To do this, we make the following change of variables:
\begin{align}
t'=-(p_1+p_4-k)^2\Longrightarrow k^0=P-\sqrt{P^{2}+t'-t}.\label{eq:COV4pt_tchannel}
\end{align}
We use primed variables to denote a Lorentz invariant that is shifted by $k$ and continue to use unprimed variables for the external invariants.

After making this change of variables, we will take the limit $P\rightarrow \infty$. 
In this limit the measure terms of \eqref{eq:4ptDispLargTimeV1} are equal to the expected Lorentz invariant quantity,
\begin{align}
\lim_{P\rightarrow \infty}\frac{dk^0}{2\pi}\frac{i}{k^0-i\epsilon}=\frac{dt'}{2\pi i}\frac{1}{t'-t-i\epsilon}.
\end{align} 

Next, we will study how the internal Lorentz invariants scale in the large $P$ limit. 
From the change of variables \eqref{eq:COV4pt_tchannel}, we find that in this limit the primed invariants become equal to the corresponding external invariants,\footnote{The momenta $p_{1,2}$ are not shifted by $k$, so the invariants $\z_{1,2}$ and $s$ are unaffected by the $k$ integral and large $P$ limit.}
\begin{align}
\lim\limits_{P\rightarrow\infty} \z_{3,4}'&=\z_{3,4},
\\
\lim\limits_{P\rightarrow\infty} \{t',u'\}&=\{t,u\}.
\end{align}
Therefore in the large $P$ limit, the contribution of $\ll \overline{T}[\f(p_2)\f(p_3+k)]T[\f(p_1)\f(p_4-k)]\rr$ to the dispersion formula \eqref{eq:4ptDispLargTimeV1} is,
\begin{align}
\lim\limits_{P\rightarrow\infty}\int \frac{d^{d}k}{(2\pi)}\frac{-i}{k^0+i\epsilon}\delta(\vec{k}) \ll \overline{T}[\f(p_2)\f(p_3+k)]T[\f(p_1)\f(p_4-k)]\rr
\nonumber
\\
= \frac{1}{2\pi i}\int\limits_{0}^{\infty}\frac{dt'}{t'-t-i\epsilon}\disc_{t'}\tau_{4}(\z_i,s,t',u).
\end{align}
This is exactly the $t$-channel contribution to the Lorentz invariant relation \eqref{eq:dispersion_4pt_pt3}.

We can repeat the same analysis for $\< \overline{T}[\f(p_2)\f(p_4+k)]T[\f(p_1)\f(p_3-k)]\>$, which amounts to taking $t\leftrightarrow u$. 
Combining the two terms yields the Lorentz invariant dispersion formula,
\begin{align}
\tau_{4}(\zeta_i,s,t,u)=\frac{1}{2\pi i}\left(\int\limits_{0}^{\infty}\frac{dt'}{t'-t-i\epsilon}\disc_{t'}\tau_{4}(\zeta_i,s,t',u)+\int\limits_{0}^{\infty}\frac{du'}{u'-u-i\epsilon}\disc_{u'}\tau_{4}(\zeta_i,s,t,u')\right).\label{eq:dispersion_4pt_LT}
\end{align}
This matches the previous result \eqref{eq:dispersion_4pt_pt3} that was derived using analyticity arguments.

In the remainder of this section, we will give a set of sufficient conditions for the remaining $6$ correlators in \eqref{eq:4ptDispLargTimeV1} to vanish at large $P$. We will leave the details to Appendix \ref{app:details}.
The first class of correlators we can consider are those where a single operator acts directly on the left or right vacuum, for example the first two correlators written out in \eqref{eq:4ptDispLargTimeV1}. To drop correlators of this form, we need to assume,
\begin{align}
\lim\limits_{t \rightarrow\infty}\ll\f(p_4)T[\f(p_1)\f(p_2)\f(p_3)]\rr=0,
\end{align} 
for $\z_i$, $s<0$ and fixed.
This condition is in line with our expectations from Section \ref{ssec:4ptanalyticity}, namely that correlation functions obey an unsubtracted four-point dispersion formula if they vanish in the Regge limit.

Next, we consider the correlator $\ll\overline{T}[\f(p_1)\f(p_4+k)]T[\f(p_2)\f(p_3-k)]\rr$, which was suppressed in \eqref{eq:4ptDispLargTimeV1}. In order to drop this correlator from the dispersion formula \eqref{eq:4ptDispLargTimeV1}, we require that the corresponding unshifted correlator, $\<\overline{T}[\f(p_1)\f(p_4)]T[\f(p_2)\f(p_3)]\>$, grows slower than $P^{2}$ if we set,
\begin{align}
u = 8P^{2}, \label{eq:4ptFunnyScaling1}
\\
\z_3= 4P^{2},
\\
\z_4= 4P^{2}, \label{eq:4ptFunnyScaling3}
\end{align}
and then send $P\rightarrow \infty$. 
There is an identical condition for the correlator \\ $\ll\overline{T}[\f(p_1)\f(p_3+k)]T[\f(p_2)\f(p_4-k)]\rr$, which was also suppressed in \eqref{eq:4ptDispLargTimeV1}, but with the replacement $u\leftrightarrow t$. If any of the above correlators do not vanish in the limit $P\rightarrow\infty$, then we need to introduce additional subtractions.

\subsection{Higher-Point Functions}
\label{sec:higher_points}
In this section we will generalize the four-point analysis to generic higher-point correlators.
To do this, we use the following kinematics,\footnote{We assume $d\geq n$ so that the momenta are linearly independent in these kinematics. We expect that this assumption can be relaxed by starting with a different configuration.}
\begin{align}
p_1&=(P,P,p_{1}^{\perp}),  \label{eq:npt_kinematics1}
\\
p_2&=(-P,-P,p_{2}^{\perp}), 
\\
p_i&=(0,0,p_i^{\perp}) \quad \text{for} \quad i=3,\ldots,n.  \label{eq:npt_kinematics3}
\end{align}
The momenta $p_i$ and their sums are all spacelike, and therefore we can use the general $n$-point formula given in \eqref{eq:dispersion_Largesttime_kspaceV2}. We will again study the large $P$ limit in order to find a Lorentz invariant dispersion formula.

In the configuration \eqref{eq:npt_kinematics1}-\eqref{eq:npt_kinematics3}, studying the large $P$ limit of the $n$-point correlator will be effectively the same as the previous four-point function analysis. 
The only new feature at $n$-points is that we have $n-4$ extra operators, $\{\f(p_{3}),\ldots,\f(p_{n-2})\}$, which are independent of $P$, are unshifted by $k$, and whose momenta lies purely in the $d-2$ transverse directions. 
We can then map the $n$-point problem to a four-point problem with shifted transverse momenta. 
The transverse momenta were arbitrary in the four-point analysis, which means the inclusion of the new $n-4$ operators is trivial.  

To write down the $n$-point dispersion formula, it will also be convenient to label the Mandelstams by how we partition the external operators. 
To do this, we will define $\Pi^{(m)}_r(n)$ as the set of partitions of $\{1+m,\ldots,n+m\}$ into two sets of size $r$ and $n-r$. Then given an element $\a \in \Pi^{(2)}_{r}(n-4)$ we use the following notation,
\begin{align}
s_{i,j,\a}=-\left(p_i+p_j+\sum\limits_{m=1}^{r}p_{\a_m}\right)^{2}.
\end{align}
To write the dispersion formula in a simple way, we will also use the notation:
\begin{align}
\ll T[\f(p_1)\ldots\f(p_n)]\rr&=\tau_{n}(\z_i,s_{ij}),
\\[3pt]
\ll\overline{T}[\f(p_{1})\ldots\f(p_r)]T[\f(p_{r+1})\ldots\f(p_n)]\rr&=(-1)^{r+1}\theta(p_{r+1}^0+\ldots+p_n^0)\text{Cut}_{s_{i_1\ldots i_r}}\tau_{n}(\z_i,s_{ij}), \label{eq:discnpoints}
\end{align}
where $s_{i_1\ldots i_n}$ were defined in \eqref{eq:multi_mand}. 
Here $\text{Cut}_{s_{i_1\ldots i_r}}$ is defined by \eqref{eq:discnpoints}, i.e., it gives the dependence of the partially time-ordered correlator on the Lorentz invariants, up to a possible overall sign.\footnote{The ``$\text{Cut}$" and ``disc" operations are generically different. The former, by definition, picks out one partially time-ordered correlator while the latter generally gives a sum of partially time-ordered correlators for different channels.}

If we take the large $P$ limit, the dispersion formula \eqref{eq:dispersion_Largesttime_kspaceV2} becomes, 
\begin{align}
\hspace{-.0in}\tau_{n}(\z_i,s_{\beta})=&\frac{1}{2\pi i}\sum\limits_{r=0}^{n-4}\sum\limits_{\a\in \Pi^{(2)}_{r}(n-4)}\bigg[
\int\limits_{0}^{\infty}\frac{ds'_{1,n,\a}}{s'_{1,n,\a}-s_{1,n,\a}-i\epsilon}\text{Cut}_{s'_{1,n,\a}}\tau_{n}(\z_i,s'_{1,n,\a},\ldots)
\nonumber
\\ &\hspace{3.5in}+(p_{n-1}\leftrightarrow p_n)
\bigg].
\label{eq:dispersion_n_pointV2}
\end{align}
On the right hand side we explicitly wrote out the Mandelstam invariant which is being integrated over and suppressed the other independent Mandelstams which are held fixed at their external value. We also recall that by using \eqref{eq:TbTtoAA}, we can rewrite the integrand in terms of a double advanced product.

To prove \eqref{eq:dispersion_n_pointV2}, we again need to assume that partially time-ordered correlators are sufficiently bounded when $P\rightarrow \infty$. The sufficient conditions are identical in form to the four-point function case and are derived in Appendix \ref{app:details}. Finally, at $n=5$ this dispersion formula agrees with the one found using analyticity arguments in Section \ref{ssec:five-point}. This can be seen by rewriting each dispersion formula in terms of the partially time-ordered correlators.

\section{Relation to Conformal Dispersion}
\label{sec:equivalence}
In this section we explain how the CFT and momentum space dispersion formulas are related.
In Section \ref{sec:Superbounded} we discuss superbounded correlators in both the position and momentum space Regge limits.
We argue that if a correlator is superbounded in position space, then the two dispersion formulas are equivalent modulo semi-local terms.
In Sections \ref{ssec:CFTPR}-\ref{ssec:comparing_dispersions} we study the Polyakov-Regge expansions generated by each dispersion formula.
In Section \ref{ssec:CFTPR} we review the CFT Polyakov-Regge expansion.
In Section \ref{sec:k_spacePRExpansion} we study the corresponding momentum space Polyakov-Regge expansion and explain its similarities to the CFT expansion.
Finally, in Section \ref{ssec:comparing_dispersions} we show that the CFT and momentum space Polyakov-Regge blocks are related by a Fourier transform.

\subsection{Superbounded Correlators}
\label{sec:Superbounded}
In this section we will study a scalar, CFT four-point function,
\begin{align}
\<\f(x_1)\f(x_2)\f(x_3)\f(x_4)\>_{\text{E}}=\frac{1}{(x_{12}x_{34})^{2\Delta_{\f}}}\mathcal{G}(z,\bar{z}),\label{eq:mG_def}
\end{align} 
where $x_{ij}=x_i-x_j$, $\Delta_{\f}$ is the dimension of $\f$ and $(z,\bar{z})$ are the conformal cross-ratios,
\begin{align}
z\bar{z}=\frac{x_{12}^{2}x_{34}^{2}}{x_{13}^{2}x_{24}^{2}},\qquad (1-z)(1-\bar{z})=\frac{x_{14}^{2}x_{23}^{2}}{x_{13}^{2}x_{24}^{2}}.
\end{align}
In Euclidean space, we can expand $\<\f(x_1)\f(x_2)\f(x_3)\f(x_4)\>_{\text{E}}$ using the operator product expansion in three different channels,
\begin{align}
\mathcal{G}(z,\bar{z})=&\sum\limits_{\O}\lambda_{\f\f\O}^{2}g^s_{\O}(z,\bar{z})=\sum\limits_{\O}\lambda_{\f\f\O}^{2}g^t_{\O}(z,\bar{z})
=\sum\limits_{\O}\lambda_{\f\f\O}^{2}g^u_{\O}(z,\bar{z}).\label{eq:blockexpansion}
\end{align}
The $s$-channel expansion corresponds to using the $\f(x_1)\f(x_2)$ OPE. The $t$- and $u$-channel expansions are found by taking $x_1\leftrightarrow x_3$ and $x_2\leftrightarrow x_3$, respectively. 
Here $\lambda_{\f\f\O}$ are the OPE coefficients and the blocks $g^{s,t,u}_{\O}(z,\bar{z})$ encode the contribution of a primary operator $\O$ and all its descendants in a given OPE channel.
The $t$- and $u$-channel blocks can be defined in terms of the $s$-channel blocks:\footnote{We absorb the powers of $z$ and $\bar{z}$ into the blocks in order to simplify the crossing equation \eqref{eq:blockexpansion}.}
\begin{align}
g^{t}_{\O}(z,\bar{z})&=\frac{(z\bar{z})^{\Delta_{\f}}}{((1-z)(1-\bar{z}))^{\Delta_{\f}}}g_{\O}^{s}(1-z,1-\bar{z}),\label{eq:gt_to_gs}
\\
g^{u}_{\O}(z,\bar{z})&=(z\bar{z})^{\Delta_{\f}}g_{\O}^{s}(1/z,1/\bar{z}).
\end{align}
We will also label blocks by the dimension and spin, $(\Delta,J)$, of the primary operator $\O$.
Only symmetric traceless operators can appear in the OPE of two scalars.

The CFT dispersion formula reconstructs $\mathcal{G}(z,\bar{z})$ from two of its double-discontinuities \cite{Carmi:2019cub}. 
The function $\mathcal{G}(z,\bar{z})$ has a $t$-channel cut for $\bar{z}\in[1,\infty)$ and an $s$-channel cut for $\bar{z}\in(-\infty,0]$.
The $t$-channel double-discontinuity is defined as,
\begin{align}
\text{dDisc}_{t}\mathcal{G}(z,\bar{z})=\mathcal{G}(z,\bar{z})-\frac{1}{2}\left(\mathcal{G}^{\circlearrowleft}(z,\bar{z})+\mathcal{G}^{\circlearrowright}(z,\bar{z})\right).\label{eq:dDisct_def_AC}
\end{align}
The notation $\mathcal{G}^{\circlearrowleft}(z,\bar{z})$ and $\mathcal{G}^{\circlearrowright}(z,\bar{z})$ means we analytically continue $\bar{z}$ around $1$ along a (counter-)clockwise path. 
The $t$-channel dDisc is equal to the following double-commutator \cite{Caron-Huot:2017vep},
\begin{align}
\text{dDisc}_{t}[\mathcal{G}(z,\bar{z})]=-\frac{1}{2}x_{12}^{2\Delta_{\f}}x_{34}^{2\Delta_{\f}}\<[\f(x_3),\f(x_2)][\f(x_1),\f(x_4)]\>.\label{eq:defDdisc}
\end{align}
The $s$- and $u$-channel double-discontinuities are defined similarly, except in \eqref{eq:dDisct_def_AC} we analytically continue around $z=0$ and $z=\infty$, respectively. The corresponding double-commutators are found by taking \eqref{eq:defDdisc} and making the replacements $2\leftrightarrow4$ and $3\leftrightarrow 4$.

The CFT dispersion formula expresses $\mathcal{G}$ as an integral over the $t$- and $u$-channel double discontinuities,\footnote{We will assume for simplicity that there are no non-normalizable contributions corresponding to scalar blocks with $\Delta<d/2$, with the exception of the identity block.}
\begin{align}
\mathcal{G}(z,\bar{z})&=1+\mathcal{G}^{t}(z,\bar{z})+\mathcal{G}^{u}(z,\bar{z}), \label{eq:CFTdispersionGfull}
\\
\mathcal{G}^{t}(z,\bar{z})&=\int\limits_{0}^{1}dwd\bar{w}K(z,\bar{z},w,\bar{w})\text{dDisc}_{t}[\mathcal{G}(w,\bar{w})], \label{eq:CFTdispersionGT}
\\
\mathcal{G}^{u}(z,\bar{z})&=\int\limits_{-\infty}^{0}dwd\bar{w}K\left(\frac{z}{z-1},\frac{\bar{z}}{\bar{z}-1},\frac{w}{w-1},\frac{\bar{w}}{\bar{w}-1}\right)\text{dDisc}_{u}\left[\mathcal{G}\left(\frac{w}{w-1},\frac{\bar{w}}{\bar{w}-1}\right)\right]. \label{eq:CFTdispersionGU}
\end{align}
The explicit form of the kernel $K$ can be found in \cite{Carmi:2019cub}.

\begin{figure}
\centering     
\subfigure[]{\label{fig:Regge_Limit_position}\includegraphics[scale=.35]{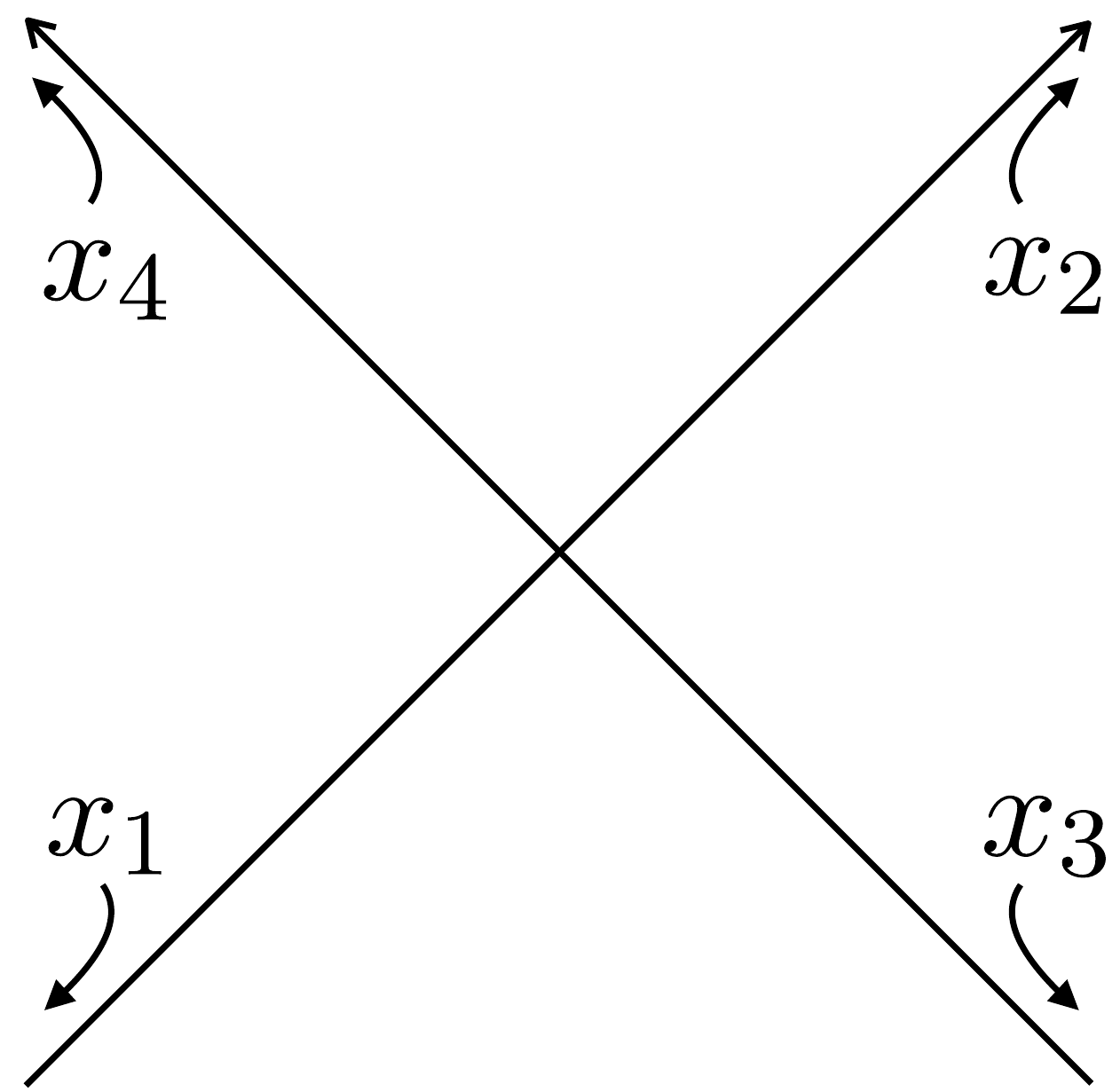}
}
\hspace{1.25in}
\subfigure[]{\label{fig:Regge_Limit}\includegraphics[scale=.35]{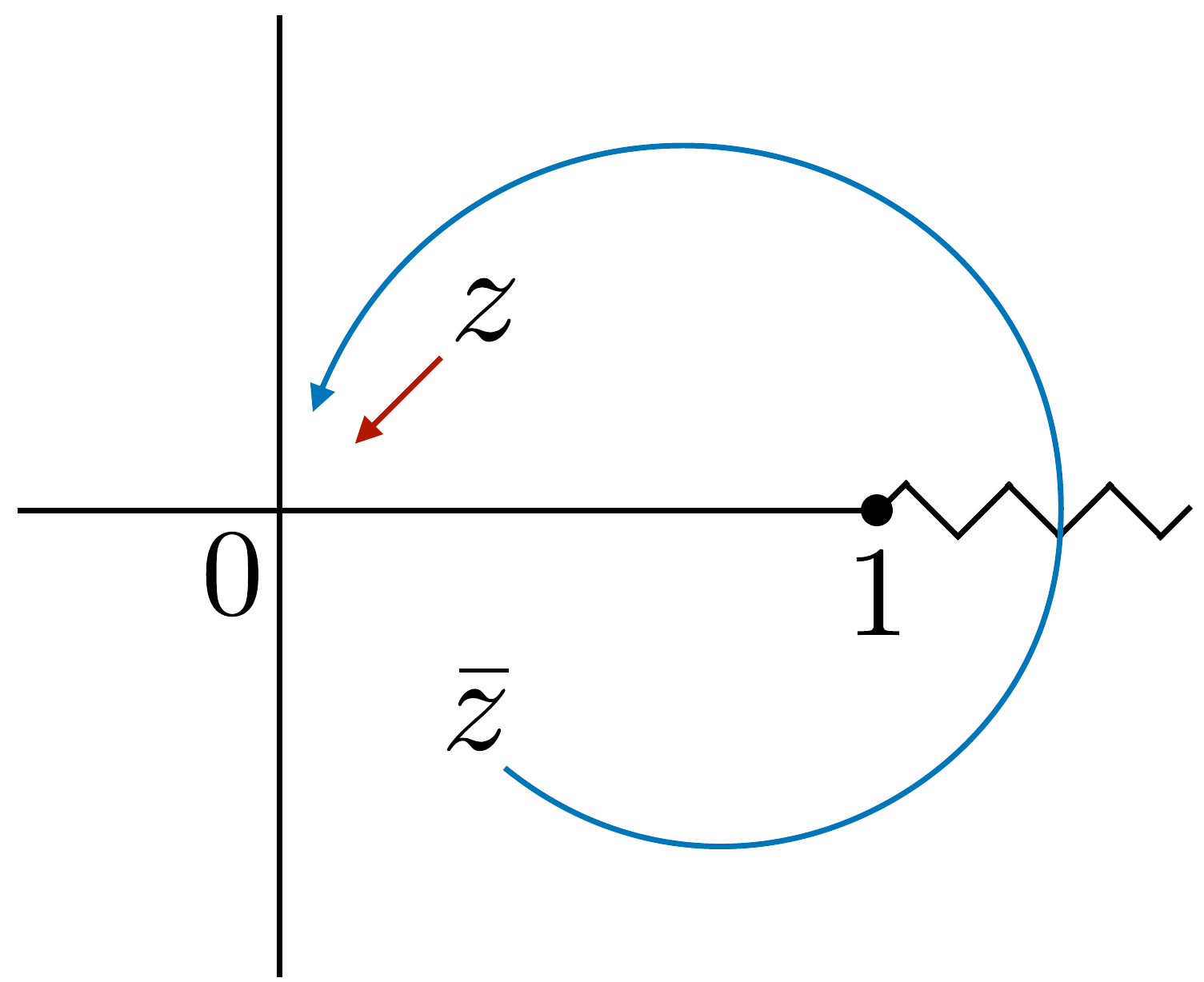}}
\caption{The figure on the left corresponds to taking the Regge limit in position space. Here we send the four operators to null infinity such that $x_{12}^2$ and $x_{34}^2$ are held fixed and spacelike. The figure on the right gives the same limit in CFT cross-ratio space. In momentum space these limits correspond to taking the large $t$ Regge limit.}
\label{fig:Regge_limit_full}
\end{figure}

In order for the unsubtracted CFT dispersion formula to be valid, the correlator $\mathcal{G}(z,\bar{z})$ must have sufficiently soft behavior in the CFT $s$-channel Regge limit.
In terms of the cross-ratios $(z,\bar{z})$, the Regge limit is defined by continuing $\bar{z}$ around 1 and then taking $z,\bar{z}\rightarrow0$ such that $z/\bar{z}$ is held fixed, see figure \ref{fig:Regge_limit_full}.\footnote{This is sometimes known as the CFT $s$-channel Regge limit because it has a similar form to the $s$-channel OPE limit. In momentum space, the CFT $s$-channel Regge limit corresponds to taking $t$ large with $s$ held fixed. To avoid confusion, in momentum space we will be explicit about which Mandelstam is taken to be large.} In this limit, the correlator $\mathcal{G}$ scales like \cite{Costa:2012cb},
\begin{align}
\lim\limits_{\substack{z\rightarrow 0\\ z/\bar{z} \text{ fixed}}}\mathcal{G}^{\circlearrowleft}(z,\bar{z})\sim (z\bar{z})^{\frac{1-J_0}{2}}.\label{eq:CFTRegge}
\end{align}
Here $J_0$ is the effective spin in the position space Regge limit. 
In a non-perturbative CFT $J_0\leq 1$ while in a holographic CFT $J_0\leq 2$ at tree-level in $1/N$ \cite{Maldacena:2015waa,Hartman:2015lfa,Caron-Huot:2017vep}.\footnote{To make holographic correlators consistent with the non-perturbative bound, one needs to study higher-loops and prove that the correlator eikonalizes in the Regge limit \cite{Cornalba:2007zb,Meltzer:2019pyl,Antunes:2020pof}.} 
In order for the unsubtracted dispersion formula to hold, we need the stronger constraint that $\mathcal{G}(z,\bar{z})$ has effective spin $J_0<0$:
\begin{align}
\lim\limits_{\substack{z\rightarrow 0\\ z/\bar{z} \text{ fixed}} }\mathcal{G}^{\circlearrowleft}(z,\bar{z})\lesssim (z\bar{z})^{\frac{1}{2}}.
\end{align}
In \cite{Mazac:2019shk,Caron-Huot:2020adz} correlators with $J_0<0$ were referred to as superbounded correlators.
Using the CFT dispersion formula, superbounded correlators are uniquely determined from their $t$- and $u$-channel double-discontinuities.

The unsubtracted momentum space dispersion formula \eqref{eq:dispersion_4pt_pt3},
\begin{align}
\tau_4(\zeta_i,s,t,u)=\frac{1}{2\pi i}\left(\int\limits_{0}^{\infty} \frac{dt'}{t'-t-i\epsilon}\disc_{t'} \tau_4(\zeta_i,s,t',u)
+\int\limits_{0}^{\infty} \frac{du'}{u'-u-i\epsilon}\disc_{u'} \tau_4(\zeta_i,s,t,u')
\right),\label{eq:disp4ptequiv}
\end{align}
is also only valid if the correlator is sufficiently bounded in the Regge limit.
In the limit $|t|\rightarrow \infty$ with $\zeta_i$, $s <0$ and fixed, the correlator has the following behavior,
\begin{align}
\lim\limits_{|t|\rightarrow \infty}\tau_4(\zeta_i,s,t,u)\sim t^{\tilde{J}_0}, \label{eq:k_space_Regge_Scaling}
\end{align}
where $\tilde{J}_0$ is the effective spin in the momentum space Regge limit.
As reviewed in Section \ref{sec:disp_from_analyticity}, \eqref{eq:disp4ptequiv} holds if $\tilde{J}_{0}<0$. In analogy with the CFT dispersion formula, we will refer to correlators with $\tilde{J}_0<0$ as superbounded in the momentum space Regge limit.
Superbounded correlators in momentum space are uniquely fixed by their $t$- and $u$-channel momentum space discontinuities.

We can now study under what conditions the two dispersion formulas, \eqref{eq:CFTdispersionGfull} and \eqref{eq:disp4ptequiv}, are equivalent.
In general, we will say two dispersion formulas are equivalent if:
\begin{enumerate}
\item Both dispersion formulas depend on the same physical data, i.e. the same set of double-commutators or discontinuities.\label{item:equiv_1}
\item Given the same physical input, both dispersion formulas produce the same correlation function.\label{item:equiv_2}
\end{enumerate}
Condition \ref{item:equiv_1} is always true for the dispersion formulas \eqref{eq:CFTdispersionGfull} and \eqref{eq:disp4ptequiv} as they both depend on the same $t$- and $u$-channel advanced double-commutators. 
This is discussed below \eqref{eq:disctToDC}.
Condition \ref{item:equiv_2} is less obvious because the two dispersion formulas can potentially yield two different correlation functions which have the same $t$- and $u$-channel double-commutators.
For example, they can differ by a sum of AdS contact diagrams \cite{Caron-Huot:2017vep,Carmi:2019cub}.
To understand when condition \ref{item:equiv_2} holds, we need to study how the two effective spins, $J_0$ and $\tilde{J}_0$, are related.

The relation between the position and momentum space Regge limits was studied in \cite{Cornalba:2009ax}.
There they proved that if the Regge limit commutes with the Fourier transform, then the two effective spins are equal, $\tilde{J}_0=J_0$.
This implies that our two notions of superboundedness are equivalent if we can take the Regge limit under the Fourier integral.
However, in general $\tilde{J}_0\neq J_0$ and the leading contribution to the momentum space Regge limit is not determined by the position space limit.
The position space limit is defined with all $x_{ij}^{2}\neq0$, see figure \ref{fig:Regge_limit_full}, and does not depend on local terms in the position space correlator.
On the other hand, the momentum space correlator is defined by taking a Fourier transform and is sensitive to local terms in position space.

Here we need to distinguish between two different kinds of contact terms: An ultra-local contact term only has support when all points are coincident while a semi-local term has support when a subset of points are coincident. In addition, we of course have non-local terms which have support when all points are non-coincident. For example, at three-points we can have \cite{Dymarsky:2014zja,Bzowski:2014qja},
\begin{align}
\text{Non-local:}& \qquad \<\f(x_1)\f(x_2)\f(x_3)\>\supset \frac{1}{x_{12}^{\Delta_{\f}}x_{23}^{\Delta_{\f}}x_{13}^{\Delta_{\f}}},
\\
\text{Semi-local:}& \qquad \<\f(x_1)\f(x_2)\f(x_3)\>\supset \delta(x_{12})\frac{1}{x_{23}^{3\Delta_{\f}-d}},
\\
\text{Ultra-local:}& \qquad \<\f(x_1)\f(x_2)\f(x_3)\>\supset \delta(x_{12})\delta(x_{13})\delta(x_{23}).
\end{align}
For generic $\Delta_{\f}$ ultra-local terms are absent and the momentum space Regge limit will depend on non-local and semi-local terms.
Schematically, at large $t$ we have:
\begin{align}
\lim\limits_{t\rightarrow \infty}\tau_{4}(\z_i,s,t,u)\approx t^{\tilde{J}_0^{(1)}}(\text{non-analytic})+t^{\tilde{J}_0^{(2)}}(\text{analytic}).\label{eq:schematic_scaling}
\end{align}
The first term in \eqref{eq:schematic_scaling} multiplies a function which is non-analytic in the remaining external momenta and comes from a non-local term in the position space correlator.
This term is fixed by the position space Regge limit, which means $\tilde{J}_0^{(1)}=J_{0}$ \cite{Cornalba:2009ax}.
The second term in \eqref{eq:schematic_scaling} is analytic in some of the remaining momenta and corresponds to a semi-local contact term in position space.
The effective spin $\tilde{J}_0^{(2)}$ is not fixed by the position space Regge limit.
The full Regge limit in momentum space is then determined by the larger effective spin,
\begin{align}
\tilde{J}_0=\text{max}(J_0,\tilde{J}_0^{(2)}).\label{eq:Jtilde_max}
\end{align} 
This means that superboundedness in momentum space implies superboundedness in position space, but that the converse is not true.

A similar phenomenon was found in \cite{Dymarsky:2014zja,Bzowski:2014qja} in the context of the Euclidean OPE. 
There they showed that the high-energy limit of a Euclidean three-point function, $\<\f\f\f\>_{\text{E}}$, in momentum space is determined by semi-local terms when $\Delta_{\f}>d/2$.
The semi-local terms appear because correlation functions of heavy operators have a divergent Fourier transform and need to be regularized.\footnote{We thank Adam Bzowski for discussions on this point.}

We then have three different cases to consider:
\begin{enumerate}
\item $\tilde{J}_0<0$: In this case the correlator is superbounded in both Regge limits.
For such correlators, the two unsubtracted dispersion formulas are equivalent.
\item $J_0<0$ and $\tilde{J}_0>0$: 
Here we can use the CFT dispersion formula \eqref{eq:CFTdispersionGfull} to compute the correlator at non-coincident points.
However, in momentum space there are semi-local terms that are not superbounded at large $t$.
If we ignore the arcs at infinity and use the unsubtracted momentum space formula \eqref{eq:disp4ptequiv}, then we reproduce the original correlator up to possible semi-local terms. 
Therefore, in this case the two dispersion formulas are equivalent modulo contact terms.
\label{item:case2_J}
\item $J_0>0$: In this case we need to include conformally invariant subtraction terms that have support at non-coincident points.
By including enough subtractions, we can reduce this case to one of the previous two cases.
\label{item:case3_J}
\end{enumerate}

To summarize, if a correlation function is superbounded in position space, then we can use either \eqref{eq:CFTdispersionGfull} or \eqref{eq:disp4ptequiv} to reconstruct the non-analytic terms in momentum space, or the non-local terms in position space, from the causal double-commutators.
To give more evidence for this claim, in the following sections we will study how each dispersion formula acts on individual conformal blocks.
In Section \ref{sec:app_to_CFT} we will also find explicit examples where $\tilde{J}_0>J_0$.

\subsection{CFT Polyakov-Regge Expansion}
\label{ssec:CFTPR}
Here we will give a brief review of the CFT Polyakov-Regge expansion \cite{Mazac:2019shk,Sleight:2019ive,Caron-Huot:2020adz}.

To derive the CFT Polyakov-Regge expansion, we expand \eqref{eq:CFTdispersionGT} and \eqref{eq:CFTdispersionGU} in $t$- and $u$-channel conformal blocks, respectively, and perform the dispersion integrals term by term \cite{Caron-Huot:2020adz}.
The $t$-channel dDisc commutes with the $t$-channel block expansion and we find \cite{Caron-Huot:2017vep},
\begin{align}
\text{dDisc}_{t}[\mathcal{G}(z,\bar{z})]&=\sum\limits_{\O}\lambda_{\f\f\O}^{2}\text{dDisc}_{t}[g^t_{\Delta,J}(z,\bar{z})],\label{eq:dDisctexpansion}
\\
\text{dDisc}_{t}[g^t_{\Delta,J}(z,\bar{z})]&=2\sin^{2}\left(\frac{\pi}{2}(\Delta-J-2\Delta_{\f})\right)g^t_{\Delta,J}(z,\bar{z}). \label{eq:dDisctblock}
\end{align}
By applying the double-discontinuity to the conformal block expansion of $\mathcal{G}(z,\bar{z})$, we have generated the conformal block expansion for its double-discontinuity.
Alternatively, we could have derived \eqref{eq:dDisctexpansion} by inserting a complete set of states on the right hand side of \eqref{eq:defDdisc}. This gives the following expression for the double-discontinuity of a conformal block,
\begin{align}
\text{dDisc}_{t}[g^t_{\Delta,J}(z,\bar{z})]=-\frac{1}{2\lambda_{\f\f\O}^{2}}x_{12}^{2\Delta_{\f}}x_{34}^{2\Delta_{\f}}\sum\limits_{\substack{\Psi = \O, P\O , \\ \quad P^{2}\O,...}}\frac{\<[\f(x_3),\f(x_2)]|\Psi\>\<\Psi|[\f(x_1),\f(x_4)]\>}{\<\Psi|\Psi\>},\label{eq:blockdDiscAlt}
\end{align}
where the sum includes the primary operator $\O$ and all its descendants.
We divided by the OPE coefficients in \eqref{eq:blockdDiscAlt} so that the blocks are purely kinematic functions.
The identity \eqref{eq:dDisctblock} does not have an analog in momentum space while \eqref{eq:blockdDiscAlt} will generalize straightforwardly.

The $t$-channel Polyakov-Regge blocks are defined by taking the dispersive transform of a single conformal block:
\begin{align}
P^{t|s}_{\O}(z,\bar{z})\equiv \int\limits_{0}^{1} dwd\bar{w}K(z,\bar{z};w,\bar{w})\text{dDisc}_{t}[g^t_{\O}(w,\bar{w})].\label{eq:DefPR_Block1}
\end{align}
Here $P^{t|s}_{\O}$ indicates that we are using the $t$-channel part of the fixed-$s$ dispersion formula.
By definition, we have,
\begin{align}
\text{dDisc}_{t}[P^{t|s}_{\O}(z,\bar{z})]&=\text{dDisc}_{t}[g^t_{\O}(w,\bar{w})],\label{eq:dDiscPt}
\\
\text{dDisc}_{u}[P^{t|s}_{\O}(z,\bar{z})]&=0.\label{eq:dDiscPu}
\end{align}
The first line indicates that only the operator $\O$ and its descendants contribute to the $t$-channel double-commutator of $P^{t|s}_{\O}$.
One can similarly define the $u$-channel Polyakov-Regge blocks, $P^{u|s}_{\O}(z,\bar{z})$, by acting with the $u$-channel piece of the dispersion formula on a single $u$-channel conformal block. 
The Polyakov-Regge expansion is then given by,
\begin{align}
\mathcal{G}(z,\bar{z})=1+\sum\limits_{\O}\lambda_{\f\f\O}^{2}\left(P^{t|s}_{\O}(z,\bar{z})+P^{u|s}_{\O}(z,\bar{z})\right). \label{eq:PRexpansion}
\end{align}

If the dispersion integral in \eqref{eq:DefPR_Block1} converges, then the resulting Polyakov-Regge block will be superbounded in the position space Regge limit.
However, if the integral diverges and needs to be defined by analytic continuation, this may no longer be true.
The defining integral \eqref{eq:DefPR_Block1} converges if \cite{Caron-Huot:2017vep,Carmi:2019cub}:
\begin{align}
\lim\limits_{\substack{z\rightarrow 0\\ z/\bar{z} \text{ fixed}} }\text{dDisc}_{t}[g^{t}_\O(z,\bar{z})]\lesssim (z\bar{z})^{\frac{1}{2}}.
\end{align}
From \eqref{eq:gt_to_gs} and \eqref{eq:dDisctblock} we find,
\begin{align}
\lim\limits_{\substack{z\rightarrow 0\\ z/\bar{z} \text{ fixed}}} \text{dDisc}_{t}[g^{t}_\O(z,\bar{z})]\sim (z\bar{z})^{\frac{1}{2}(1-d/2+2\Delta_{\f})}.\label{eq:Regge_limit_dDisctg}
\end{align}
Here we used that $g_{\O}^{s}(1-z,1-\bar{z})$ scales like $(z\bar{z})^{\frac{1}{2}(1-d/2)}$ in the above limit \cite{Liu:2020tpf}.
Therefore, the integral \eqref{eq:DefPR_Block1} converges if,
\begin{align}
\Delta_{\f}\geq \frac{d}{4}.\label{eq:conv_PR}
\end{align}
In interacting CFTs the unitarity bound implies $\Delta_{\f}>\frac{1}{2}(d-2)$, which means the inequality \eqref{eq:conv_PR} can only be violated if $d\leq 3$.
When $\Delta_{\f}\leq d/4$, the CFT Polyakov-Regge blocks are uniquely defined by analytically continuing in $\Delta_{\f}$.

Finally, we can also relate the $t$-channel Polyakov-Regge blocks to a sum of Witten diagrams.
The following $t$-channel, exchange Witten diagram, 
\begin{align}
\includegraphics[scale=.3]{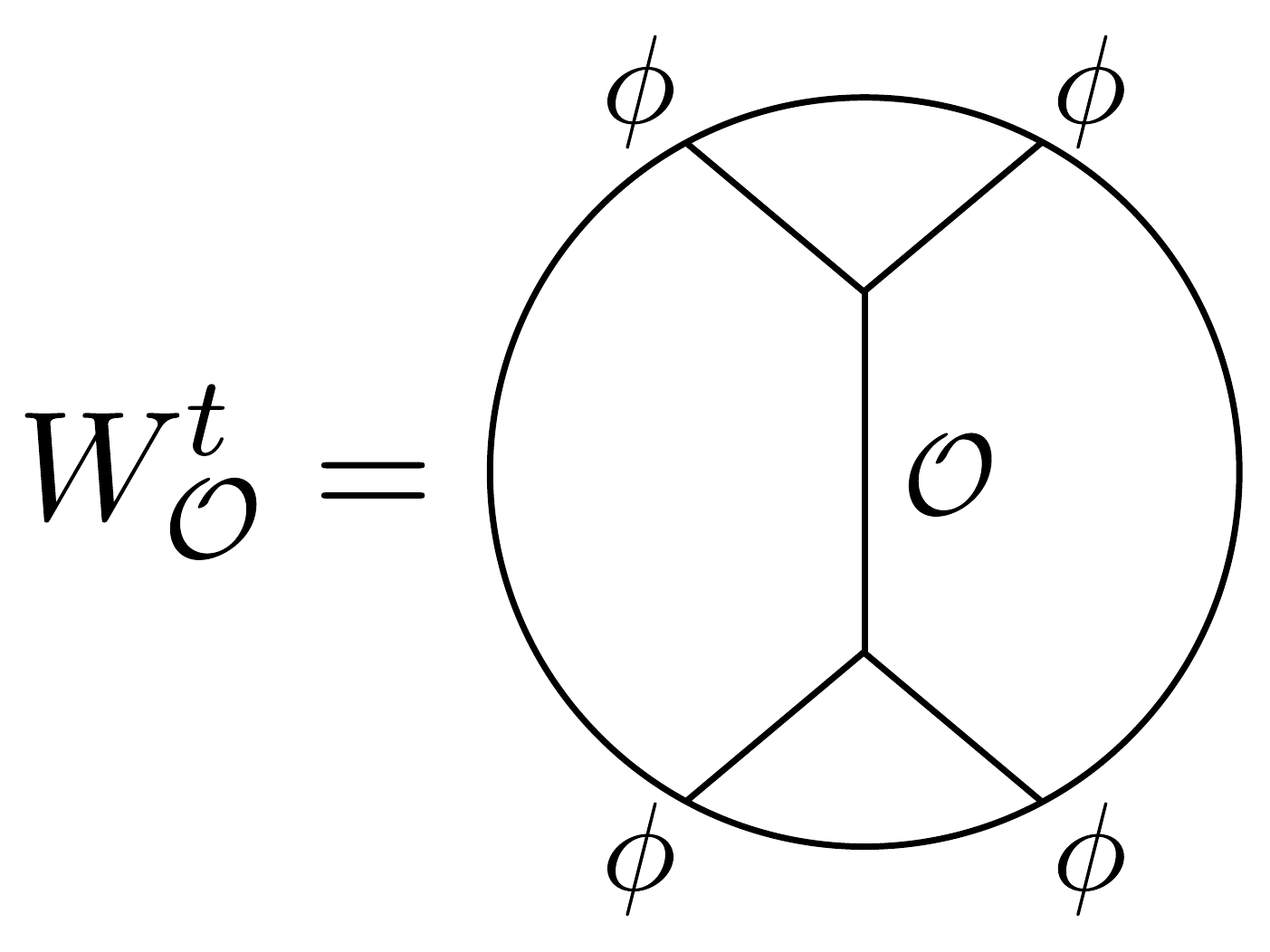}, \label{eq:Witten_Diagram}
\end{align}
has the same double-discontinuity as a single conformal block \cite{Caron-Huot:2017vep},
\begin{align}
\text{dDisc}_{t}[W^{t}_{\O}(z,\bar{z})]&=\text{dDisc}_{t}[g^t_{\O}(z,\bar{z})]. \label{eq:dDiscW_eq_dDiscg}
\end{align}
Throughout this work, we will label the internal lines of the Witten diagram using the dual, boundary operator.
In \eqref{eq:dDiscW_eq_dDiscg} we assumed the bulk couplings are normalized such that  $g^{t}_\O$ appears with unit coefficient in the conformal block expansion of $W^{t}_{\O}$.
The $t$-channel Polyakov-Regge block is then equal to the dispersive transform of this exchange diagram,
\begin{align}
P^{t|s}_{\O}(z,\bar{z})\equiv \int\limits_{0}^{1} dwd\bar{w}K(z,\bar{z};w,\bar{w})\text{dDisc}_{t}[W^{t}_{\O}(w,\bar{w})].\label{eq:DefPR_Block2}
\end{align}

\begin{figure}
	\centering
		\includegraphics[scale=.3]{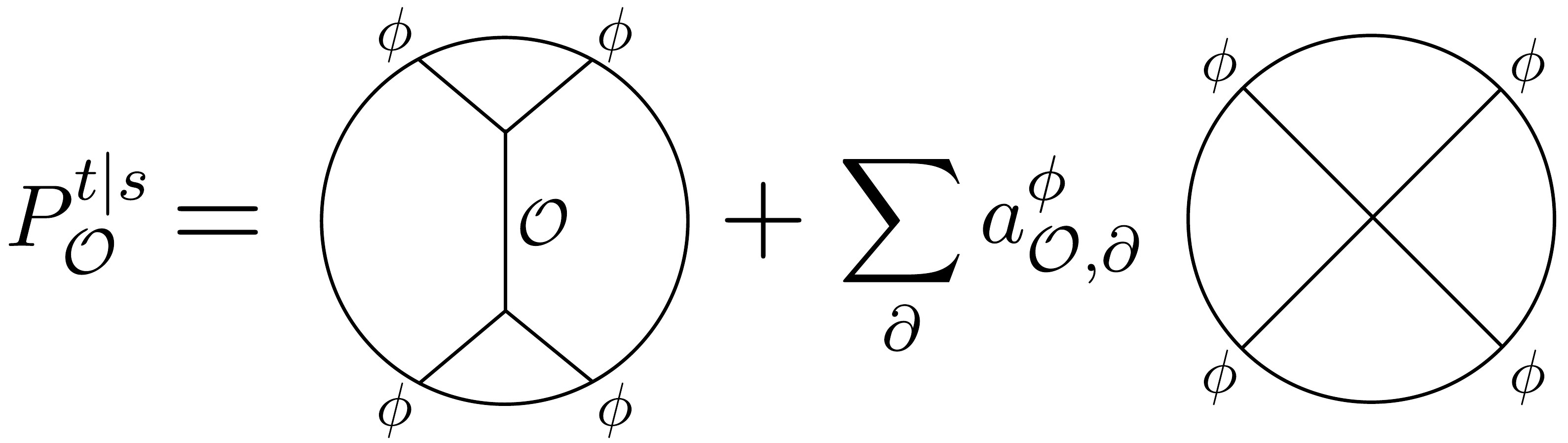}
		\caption{The $t$-channel Polyakov Regge block is a $t$-channel exchange Witten diagrams plus a finite sum of quartic, contact diagrams.}
\label{fig:PR_Block_Witten}
\end{figure}

The identity \eqref{eq:DefPR_Block2} implies that the difference between the Polyakov-Regge block and the exchange Witten diagram must be a sum of functions which have a vanishing $t$- and $u$-channel double-discontinuity. 
From the uniqueness arguments of \cite{Heemskerk:2009pn}, one can show that the difference between $P^{t|s}_{\O}$ and $W^t_{\O}$ is a sum of quartic, contact diagrams with a bounded number of derivatives \cite{Mazac:2019shk,Sleight:2019ive,Caron-Huot:2020adz},\footnote{Although $s$-channel exchange diagrams have a vanishing $t$- and $u$-channel dDisc, they cannot be used to improve the CFT $s$-channel Regge growth. This is clearest to see in Mellin space \cite{Mazac:2019shk,Sleight:2019ive}.}
\begin{align}
P^{t|s}_{\O}(z,\bar{z})=W^t_{\O}(z,\bar{z})+\sum\limits_{\partial} a^{\f}_{\O,\partial}W_{\partial,\text{cont}}(z,\bar{z}). \label{eq:PintermsofW}
\end{align}
This is shown in figure \ref{fig:PR_Block_Witten}.
The coefficients $a^{\f}_{\O,\partial}$ are fixed by requiring $P^{t|s}_{\O}(z,\bar{z})$ has effective spin $J_0<0$ in the CFT $s$-channel Regge limit for $\Delta_{\f}>d/4$.
The coefficients $a^{\f}_{\O,\partial}$ are determined for $\Delta_{\f}\leq d/4$ by analytic continuation.

\subsection{Momentum Space Polyakov-Regge Expansion}
\label{sec:k_spacePRExpansion} 
In this section we will derive and study the Polyakov-Regge expansion associated to the momentum space dispersion formula.
One difference in comparison to the previous section is that while the time-ordered correlator in position space has a conformal block expansion, the time-ordered correlator in momentum space, $\tau_{4}(\z_i,s,t,u)$, does not.
Heuristically, the Euclidean OPE in a given channel does not commute with the Fourier transform because we are integrating the four external operators over the entire spacetime \cite{Dymarsky:2014zja,Bzowski:2014qja}.
However, while $\tau_{4}(\z_i,s,t,u)$ itself does not have an OPE, the discontinuity $\disc_{t} \tau_4(\zeta_i,s,t,u)$ does have a convergent $t$-channel expansion.
To see this, we use
\eqref{eq:disctToDC} to relate the discontinuity of a time-ordered correlator to a sum of advanced commutators:
\begin{align}
\disc_{t}\ll T[\f(p_1)\f(p_2)\f(p_3)\f(p_4)]\rr=-\big(\<A[\f(p_2);\f(p_3)]A[\f(p_1);\f(p_4)]\> +(p_2,p_3)\leftrightarrow (p_1,p_4)\big).\label{eq:disc_to_AA}
\end{align} 
The $t$-channel Lorentzian OPE of \eqref{eq:disc_to_AA} is found by using the following completeness relation \cite{Gillioz:2016jnn,Gillioz:2018kwh,Gillioz:2020wgw}:
\begin{align}
\mathbb{I}=|0\>\<0|+\sum\limits_{\O}\int\limits_{\overline{V}_+}\frac{d^{d}k}{(2\pi)^d} \ \mathcal{P}_{\O}(k,\epsilon_1,\epsilon_2)|\O(k,\epsilon_1)\>\<\O(-k,\epsilon_2)|.\label{eq:completeness}
\end{align} 
This resolution of the identity is a reorganization of the one used in \eqref{eq:blockdDiscAlt}. 
Instead of summing over descendants, we are integrating $|\O(k)\>\<\O(-k)|$ over the forward lightcone.
In \eqref{eq:completeness} we have adopted an index-free notation to make expressions more compact.
The operators $\O(k,\epsilon)$ are defined by,
\begin{align}
\O(k,\epsilon)\equiv\O^{\mu_1\ldots\mu_J}(k)\epsilon_{\mu_1}\ldots\epsilon_{\mu_J},
\end{align}
where $\O$ is a symmetric traceless, primary operator and $\epsilon_{\mu}$ is a null polarization vector.
The projector $\mathcal{P}_\O(k,\epsilon_1,\epsilon_2)$ is a differential operator in $\epsilon_{1,2}$ which implements index contractions between $\O^{\mu_1\ldots\mu_J}$, $\eta_{\mu\nu}$ and $k_{\mu}$. 
The explicit form of the completeness relation can be found in \cite{Gillioz:2016jnn,Gillioz:2018kwh,Gillioz:2020wgw}.

The contribution of a single operator to \eqref{eq:disc_to_AA} is by definition the $t$-channel conformal block, $\hat{g}^t_{\O}(p_i)$, for $\disc_{t}\ll T[\f(p_1)\ldots\f(p_4)]\rr$:
\begin{align}
\hat{g}^t_{\O}(\z_i,s,t,u)&\equiv\frac{1}{\lambda_{\f\f\O}^{2}}\disc_{t}\ll T[\f(p_1)\f(p_2)\f(p_3)\f(p_4)]\rr\bigg|_{\O}
\nonumber
\\
&=-\frac{1}{\lambda_{\f\f\O}^{2}}\mathcal{P}_{\O}(p_{14},\epsilon_1,\epsilon_2)\ll A[\f(p_2);\f(p_3)]\O(p_{14},\epsilon_1)\rr
\ll\O(p_{23},\epsilon_2)A[\f(p_1);\f(p_4)]\rr
\nonumber \\ &\hspace{3.3in} +(p_1,p_4)\leftrightarrow(p_2,p_3),\label{eq:CBblockdisctPt1}
\end{align}
where $p_{ij}=p_i+p_j$.
The momentum space conformal blocks are then just a product of CFT three-point functions.
After taking into account a change of notation, \eqref{eq:CBblockdisctPt1} are exactly the conformal blocks for $\disc_{t}\tau_{4}(\z_i,s,t,u)$,
\begin{align}
\disc_{t}\tau_{4}(\z_i,s,t,u)=\sum\limits_{\O}\lambda_{\f\f\O}^{2}\hat{g}^t_{\O}(\z_i,s,t,u).
\end{align}
By expanding the integrand of \eqref{eq:disp4ptequiv} in conformal blocks and performing the dispersion integrals term by term, we find the following momentum space, Polyakov-Regge expansion,
\begin{align}
\tau_{4}(\z_i,s,t,u)&=\sum\limits_{\O}\lambda_{\f\f\O}^{2}\left(\widehat{P}^{t|s}_{\O}(\z_i,s,t,u)+\widehat{P}^{u|s}_{\O}(\z_i,s,t,u)\right)
\label{eq:blockexpansion_integrand}
\\
\widehat{P}^{t|s}_{\O}(\z_i,s,t,u)&= \int\limits_{0}^{\infty}\frac{dt'}{t'-t-i\epsilon}\hat{g}^t_{\O}(\z_i,s,t',u).\label{eq:P_hat_def}
\end{align}
By definition, the momentum space Polyakov-Regge blocks have the following discontinuities,
\begin{align}
\disc_{t}\widehat{P}^{t|s}_{\O}(\z_i,s,t,u)&=\hat{g}^t_{\O}(\z_i,s,t,u), \label{eq:P_hat_defdiscs_t}
\\
\disc_{u}\widehat{P}^{t|s}_{\O}(\z_i,s,t,u)&=0. \label{eq:P_hat_defdiscs_u}
\end{align}

If the integral in \eqref{eq:P_hat_def} converges, then $\widehat{P}^{t|s}_{\O}$ will be superbounded in the momentum space Regge limit.
However, as we saw in the previous section, dispersion integrals often diverge and need to be defined by analytic continuation.
The integral in \eqref{eq:P_hat_def} only converges if $\hat{g}^t_{\O}(\z_i,s,t,u)$ vanishes at large $t$.

To determine the large $t$ scaling of $\hat{g}^t_{\O}(\z_i,s,t,u)$, it will be useful to write this conformal block in terms of partially time-ordered correlators.
Assuming the external momenta are spacelike, $\z_i<0$, we can use the operator identities \eqref{eq:AtoT} and \eqref{eq:AtoTb} to rewrite \eqref{eq:CBblockdisctPt1} as:
\begin{align}
\hat{g}^t_{\O}(\z_i,s,t,u)&=-\frac{1}{\lambda_{\f\f\O}^{2}}\mathcal{P}_{\O}(p_{14},\epsilon_1,\epsilon_2)\ll \overline{T}[\f(p_2)\f(p_3)]\O(p_{14},\epsilon_1)\rr
\ll\O(p_{23},\epsilon_2)T[\f(p_1)\f(p_4)]\rr
\nonumber \\ 
&\hspace{3.3in} +(p_1,p_4)\leftrightarrow(p_2,p_3).\label{eq:CBblockdisctPt2}
\end{align}
Conformal blocks of this form were computed in \cite{Gillioz:2020wgw} and we will give their explicit form in Appendix \ref{app:Regge_Blocks}.
Using the results of \cite{Gillioz:2020wgw}, we find:\footnote{In comparison to \cite{Gillioz:2020wgw}, we do not remove the overall power of the Mandelstams when defining the block, see equation 1.2 there.}
\[
    \lim\limits_{t\rightarrow \infty}\hat{g}^t_{\O}(\z_i,s,t,u)\sim 
\begin{cases}
    t^{d/2-2\Delta_{\f}}\prod\limits_{i=1}^{4}(-\z_i)^{\Delta_{\f}-d/2},& \text{ if } \Delta_{\f}<d/2,\\
    t^{2\Delta_{\f}-3d/2},&  \text{ if } \Delta_{\f}>d/2.\numberthis \label{eq:cond_conv_PH}
\end{cases}
\]
with $\z_i$, $s<0$.
For $\Delta_{\f}<d/2$, the dominant term at large $t$ is non-analytic in all of the momenta and corresponds to a non-local term in position space.
In this case we have $\tilde{J}_0=J_0=d/2-2\Delta_{\f}$, which agrees with the scaling found in \eqref{eq:Regge_limit_dDisctg}.
For $\Delta_{\f}>d/2$, the dominant term is non-analytic in $t$ and analytic in the remaining momenta. 
This corresponds to the following semi-local term,
\begin{align}
\int d^{d}x_1\ldots d^{d}x_4e^{i(p_1\cdot x_1+\ldots +p_4\cdot x_4)}\delta(x_{14})\delta(x_{23})\frac{1}{x_{13}^{4\Delta_{\f}-d}}\propto t^{2\Delta_{\f}-3d/2}.
\end{align}

From \eqref{eq:cond_conv_PH}, we find that $\widehat{P}^{t|s}_{\O}$ is superbounded in momentum space if,
\begin{align}
\frac{d}{4}<\Delta_{\f}<\frac{3d}{4}.\label{eq:conv_PH}
\end{align}
For general $\Delta_{\f}$, the momentum space Polyakov-Regge block can be defined by analytic continuation.
In practice, the Polyakov-Regge block is defined for generic $\Delta_{\f}$ by adding and subtracting a finite number of terms from $\hat{g}^{t}_{\O}$ and defining the dispersion integral \eqref{eq:P_hat_def} by analytic continuation.
We will carry out this procedure in Appendix \ref{app:Regge_Blocks}.
Here we will only quote the results for the effective spins of $\widehat{P}^{t|s}_{\O}$ in the position and momentum space Regge limits:
\begin{align}
& \ \Delta_{\f}\leq \frac{d}{4} \quad \hspace{.2cm} \Longrightarrow \quad  J_0 \geq 0, \ \tilde{J}_0\geq 0, \label{eq:spinPH1}
\\[5pt]
\frac{d}{4}<&\ \Delta_{\f}<\frac{3d}{4} \quad \Longrightarrow \quad  J_0 < 0, \ \tilde{J}_0< 0, \label{eq:spinPH2}
\\[5pt]
    \frac{3d}{4}\leq & \ \Delta_{\f} \quad \hspace{1.06cm} \Longrightarrow \quad  J_0 <0, \ \tilde{J}_0\geq 0. \label{eq:spinPH3}
\end{align}
In the first case \eqref{eq:spinPH1}, $\widehat{P}^{t|s}_{\O}$ is not superbounded in position or momentum space.
In the second case \eqref{eq:spinPH2}, the original integral \eqref{eq:P_hat_def} converges and $\widehat{P}^{t|s}_{\O}$ is superbounded in both Regge limits.
In the third case \eqref{eq:spinPH3}, $\widehat{P}^{t|s}_{\O}$ is superbounded in position space, but is not superbounded in momentum space.
In other words, in the third case the non-analytic terms of $\widehat{P}^{t|s}_{\O}$ are superbounded while the semi-local terms will grow at large $t$.
Finally, using the results from the previous section, we can note that both the CFT and momentum space Polyakov-Regge blocks, $P^{t|s}_{\O}$ and $\widehat{P}^{t|s}_{\O}$, are superbounded in position space whenever $\Delta_{\f}> d/4$.

We can further generalize the results of the previous section by showing how $\widehat{P}^{t|s}_{\O}$ is related to tree-level Witten diagrams.
In Appendix \ref{app:holoblocks} we prove that $\hat{g}^{t}_{\O}$ is equal to a cut Witten diagram, 
\begin{align}
\hat{g}^t_{\O}(\z_i,s,t,u)=\text{disc}_{t}W^{t}_{\O}(\z_i,s,t,u), \label{eq:gblocktoWitten}
\end{align}
where $W^{t}_{\O}(\z_i,s,t,u)$ is the Witten diagram \eqref{eq:Witten_Diagram} in momentum space.
The identity \eqref{eq:gblocktoWitten} implies that $\widehat{P}^{t|s}_{\O}$ is equal to the dispersive transform of an exchange diagram,
\begin{align}
\widehat{P}^{t|s}_{\O}(\z_i,s,t,u)&= \int\limits_{0}^{\infty}\frac{dt'}{t'-t-i\epsilon}\text{disc}_{t'}W^t_{\O}(\z_i,s,t',u).\label{eq:PH_disp_W}
\end{align}

Given the relation \eqref{eq:PH_disp_W}, the natural ansatz is that $\widehat{P}^{t|s}_{\O}$ is given by the Fourier transform of \eqref{eq:PintermsofW},
\begin{align}
\widehat{P}^{t|s}_{\O}(\z_i,s,t,u)=W^t_{\O}(\z_i,s,t,u)+\sum\limits_{\partial} a^{\f}_{\O,\partial}W_{\partial,\text{cont}}(\z_i,s,t,u).\label{eq:Ph_to_Witten}
\end{align}
The expression \eqref{eq:Ph_to_Witten} is consistent with \eqref{eq:PH_disp_W} because contact Witten diagrams have a vanishing momentum space discontinuity in all three channels \cite{Meltzer:2020qbr}.
The expression \eqref{eq:Ph_to_Witten} is also superbounded in position space when $\Delta_{\f}>d/4$, which agrees with \eqref{eq:spinPH1}-\eqref{eq:spinPH3}.
However, it is possible that we need to add local terms to \eqref{eq:Ph_to_Witten} in order for it to be superbounded in momentum space when $d/4<\Delta_{\f}<3d/4$.
In the next section we will use uniqueness arguments to show that the two Polyakov-Regge blocks are related by a Fourier transform.

\subsection{Comparing Expansions} 
\label{ssec:comparing_dispersions}

In this section we will prove that the two Polyakov-Regge blocks, \eqref{eq:DefPR_Block2} and \eqref{eq:P_hat_def}, agree in position space at non-coincident points.
We restrict to non-local terms in position space because we do not have an independent definition for the CFT Polyakov-Regge blocks at coincident points.

From our previous analysis, we have shown the following:
\begin{enumerate}
\item When $\Delta_{\f}>d/4$, the momentum space Polyakov-Regge block, $\widehat{P}^{t|s}_{\O}$, is superbounded in position space, see \eqref{eq:spinPH1}-\eqref{eq:spinPH3}.\label{eq:prop1PH}
\item 
By definition, $\disc_{t}\widehat{P}^{t|s}_{\O}(\z_i,s,t,u)=\hat{g}_{\O}^{t}(\z_i,s,t,u)$. 
To make a connection with position space formulas, it is useful to rewrite this in terms of correlators.
If we assume our correlation function is given by a single Polyakov-Regge block,
\begin{align}
\ll T[\f(p_1)\f(p_2)\f(p_3)\f(p_4)]\rr = \widehat{P}^{t|s}_{\O}(\z_i,s,t,u),
\end{align}
then we can use \eqref{eq:disc_to_AA} and \eqref{eq:CBblockdisctPt1} to show,
\begin{align}
\ll A[\f(p_2);&\f(p_3)]A[\f(p_1);\f(p_4)]\rr 
\nonumber
\\ &=\mathcal{P}_{\O}(p_{14},\epsilon_1,\epsilon_2)\ll A[\f(p_2);\f(p_3)]\O(p_{14},\epsilon_1)\rr
\ll\O(p_{23},\epsilon_2)A[\f(p_1);\f(p_4)]\rr,
\end{align} 
or that the double-commutator of $\widehat{P}^{t|s}_{\O}$ is fixed by $\O$ exchange in the $t$-channel.
\label{eq:prop2PH}
\item The momentum-space Polyakov-Regge block has a vanishing $u$-channel double-commutator, $\text{disc}_u\widehat{P}^{t|s}_{\O}(\z_i,s,t,u)=0$.\label{eq:prop3PH}
\end{enumerate}
The properties \eqref{eq:prop1PH}-\eqref{eq:prop3PH} are shared by $P^{t|s}_{\O}$. 
The CFT Polyakov-Regge block is also superbounded for $\Delta_{\f}>d/4$, see \eqref{eq:conv_PR}.
In addition, its $t$-channel double-commutator is fixed by $\O$ exchange and its $u$-channel double-commutator vanishes, see \eqref{eq:dDiscPt} and \eqref{eq:dDiscPu} respectively.
Functions which are superbounded in position space are uniquely fixed by their double-discontinuities \cite{Caron-Huot:2017vep,Carmi:2019cub} and therefore the two Polyakov-Regge blocks are related by a Fourier transform,
\begin{align}
\widehat{P}_{\O}^{t|s}(\z_i,s,t,u) \quad \xleftrightarrow{\text{Fourier Transform}} \quad \frac{1}{(x_{12}x_{34})^{2\Delta_{\f}}}P_{\O}^{t|s}(z,\bar{z})  \quad \text{if } \ \Delta_{\f}>\frac{d}{4}.\label{eq:equiv_PhP_smalldim}
\end{align}
On the right hand side we restored the overall powers of $x_{ij}$ that were factored out in \eqref{eq:mG_def}.
Both Polyakov-Regge blocks are analytic functions of $\Delta_{\f}$, which allows us to analytically continue the relation \eqref{eq:equiv_PhP_smalldim} to arbitrary external dimension:
\begin{align}
\widehat{P}_{\O}^{t|s}(\z_i,s,t,u) \quad \xleftrightarrow{\text{Fourier Transform}} \quad \frac{1}{(x_{12}x_{34})^{2\Delta_{\f}}}P_{\O}^{t|s}(z,\bar{z})  \quad \text{for all} \ \Delta_{\f}.\label{eq:gen_equivalence}
\end{align}
Both relations \eqref{eq:equiv_PhP_smalldim} and \eqref{eq:gen_equivalence} should be understood to hold at non-coincident points in position space, or for non-analytic terms in momentum space.

The relation \eqref{eq:gen_equivalence} proves that if we expand the integrand of the CFT and momentum space dispersion formulas using the OPE and then perform the dispersion integrals, the resulting sums agree by term by term.
It also proves that the momentum space Polyakov-Regge blocks have the holographic representation \eqref{eq:Ph_to_Witten}.
In other words, if we apply both dispersion formulas to the same $t$-channel exchange Witten diagram, $W^{t}_{\O}$, then they reproduce the original exchange diagram plus the same sum of quartic, contact Witten diagrams.


\section{Applications to CFT Correlators}
\label{sec:app_to_CFT}
In this section we will test the momentum space dispersion formula on various CFT correlation functions.
In Section \ref{sec:check23} we consider basic consistency checks for scalar CFT two- and three-point functions.
In Section \ref{sec:scalarfourpoint} we study the dispersion formula for Witten diagrams with four external scalars.
In Section \ref{sec:addspin} we consider the dispersion formula for external spinning operators.
\subsection{Two and Three-Point Functions}
\label{sec:check23}
The simplest correlator we can consider is the scalar, CFT two-point function. In position space the time-ordered two-point function is,
\begin{align}
\tau_{2}(x)=\frac{1}{(x^2+i\epsilon)^{\Delta}}.
\end{align}
Taking the Fourier transform yields,
\begin{align}
\tau_{2}(\z)=-i\frac{\pi^d\Gamma\left(\frac{d}{2}-\Delta\right)}{2^{2\Delta-d}\Gamma(\Delta)}(-\z-i\epsilon)^{\Delta-d/2},\label{eq:tau_2_k_space}
\end{align}
where $\z=-p^{2}$. 
This Fourier transform needs to be defined by analytic continuation for $\Delta\geq \frac{d}{2}$.
The dispersive representation \eqref{eq:2ptdispfromanalyticity} for the CFT two-point function is then simple to write down \cite{Mack:1973mq}:
\begin{align}
\tau_2(\z)=\frac{\pi ^{d+1} 2^{d-2 \Delta +1}}{\Gamma (\Delta ) \Gamma \left(-\frac{d}{2}+\Delta +1\right)}\int\limits_0^{\infty} \frac{d\z'}{2\pi i}\frac{\z'^{\Delta-d/2}}{\z'-\z-i\epsilon}.\label{eq:two_pt_KL_CFT}
\end{align}
The above integral only converges when $\frac{d-2}{2}<\Delta<\frac{d}{2}$.\footnote{When $\Delta=\frac{d-2}{2}$ we can use $\lim\limits_{\epsilon\rightarrow 0}\Gamma^{-1}(\epsilon) \z^{\epsilon}\theta(\z) =\delta(\z)$ to recover the free-field two-point function \cite{Mack:1973mq}.} 
For $\Delta>d/2$ we can define the dispersive integral by analytic continuation. 
When $\Delta=d/2+n$ the momentum space correlator \eqref{eq:tau_2_k_space} diverges and needs further regularization and renormalization \cite{Bzowski:2013sza,Bzowski:2015pba}.

We can also test the three-point dispersion formula on a scalar, CFT three-point function. Correlators of the form $\<T[\f(p_1)\f(p_2)\O_{\Delta,J}(p_3)]\>$, where $\O$ is a generic spinning operator, are completely fixed by conformal symmetry. 
This implies that the time-ordered, three-point function of scalars is proportional to a three-point Witten diagram in AdS $\Phi^3$ theory,\footnote{Adding derivative interactions does not affect the final form of the three-point Witten diagram.}
\begin{align}
\includegraphics[scale=.33]{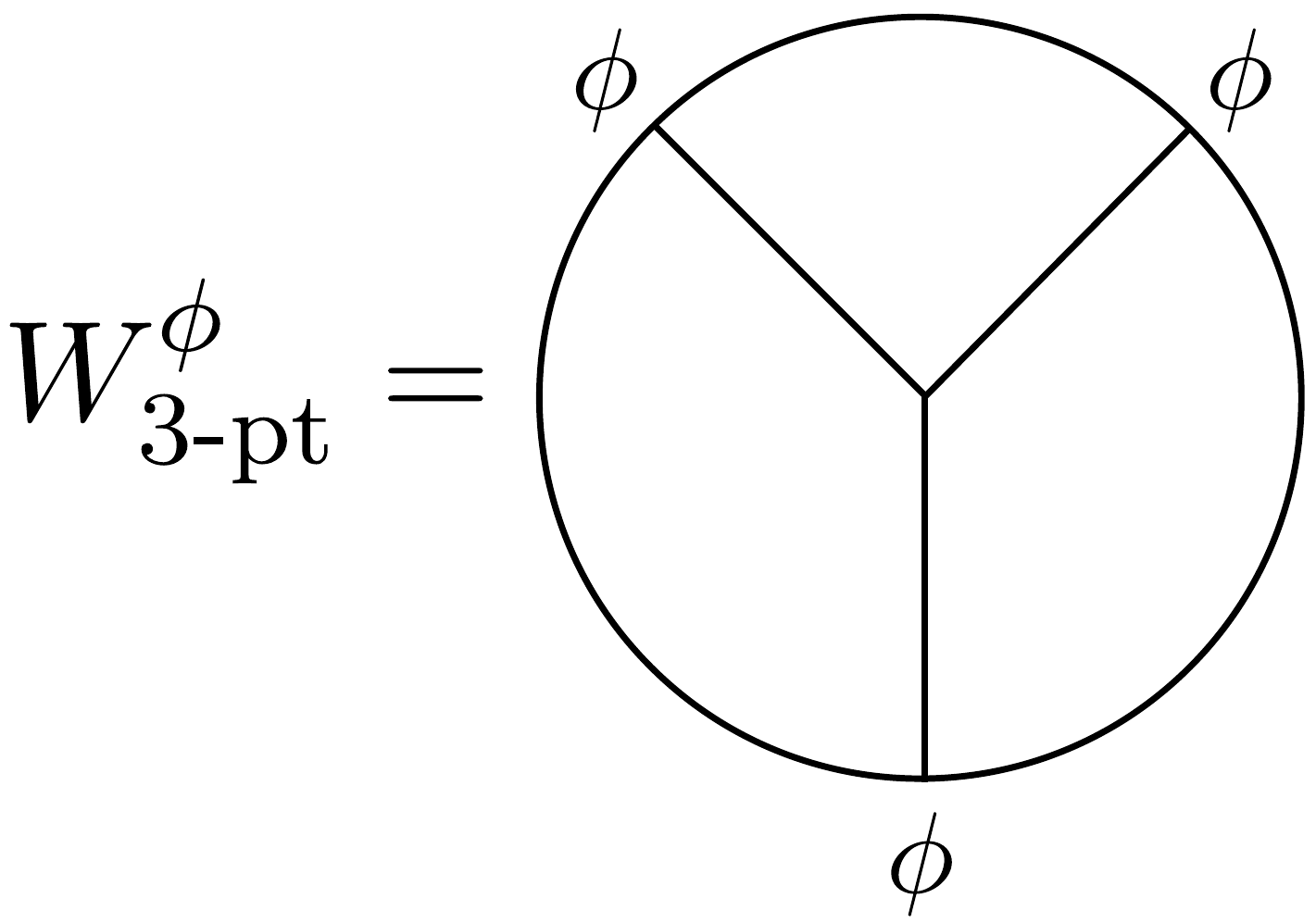},
\label{eq:3pt_Witten}
\end{align}
which is given by,
\begin{align}
W^{\phi}_{\text{3-pt}}(\z_1,\z_2,\z_3)=ig \int\frac{dz}{z^{d+1}}K_{\Delta,0}(p_1,z)K_{\Delta,0}(p_2,z)K_{\Delta,0}(p_3,z).\label{eq:three_pt_int}
\end{align}
Here $g$ is the bulk coupling and $K_{\Delta,0}$ is the boundary-to-bulk propagator for a scalar of dimension $\Delta$.
We then want to show that \eqref{eq:three_pt_int} has the dispersive representation:
\begin{align}
W^{\phi}_{\text{3-pt}}(\z_1,\z_2,\z_3)=\frac{1}{2\pi i}\int\limits_{0}^{\infty} \frac{d\z'_3}{\z'_3-\z_3-i\epsilon}\disc_{\z'_3}W^{\phi}_{\text{3-pt}}(\z_1,\z_2,\z'_3). \label{eq:three_pointWittenTest}
\end{align}
To prove \eqref{eq:three_pointWittenTest}, it is sufficient to show that the boundary-to-bulk propagator obeys a two-point dispersion formula,
\begin{align}
K_{\Delta,0}(p,z)=\frac{1}{2\pi i}\int\limits_{0}^{\infty} \frac{d\z'}{\z'-\z-i\epsilon}\disc_{\z'}K_{\Delta,0}(p',z),\label{eq:bulkbdy_two_pt_dispersion}
\end{align}
where $\z'=-p'^2$. By using an AdS representation for the correlator, we have reduced a boundary three-point dispersion formula to a simpler bulk two-point dispersion formula.

One can recognize that \eqref{eq:bulkbdy_two_pt_dispersion} is the K\"all\'en-Lehmann spectral representation for the boundary-to-bulk propagator. 
To prove \eqref{eq:bulkbdy_two_pt_dispersion}, we can export the arguments of Section \ref{sec:2_3_ptanalyticity} to AdS in order to derive the dispersive representation of the bulk-to-bulk propagator. 
If we then take one point to the boundary, we find the dispersive form of the boundary-to-bulk propagator \eqref{eq:bulkbdy_two_pt_dispersion}. To be concrete, in this section we will instead show \eqref{eq:bulkbdy_two_pt_dispersion} follows from the explicit form of the scalar boundary-to-bulk propagator \cite{Muck:1998rr,Liu:1998ty},
\begin{align}
K_{\Delta,0}(p,z)=-i\frac{1}{2^{\nu}\Gamma(1+\nu)}z^{\frac{d}{2}}(-\z-i\epsilon)^{\nu/2}\mathcal{K}_{\nu}(z\sqrt{-\z-i\epsilon}\hspace{.1cm}).\label{eq:scalarBbdy}
\end{align}
Here $\nu=\Delta-d/2$ and $\mathcal{K}$ is the modified Bessel function of the second kind. We can directly rewrite the boundary-to-bulk propagator in a dispersive form,
\begin{align}
K_{\Delta,0}(p,z)=\frac{\pi}{2^{\nu}\Gamma(1+\nu)}\int\limits_{0}^{\infty}\frac{d\z'}{2\pi i}\frac{\z'^{\nu/2}\mathcal{J}_{\nu}(\sqrt{\z'}z)}{\z'-\z-i\epsilon},\label{eq:K_bulk_bdy_dispersive}
\end{align}
where $\mathcal{J}_{\nu}$ is the Bessel function of the first kind.
Taking the discontinuity in $\z$, it is manifest \eqref{eq:K_bulk_bdy_dispersive} obeys the bulk dispersion formula \eqref{eq:bulkbdy_two_pt_dispersion}. 
This in turn implies that the original CFT three-point function satisfies the boundary dispersion formula \eqref{eq:three_pointWittenTest}. 

The above argument requires commuting the $\z'$ and $z$ integrals. 
If the $z$ integral diverges, then we can apply the same analysis to a regularized Witten diagram where we take the $z$ integral to run over $(z_{c},\infty)$ with $z_{c}>0$.\footnote{The Bessel function $\mathcal{K}_{\nu}(z\sqrt{-\z-i\epsilon})$ decays exponentially at large $z$, so the large $z$ region of integration does not introduce any divergences.} 
In addition, the dispersion integral \eqref{eq:K_bulk_bdy_dispersive} only converges for $|\nu|<1/3$ and can be defined for general values of $\nu$ by analytic continuation.
This is analogous to how the dispersive form of the CFT two-point function \eqref{eq:two_pt_KL_CFT} is defined for general $\Delta$.

\subsection{Scalar Witten Diagrams}
\label{sec:scalarfourpoint}
In this section we test the momentum space dispersion formula on exchange Witten diagrams with external scalars. 
From the arguments of Section \ref{sec:equivalence}, we know that acting with the momentum space dispersion formula on a $t$-channel exchange diagram reproduces the same diagram, up to a possible linear combination of contact diagrams. 
The goal of this section is to demonstrate how this works using explicit examples. 
For the remainder of this section we will use $\phi$ to denote a boundary scalar with general dimension $\Delta$ and $\varphi$ to denote a conformally coupled scalar with dimension $\Delta_{\varphi}=\frac{1}{2}(d+1)$. 
We also use $J^{\mu}$ and $T^{\mu\nu}$ to denote spin-one and spin-two conserved operators, respectively.\footnote{We hope it is clear from context where $J$ refers to the conserved current versus the spin of a general operator.}
\subsubsection*{Scalar Exchange Diagrams}
One simple Witten diagram is the scalar exchange diagram in AdS $\Phi^3$ theory. 
The bulk field $\Phi$ is dual to a boundary scalar $\f$ and the $t$-channel Witten diagram is,
\begin{align}
\includegraphics[scale=.33]{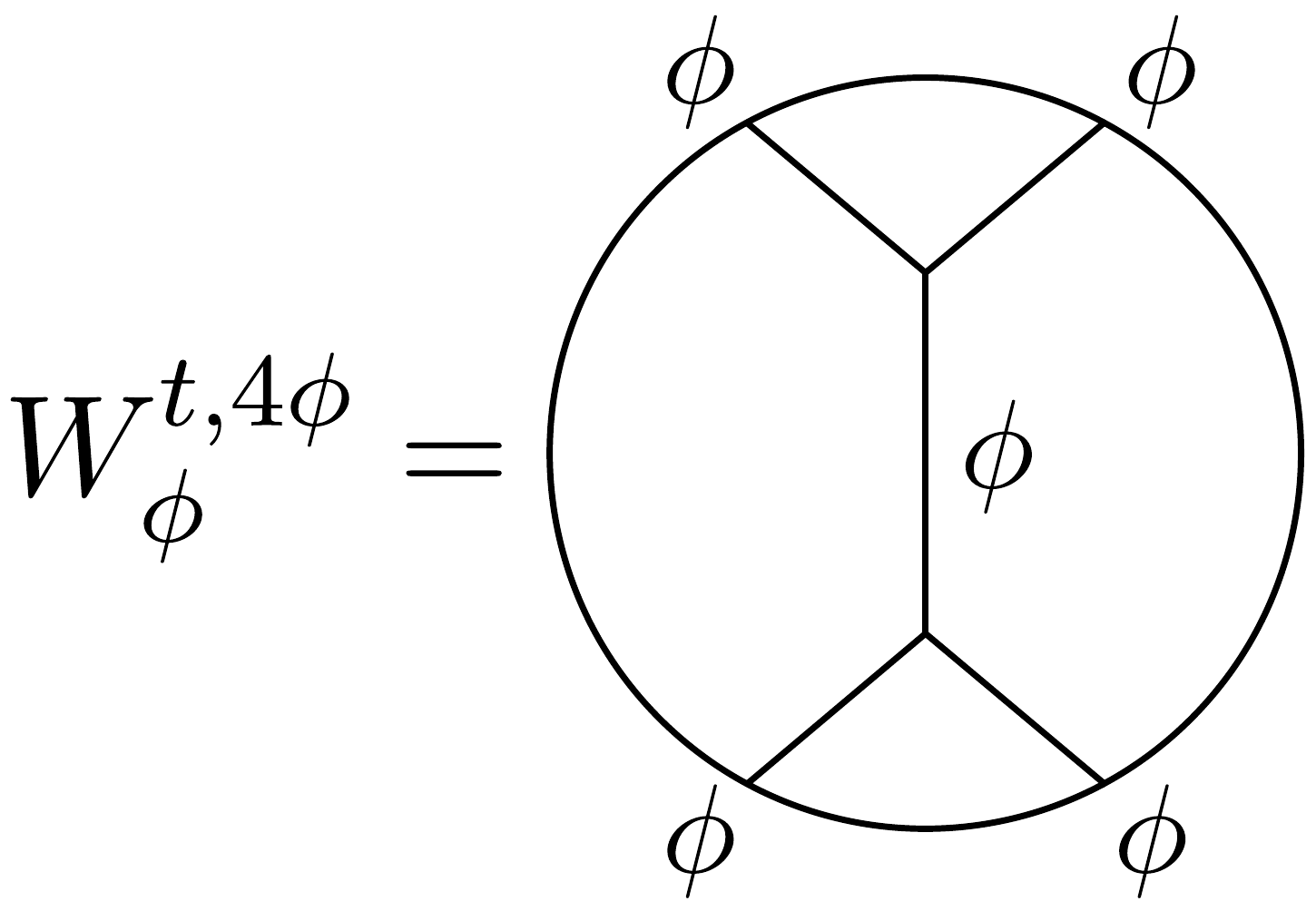}
\end{align}
or,
\begin{align}
W^{t,4\f}_{\f}(\z_i,s,t,u)=-g^{2}\int\frac{dz_1dz_2}{z_1^{d+1}z_2^{d+1}}&K_{\f}(p_1,z_1)K_{\f}(p_4,z_1)G_{\f}(p_{14},z_1,z_2)
\nonumber
\\
&\times K_{\f}(p_2,z_2)K_{\f}(p_3,z_2). \label{eq:scalar_exchange_diagram}
\end{align}
Here $G_{\f}(p_{14},z_1,z_2)$ is the bulk-to-bulk propagator of $\Phi$.
To prove that the four-point Witten diagram obeys the dispersion formula,
\begin{align}
W^{t,4\f}_{\f}(\z_i,s,t,u)=\frac{1}{2\pi i}\int\limits_{0}^{\infty}\frac{dt'}{t'-t-i\epsilon}\disc_{t'}W^{t,4\f}_{\f}(\z_i,s,t',u),\label{eq:disp_sc_exchange}
\end{align}
it is sufficient to show that the bulk-to-bulk propagator $G_{\Delta,0}$ can be computed using a bulk dispersion formula,
\begin{align}
G_{\Delta,0}(p,z_1,z_2)=\frac{1}{2\pi i}\int\limits_{0}^{\infty}\frac{dt'}{t'-t-i\epsilon}\disc_{t'}G_{\Delta,0}(p',z_1,z_2), \label{eq:disp_bulk_twopt}
\end{align}
where here $t'=-p'^2$. 
This is the bulk K\"all\'en-Lehmann spectral representation and, as we discussed after \eqref{eq:bulkbdy_two_pt_dispersion}, can be proven using the arguments of Section \ref{sec:2_3_ptanalyticity} in AdS. 
Here we will instead show \eqref{eq:disp_bulk_twopt} holds using the known form of the bulk-to-bulk propagator.
In fact, in \cite{Liu:1998ty} the scalar bulk-to-bulk propagator is already given in a dispersive form:
\begin{align}
G_{\Delta,0}(p,z_1,z_2)=-\frac{i}{2}(z_1z_2)^{\frac{d}{2}}\int\limits_{0}^{\infty}dt'\frac{\mathcal{J}_\nu(\sqrt{t'}z_1)\mathcal{J}_\nu(\sqrt{t'}z_2)}{t'-t-i\epsilon}, \label{eq:Gbulkscalar}
\end{align}
where $t=-p^2$ and $\nu=\Delta-d/2$. 
The $t'$ integral in \eqref{eq:Gbulkscalar} converges for all $\nu$. 
From \eqref{eq:Gbulkscalar} we see that the dispersive representation \eqref{eq:disp_sc_exchange} is valid for all scalar exchange diagrams.\footnote{Dispersive representations of scalar exchange Witten diagrams were found in a purely CFT context in \cite{Polyakov:1974gs,Ferrara:1974nf}. There the motivation was to write down simple crossing symmetric, unitary correlators.} 
It is simple to generalize this to distinct external scalars $\<\f_1\f_2\f_3\f_4\>$ exchanging a generic scalar field.
Proving \eqref{eq:disp_sc_exchange} only depended on knowing the dispersive form of the bulk-to-bulk propagator, which we can insert in any Witten diagram. 
The analysis also carries over to higher-point Witten diagrams with non-derivative interactions.

One subtlety, as we discussed in the previous section, is that in general the radial integrals diverge at small $z$.
If the $z$ integral diverges, we can again regularize the Witten diagram by taking the $z$ integral to run over $(z_c,\infty)$ with $z_c>0$.
Cutting off the $z$-integral removes semi-local terms that grow at large $t$, and this explains why regularized scalar exchange diagrams satisfy an unsubtracted dispersion formula for generic $\Delta_{\f}$.

Next, we will consider cases where the Witten diagrams are known in closed form. 
One such Witten diagram is the exchange diagram \eqref{eq:scalar_exchange_diagram} for conformally coupled scalars $\varphi$ in AdS$_{6}$ \cite{Arkani-Hamed:2017fdk,Albayrak:2020isk}:
\begin{align}
W^{t,4\varphi}_{\varphi}(p_i)\bigg|_{d=5}=\frac{1}{E_{\text{tot}}}\frac{ig^{2}}{\left(\sqrt{-t}+\sqrt{-\z_1}+\sqrt{-\z_4}\right)\left(\sqrt{-t}+\sqrt{-\z_2}+\sqrt{-\z_3}\right)},\label{eq:WittenAdS6conf}
\end{align}
where $d$ is the boundary spacetime dimension. 
We also defined the total ``energy",\footnote{The ``energy" variables $|p_i|\equiv\sqrt{-\z_i}=\sqrt{p_i^2}$ are often used when studying the flat-space limit $E_{\text{tot}}\rightarrow0$. In this limit $|p_i|$ become the energy components of the $(d+1)$-dimensional momenta \cite{Raju:2012zr}.}
\begin{align}
E_{\text{tot}}=\sum\limits_{i}\sqrt{-\z_i}.\label{eq:totEnergy}
\end{align}
Expanding \eqref{eq:WittenAdS6conf} at large $t$ we find,
\begin{align}
W^{t,4\varphi}_{\varphi}(p_i)\bigg|_{d=5}\approx -ig^2\left(\frac{1}{E_{\text{tot}}t}+\frac{1}{(-t)^{\frac{3}{2}}}+\ldots\right).\label{eq:conf_scale_larget}
\end{align}
The first term in \eqref{eq:conf_scale_larget} is non-analytic in the momenta and decays like $t^{-1}$ in the Regge limit.
This comes from a non-local term in position space, which means $\tilde{J}_0=J_0=-1$.
This is consistent with previous results that show $t$-channel, spin-$J$ exchange diagrams have $J_0=J-1$ \cite{Costa:2012cb}.\footnote{The more familiar scaling $t^J$ comes from studying the large $t$ limit of $s$-channel, spin-$J$ exchange diagrams.}
The second term in \eqref{eq:conf_scale_larget} is semi-local and agrees with the scaling predicted from the second line of \eqref{eq:cond_conv_PH} when we set $d=5$ and $\Delta_{\f}=3$.
In \eqref{eq:conf_scale_larget} we see that the semi-local terms are subleading to the non-local terms in the momentum space Regge limit.

The same diagram in $d=3$, or in AdS$_{4}$, will have a slower decay at large $t$ due to the semi-local terms.
This Witten diagram has been computed in closed form \cite{Baumann:2020dch} and is given by,\footnote{In \cite{Baumann:2020dch} they are computing wavefunction coefficients in dS. To convert notation, $s_{\text{here}}=-s_{\text{there}}^{2}$ and similarly for $t$ and $u$. In addition $p_{ij,\text{there}}=|p_i|+|p_j|=\sqrt{-\z_i}+\sqrt{-\z_j}$. \label{footnote:conventions}}
\begin{align}
W^{t,4\varphi}_{\varphi}(p_i)\bigg|_{d=3}=\frac{ig^{2}}{2\sqrt{-t}}\bigg[&\text{Li}_{2}\left(\frac{\sqrt{-\z_1}+\sqrt{-\z_4}-\sqrt{-t}}{E_{\text{tot}}}\right)+\text{Li}_{2}\left(\frac{\sqrt{-\z_2}+\sqrt{-\z_3}-\sqrt{-t}}{E_{\text{tot}}}\right)
\nonumber
\\
&\hspace{-.18in}+\log\left(\frac{\sqrt{-\z_1}+\sqrt{-\z_4}+\sqrt{-t}}{E_{\text{tot}}}\right)\log\left(\frac{\sqrt{-\z_2}+\sqrt{-\z_3}+\sqrt{-t}}{E_{\text{tot}}}\right)\bigg].\label{eq:AdS4Confscal}
\end{align}
The Witten diagram \eqref{eq:AdS4Confscal} has a branch cut starting at $t=0$ as expected. 
In the limit $t\gg1$ we find,
\begin{align}
W^{t,4\varphi}_{\varphi}(p_i)\bigg|_{d=3}\approx -\frac{i \pi ^2 g^2}{6 \sqrt{-t}}.
\end{align}
We see that a semi-local term gives the leading contribution to the momentum space Regge limit.
The $t^{-1/2}$ scaling also matches the scaling predicted from \eqref{eq:cond_conv_PH} when we set $d=3$ and $\Delta_{\f}=2$.
The non-analytic terms are subleading and first appear at order $t^{-1}$.
The AdS$_4$ Witten diagram $W^{t,4\varphi}_{\varphi}(p_i)$ then has effective spin $\tilde{J}_0=-1/2$ in the momentum space Regge limit and $J_0=-1$ in the position space Regge limit.
This is an example where $\tilde{J}_0>J_0$ and the momentum space Regge limit is not equal to the Fourier transform of the position space Regge limit.

As a further consistency check, we can study the exchange of a massless scalar $\chi$ between conformally coupled scalars $\varphi$ in AdS$_4$.
The scalar $\chi$ has dimension $\Delta_{\chi}=3$ and the $t$-channel exchange Witten diagram is \cite{Baumann:2020dch}:
\begin{align}
W^{t,4\varphi}_{\chi}(p_i)\bigg|_{d=3}=\frac{-i}{\sqrt{-t}}\frac{1}{2 v w}\bigg[&\text{Li}_2\left(\frac{v-v w}{v+w}\right)+\text{Li}_2\left(\frac{w-v w}{v+w}\right)+2 w \left(v-\log \left(\frac{(v+1) w}{v+w}\right)\right)
\nonumber
\\
&+\left(\log \left(\frac{(v+1) w}{v+w}\right)-2 v\right) \log \left(\frac{v (w+1)}{v+w}\right)\bigg],\label{eq:massless_exchange}
\end{align}
where the cross-ratios $v$ and $w$ are defined by,
\begin{align}
v=\frac{\sqrt{-t}}{\sqrt{-\z_2}+\sqrt{-\z_3}},
\qquad \quad
w=\frac{\sqrt{-t}}{\sqrt{-\z_1}+\sqrt{-\z_4}}.
\end{align}
Up to the overall power of $t$, \eqref{eq:massless_exchange} is only a function of the cross-ratios $(v,w)$.
In the large $t$ limit we find,
\begin{align}
W^{t,4\varphi}_{\chi}(p_i)\approx -\frac{i}{\sqrt{-t}}.
\end{align}
This is a semi-local term whose scaling again agrees with the second line of \eqref{eq:cond_conv_PH} for $\Delta_{\f}=2$ and $d=3$.
The non-analytic terms are subleading and first appear at $t^{-1}$. 
This is also consistent with the effective spin $J_0=-1$ found in the position space Regge limit \cite{Costa:2012cb}.
\subsubsection*{Gauge Boson Exchange}
Here we will study $U(1)$ gauge boson exchange between external conformally coupled scalars $\varphi$ in AdS$_{4}$,
\begin{align}
\includegraphics[scale=.33]{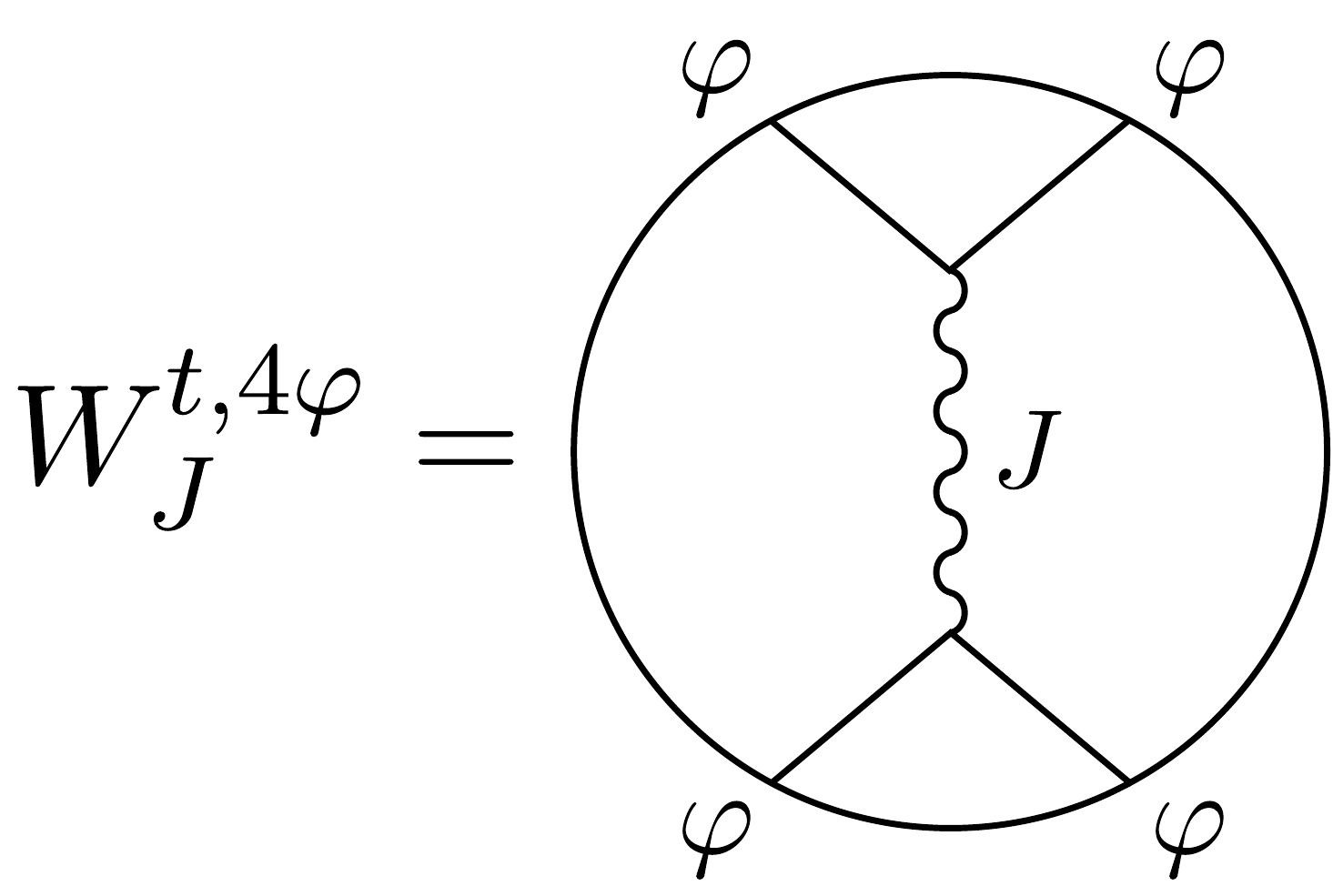}.
\end{align}
This Witten diagram has been computed in \cite{Baumann:2020dch} and we will quote the final answer here. After taking into account differences in convention, see footnote \ref{footnote:conventions}, the $t$-channel exchange diagram is:
\begin{align}
W^{t,4\varphi}_{J}(p_i)\bigg|_{d=3}&=ig_{\text{YM}}^{2}\frac{t\Pi_{1,1}+E_LE_R\Pi_{1,0}}{E_{\text{tot}}E_LE_R}.\label{eq:gaugebosonAdS4}
\end{align}
Here the left and right ``energies" are defined by:
\begin{align}
E_{L}&=\sqrt{-\z_1}+\sqrt{-\z_4}+\sqrt{-t},\label{eq:EL}
\\
E_{R}&=\sqrt{-\z_2}+\sqrt{-\z_3}+\sqrt{-t},\label{eq:ER}
\end{align}  
and $E_{\text{tot}}$ is the total energy \eqref{eq:totEnergy}. The terms $\Pi_{1,1}$ and $\Pi_{1,0}$ are,
\begin{align}
\Pi_{1,0}&=\frac{(\sqrt{-\z_2}-\sqrt{-\z_3})(\sqrt{-\z_1}-\sqrt{-\z_4})}{t},
\\
\Pi_{1,1}&=\frac{t (u-s)-(\z_2-\z_3) (\z_1-\z_4)}{t^2}.
\end{align}
These terms originate from contracting the bulk vertices with the AdS gauge boson propagator.
 
Unlike the scalar exchange diagrams, the gauge boson exchange diagram does not vanish in the limit $|t|\rightarrow\infty$. Instead we have,
\begin{align}
\lim\limits_{|t|\rightarrow\infty} W^{t,4\varphi}_{J}(p_i)\bigg|_{d=3}=\frac{ig_{\text{YM}}^{2}}{E_{\text{tot}}}.
\end{align}
We can recognize that the right hand side is equal to the $\Phi^4$ contact diagram for conformally coupled scalars in AdS$_{4}$, 
\begin{align}
\includegraphics[scale=.33]{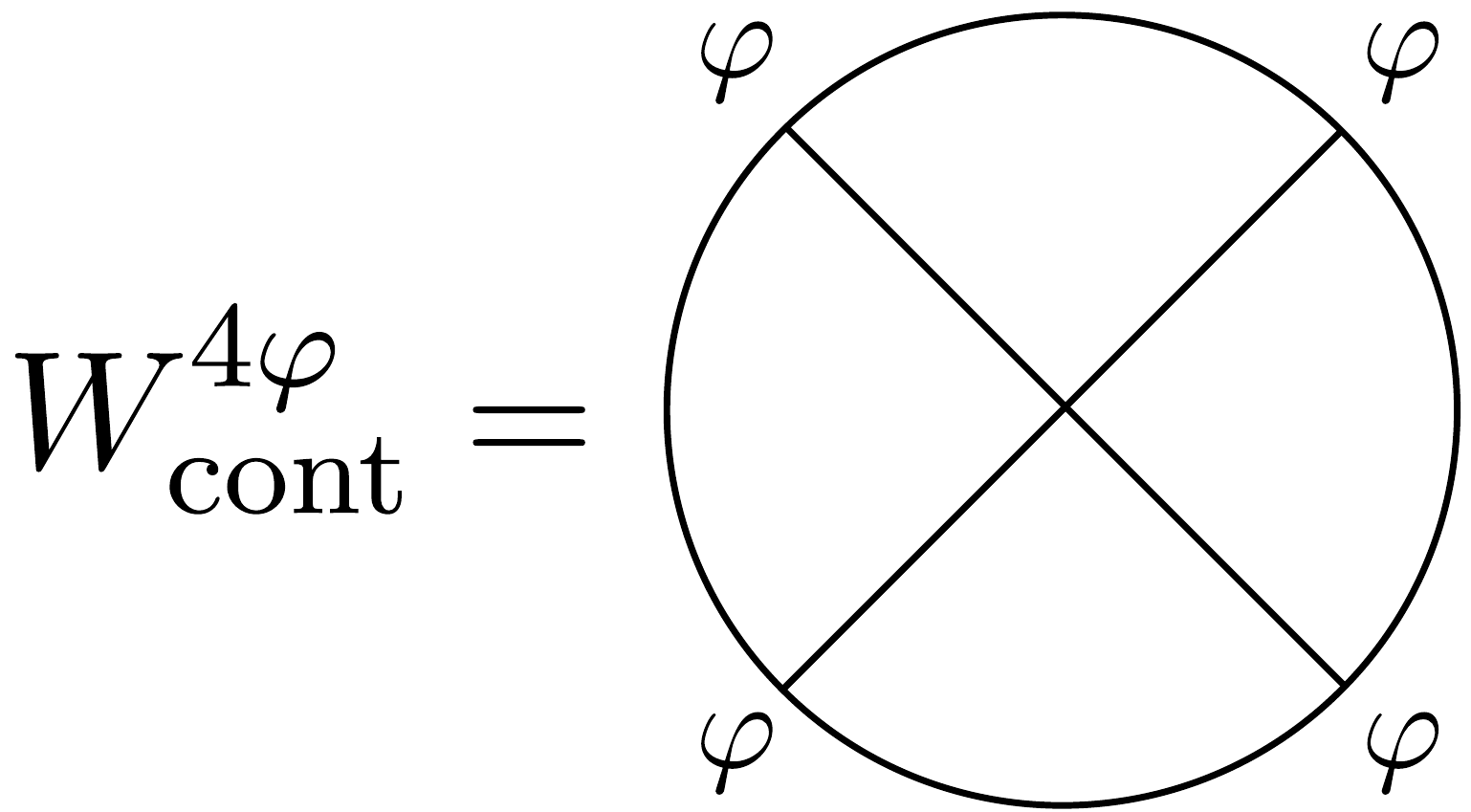}.
\end{align}
This diagram is given by,\footnote{We will not include bulk couplings for contact diagrams that are used as subtraction terms.}
\begin{align}
W^{4\varphi}_{\text{cont}}(p_i)\bigg|_{d=3}&=\int\limits_{0}^{\infty}\frac{dz}{z^{d+1}}K_{\varphi}(p_1,z)K_{\varphi}(p_2,z)K_{\varphi}(p_3,z)K_{\varphi}(p_4,z)\bigg|_{d=3}
\nonumber \\
&=\frac{1}{E_{\text{tot}}}.\label{eq:contactAdS4}
\end{align}
We can then construct the following Polyakov-Regge block for $\<\varphi\varphi\varphi\varphi\>$ in $3d$ CFTs,
\begin{align}
\widehat{P}^{t|s}_{J}(p_i)\bigg|_{d=3}= W^{t,4\varphi}_{J}(p_i)-ig_{\text{YM}}^{2}W^{4\varphi}_{\text{cont}}(p_i),
\end{align}
which vanishes in the large $t$ Regge limit.
This is an example where the four-point dispersion formula reproduces the original exchange diagram, up to quartic, contact diagrams:
\begin{align}
\widehat{P}^{t|s}_{J}(\z_i,s,t,u)=\frac{1}{2\pi i}\int\limits_{0}^{\infty}\frac{dt'}{t'-t-i\epsilon}\disc_{t'}W^{t,4\varphi}_{J}(\z_i,s,t',u).
\end{align}
\subsubsection*{Graviton Exchange}
Here we will study $t$-channel graviton exchange between external conformally coupled scalars in AdS$_{4}$,
\begin{align}
\includegraphics[scale=.33]{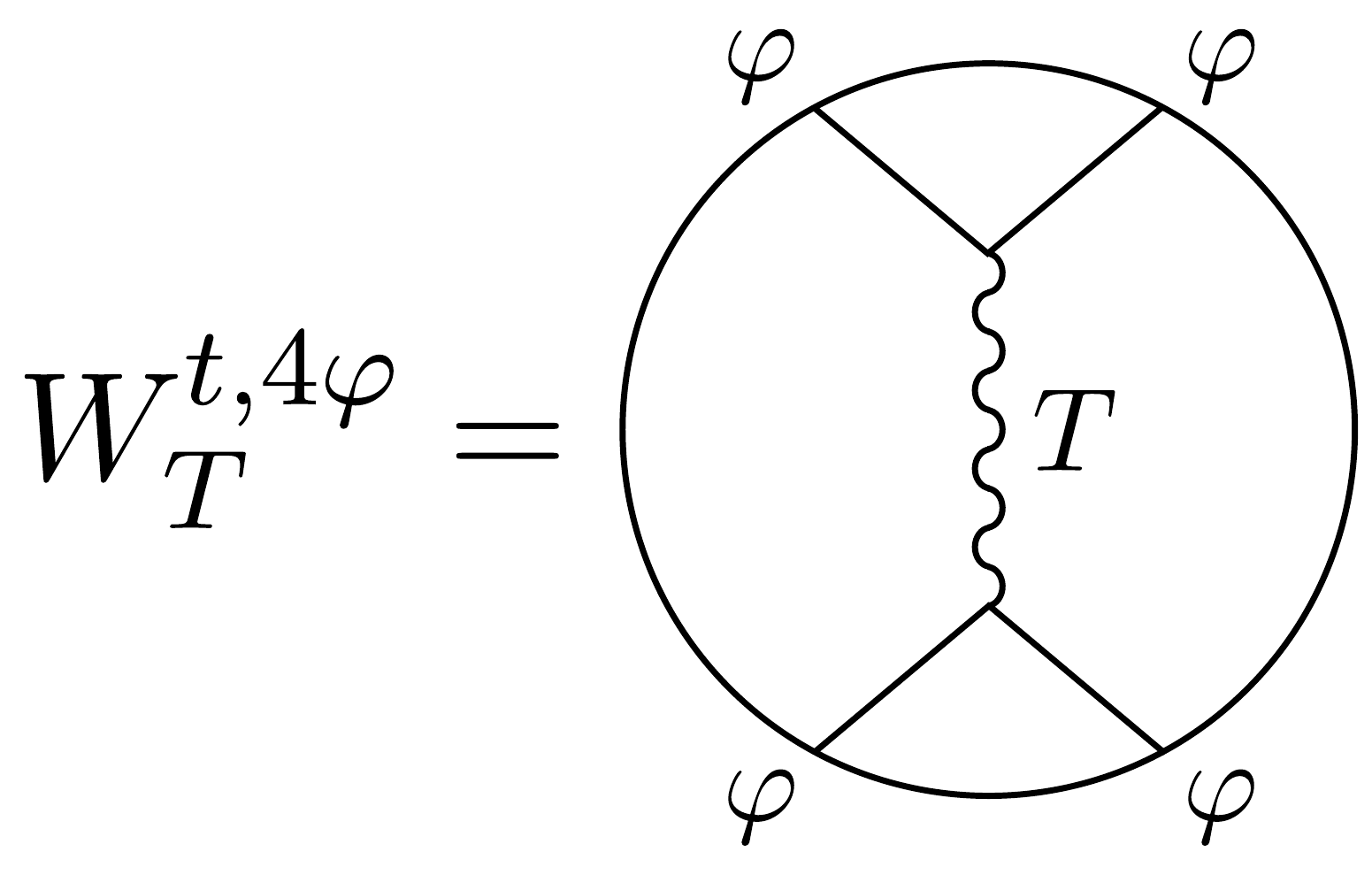}.
\end{align}
This diagram can be expressed in terms of the massless scalar exchange diagram \cite{Baumann:2020dch},
\begin{align}
W^{t,4\varphi}_T(p_i)=\frac{1}{\sqrt{-t}}\left(\Pi_{2,2}D^{2}_{wv}+\Pi_{2,1}D_{wv}(\Delta_{w}-2)+\Pi_{2,0}\Delta_{w}(\Delta_{w}-2)\right)\sqrt{-t}W^{t,4\varphi}_{\chi}(p_i),\label{eq:graviton_exchange}
\end{align}
where $W^{t,4\varphi}_{\chi}(p_i)$ is defined in \eqref{eq:massless_exchange}. 
The differential operators $D_{wv}$ and $\Delta_{w}$ are,
\begin{align}
D_{wv}&=w^{2}v^{2}\partial_{w}\partial_{v},
\\
\Delta_{w}&=w^{2}(1-w^{2})\partial_{w}^{2}-2w^{3}\partial_{w}.
\end{align}
We multiplied by $\sqrt{-t}$ in \eqref{eq:graviton_exchange} so the differential operators act on a function of $(v,w)$ alone.
The $\Pi_{i,j}$ are polarization sums which are defined in Appendix E of \cite{Baumann:2020dch}. 

The full form of the Witten diagram $W^{t,4\varphi}_T(p_i)$ is complicated, so we will not write out the full expression here. 
Instead, we will only write down the leading terms in the large $t$ Regge limit,
\begin{align}
\lim\limits_{t\rightarrow \infty}W^{t,4\varphi}_T(p_i)\hspace{.1cm} \approx & \hspace{.2cm}\frac{t-\left(\sqrt{-\z_1}+\sqrt{-\z_4}\right) \left(\sqrt{-\z_2}+\sqrt{-\z_3}\right)}{2 \left(\sqrt{-\z_1}+\sqrt{-\z_2}+\sqrt{-\z_3}+\sqrt{-\z_4}\right)^3}.\label{eq:remainderT}
\end{align}
We can recognize that \eqref{eq:remainderT} is equal to a two-derivative contact diagram for conformally-coupled scalars in AdS$_{4}$.\footnote{If we impose permutation symmetry, two derivative contact interactions for four identical scalars are trivial. However, here the two-derivative contact diagram is a subtraction term, and not a genuine interaction in our theory, so we do not impose permutation symmetry.} 
Following \cite{Baumann:2020dch}, we can construct the higher-derivative quartic diagram by acting with $\Delta_{w}$ on the non-derivative quartic diagram $W^{4\varphi}_{\text{cont}}(p_i)$,
\begin{align}
W^{4\varphi}_{\text{2$\partial$-cont}}(p_i)&=\frac{1}{\sqrt{-t}}\Delta_{w}(\sqrt{-t}W^{4\varphi}_{\text{cont}}(p_i))
\nonumber \\
&=\frac{2 \left(t-\left(\sqrt{-\z_1}+\sqrt{-\z_4}\right) \left(\sqrt{-\z_2}+\sqrt{-\z_3}\right)\right)}{\left(\sqrt{-\z_1}+\sqrt{-\z_2}+\sqrt{-\z_3}+\sqrt{-\z_4}\right)^3}. \label{eq:subterm_spin1}
\end{align}
The functional form of \eqref{eq:subterm_spin1} matches \eqref{eq:remainderT} exactly.
Using this contact diagram, we can construct the $d=3$, stress-tensor Polyakov-Regge block for conformally coupled scalars,
\begin{align}
\widehat{P}^{t|s}_{T}(p_i)=W^t_T(p_i)-\frac{1}{4}W^{4\varphi}_{\text{2$\partial$-cont}}(p_i).
\end{align}
\subsubsection*{Loop Diagrams}
In this section we will study the dispersive representation of loop-level diagrams. 
We will first consider the one-loop bubble diagram in AdS $\Phi^4$ theory,
\begin{align}
\includegraphics[scale=.31]{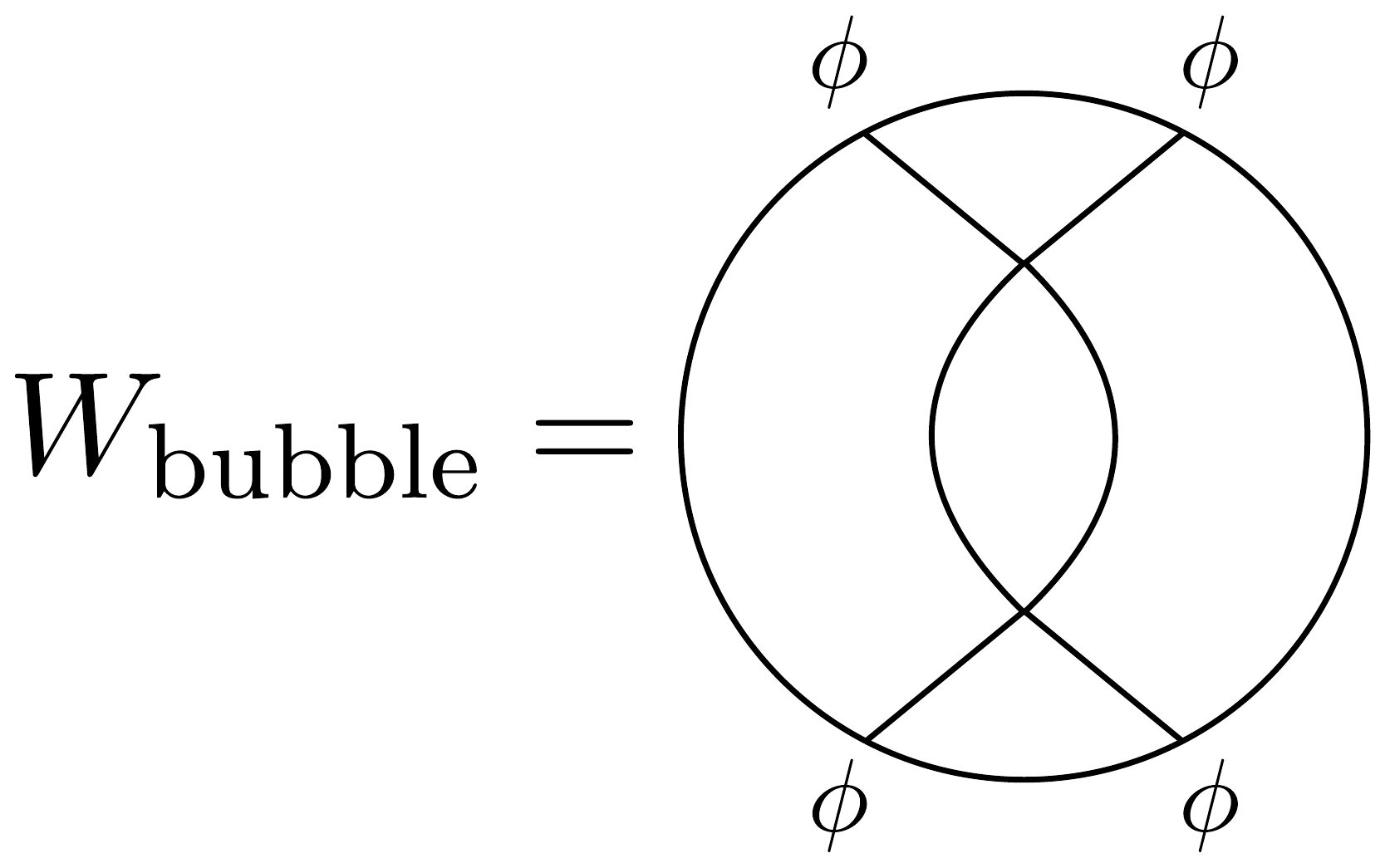} \label{eq:fig_witten_bubble}
\end{align}
or,
\begin{align}
W_{\text{bubble}}(p_i)=-g^2\int \frac{dz_1dz_2}{z_1^{d+1}z_2^{d+1}}\int \frac{d^{d}\ell}{(2\pi)^d} &K_{\f}(p_1,z_1)K_{\f}(p_4,z_1)G_{\f}(\ell,z_1,z_2)G_{\f}(p_{14}-\ell,z_1,z_2)
\nonumber 
\\
&K_{\f}(p_2,z_2)K_{\f}(p_3,z_2).\label{eq:witten_bubble}
\end{align}
To prove \eqref{eq:fig_witten_bubble} has a dispersive representation, we will use the following position space identity \cite{Fitzpatrick:2011dm},
\begin{align}
G_{\Delta,0}(y_1,z_1;y_2,z_2)^{2}=\sum\limits_{n=0}^{\infty}a_{\Delta}(n)G_{2\Delta+2n,0}(y_1,z_1;y_2,z_2),\label{eq:KL_Global}
\\
a_{\Delta}(n)=\frac{(d/2)_n(2\Delta+2n)_{1-d/2}(2\Delta+n-d+1)_n}{2\pi^{d/2}n!(\Delta+n)^{2}_{1-d/2}(2\Delta+n-d/2)_n},
\end{align}
where $(x)_n$ is the Pochhammer symbol.
The identity \eqref{eq:KL_Global} implies that the bubble diagram can be written as an infinite sum of exchange diagrams.
Going back to momentum space, we find,
\begin{align}
W_{\text{bubble}}(p_i)=\sum\limits_{n=0}^{\infty}a_{\Delta}(n)W^{t,4\phi}_{[\f\f]_{n,0}}(p_i).\label{eq:KL_rep_bubble}
\end{align}
The double-trace operators $[\f\f]_{n,J}$ have dimension $\Delta_{n,J}=2\Delta+2n+J$ and spin $J$.
As we showed earlier, regularized scalar exchange diagrams have a dispersive representation, and therefore the bubble diagram \eqref{eq:fig_witten_bubble} can also be written in a dispersive form. 
In comparison to tree-level diagrams, here we additionally need to regularize the $d$-dimensional loop integral over $\ell$ using a hard cut-off or dimensional regularization.

To prove that tree-level, scalar exchange diagrams obeyed the unsubtracted dispersion formula, we used a tree-level, dispersive representation for the bulk-to-bulk propagator $G_{\f}$. 
Bulk dispersion formulas can also be used to study higher-loop diagrams.
For example, we can consider the following loop corrected exchange diagram,
\begin{align}
\includegraphics[scale=.33]{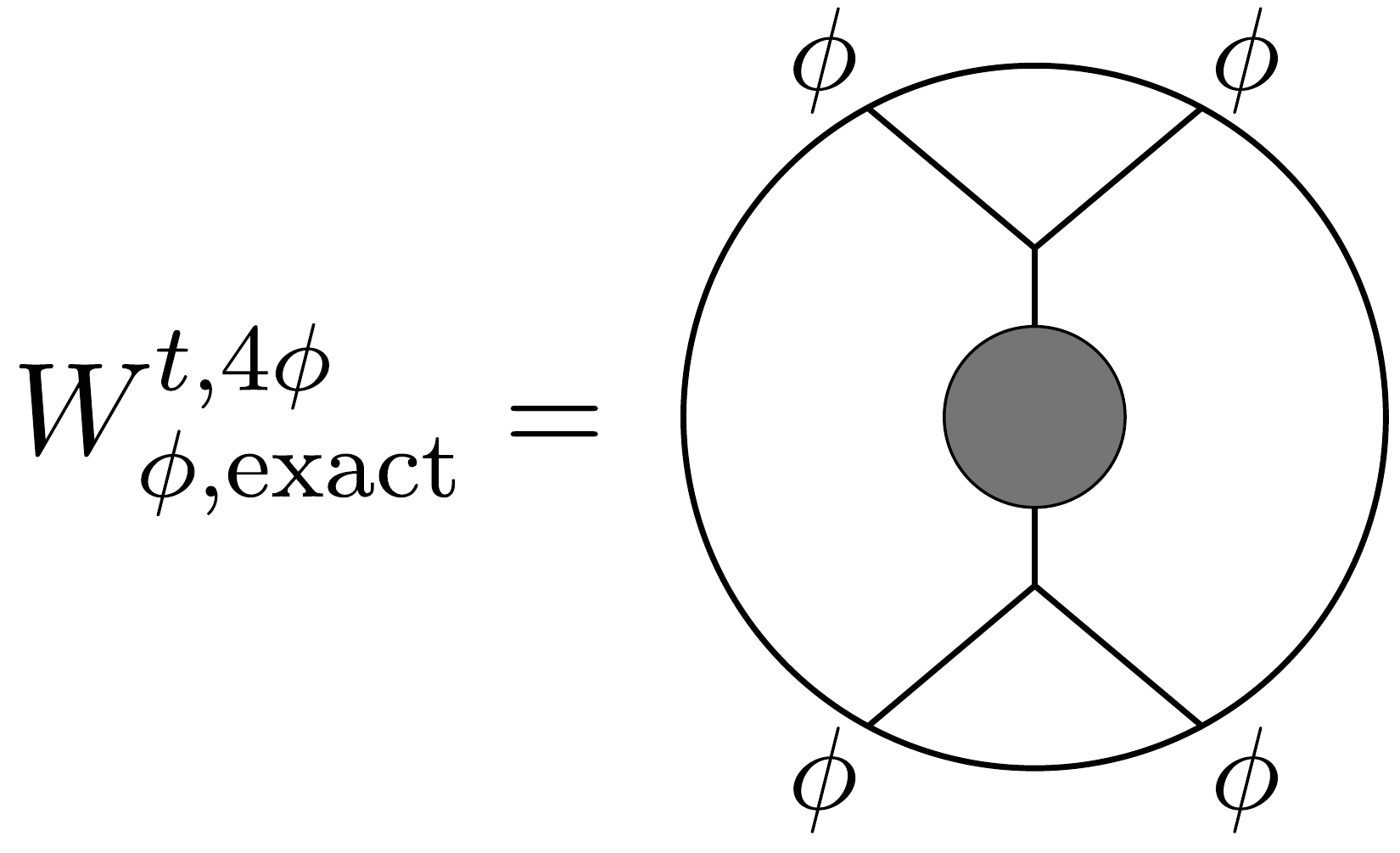}\label{eq:exact_exchange}
\end{align}
where the grey disk means that we include all possible loop corrections for the propagator.
This amounts to replacing the bulk-to-bulk propagator $G_{\f}(p,z_1,z_2)$ in \eqref{eq:scalar_exchange_diagram} with the exact time-ordered two-point function for the bulk field $\Phi$,
\begin{align}
G_{\f,\text{exact}}(p,z_1,z_2)=\ll T[\Phi(-p,z_2)\Phi(p,z_1)]\rr.
\end{align}
If the exact two-point function obeys the unsubtracted, bulk dispersion formula,
\begin{align}
G_{\f,\text{exact}}(p,z_1,z_2)=\frac{1}{2\pi i}\int\limits_{0}^{\infty} \frac{dt'}{t'-t-i\epsilon}\disc_{t'}G_{\f,\text{exact}}(p',z_1,z_2), \label{eq:dispbulk_twopt}
\end{align}
where $t'=-p'^2$, then the diagram \eqref{eq:exact_exchange} obeys the unsubtracted, boundary, four-point dispersion formula. 
Subtraction terms for the bulk two-point dispersion formula \eqref{eq:dispbulk_twopt} will map to subtraction terms for the boundary four-point dispersion formula \eqref{eq:exact_exchange}.
It would be interesting to study dispersive representations for more complicated loop diagrams, such as the triangle or box diagram, following the flat-space procedure given in Appendix B of \cite{Arkani-Hamed:2020blm}.
\subsection{Spinning Witten Diagrams}
\label{sec:addspin}
In this work we have focused on studying dispersion formulas for scalar correlators $\<\f(p_1)\ldots \f(p_n)\>$. 
In this section we will slightly generalize this by studying the Regge limit for AdS$_{4}$ Witten diagrams involving external conserved currents, $J(p,\epsilon)\equiv \epsilon_{\mu}J^{\mu}(p)$.\footnote{We will study the large $t$ limit for fixed $\epsilon_i\cdot p_j$. It would also be interesting to study the Regge limit using a helicity formalism \cite{Bzowski:2013sza,Isono:2018rrb,Isono:2019wex}.} 
To make the notation compact, we will suppress the color indices for $J^{\mu}$ throughout. 
We will also assume the polarization vector is transverse, $\epsilon\cdot p=0$. 
\subsubsection*{Two Currents and Two Scalars}

First we consider the correlator $\<JJ\varphi\varphi\>$ and study $t$-channel $\varphi$ exchange,
\begin{align}
\includegraphics[scale=.32]{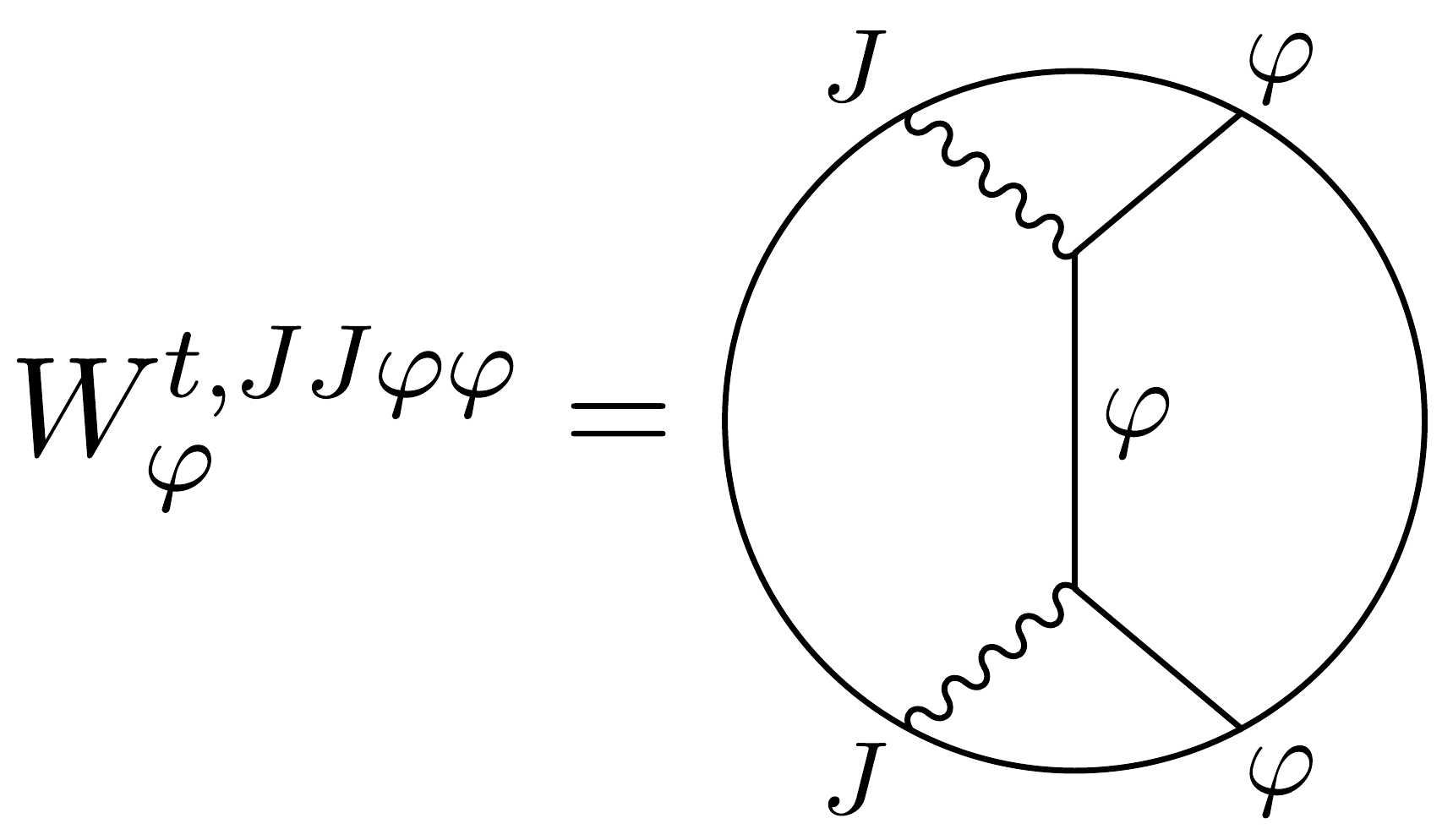},
\end{align}
which gives the following result \cite{Baumann:2020dch},
\begin{align}
W^{t,JJ\varphi\varphi}_{\varphi}(p_i)\bigg|_{d=3}=4ig_{\text{YM}}^2\frac{(\epsilon_1\cdot p_4)(\epsilon_2\cdot p_3)}{E_{\text{tot}}E_LE_R}.
\end{align}
Using the definition of the left and right energies, \eqref{eq:EL} and \eqref{eq:ER}, we find that this correlator decays like $1/t$ in the limit $t\rightarrow \infty$. This implies that the scalar exchange diagram obeys the unsubtracted dispersion formula.

Next we can consider the correlator $\<J\varphi\varphi J\>$ and study $J$ exchange in the $t$-channel,
\begin{align}
\includegraphics[scale=.33]{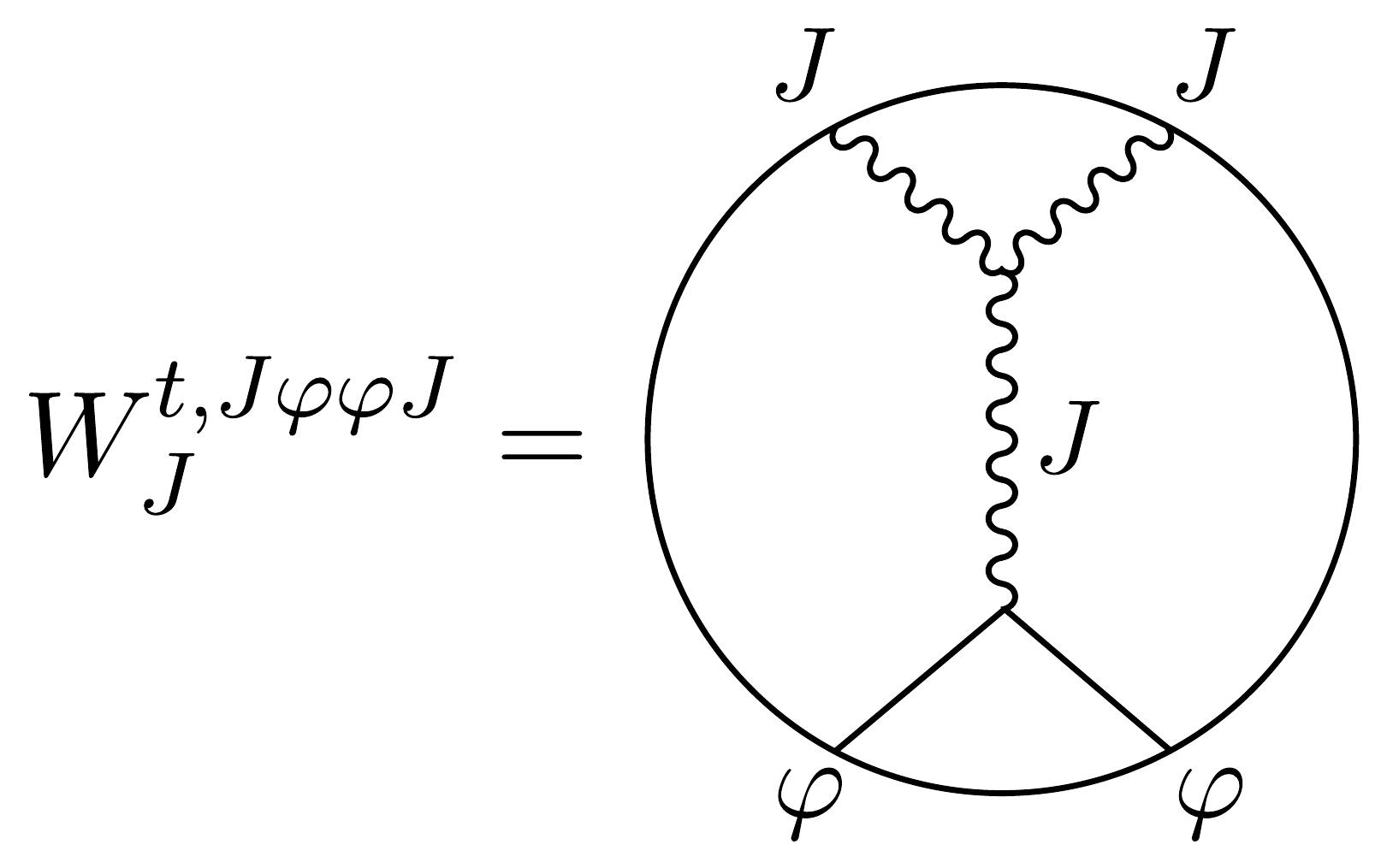}.
\end{align}
Here the result is also known in closed form \cite{Baumann:2020dch},
\begin{align}
W^{t,J\varphi\varphi J}_{J}(p_i)\bigg|_{d=3}=\frac{ig^{2}_{\text{YM}}}{E_{\text{tot}}E_LE_R}\left[(t\Pi_{1,1}+E_LE_R\Pi_{1,0})\epsilon_1\cdot \epsilon_4-4\left(\epsilon_1\cdot p_2 \epsilon_4\cdot p_3-\epsilon_1\cdot p_3\epsilon_4\cdot p_2\right)\right].
\end{align}
If we take the large $t$ limit, then the second term vanishes and we are left with the first term.
The first term has the same form as \eqref{eq:gaugebosonAdS4} and we find,
\begin{align}
\lim\limits_{|t|\rightarrow\infty}W^{t,J\varphi\varphi J}_{J}(p_i)\bigg|_{d=3}=ig^{2}_{\text{YM}}\frac{\epsilon_1\cdot \epsilon_4}{E_{\text{tot}}}.
\end{align}

To apply the momentum space dispersion formula we need to subtract off this total energy pole. 
This total energy pole is proportional to the contact Witten diagram which appears in $\<J\varphi\varphi J\>$,
\begin{align}
\includegraphics[scale=.33]{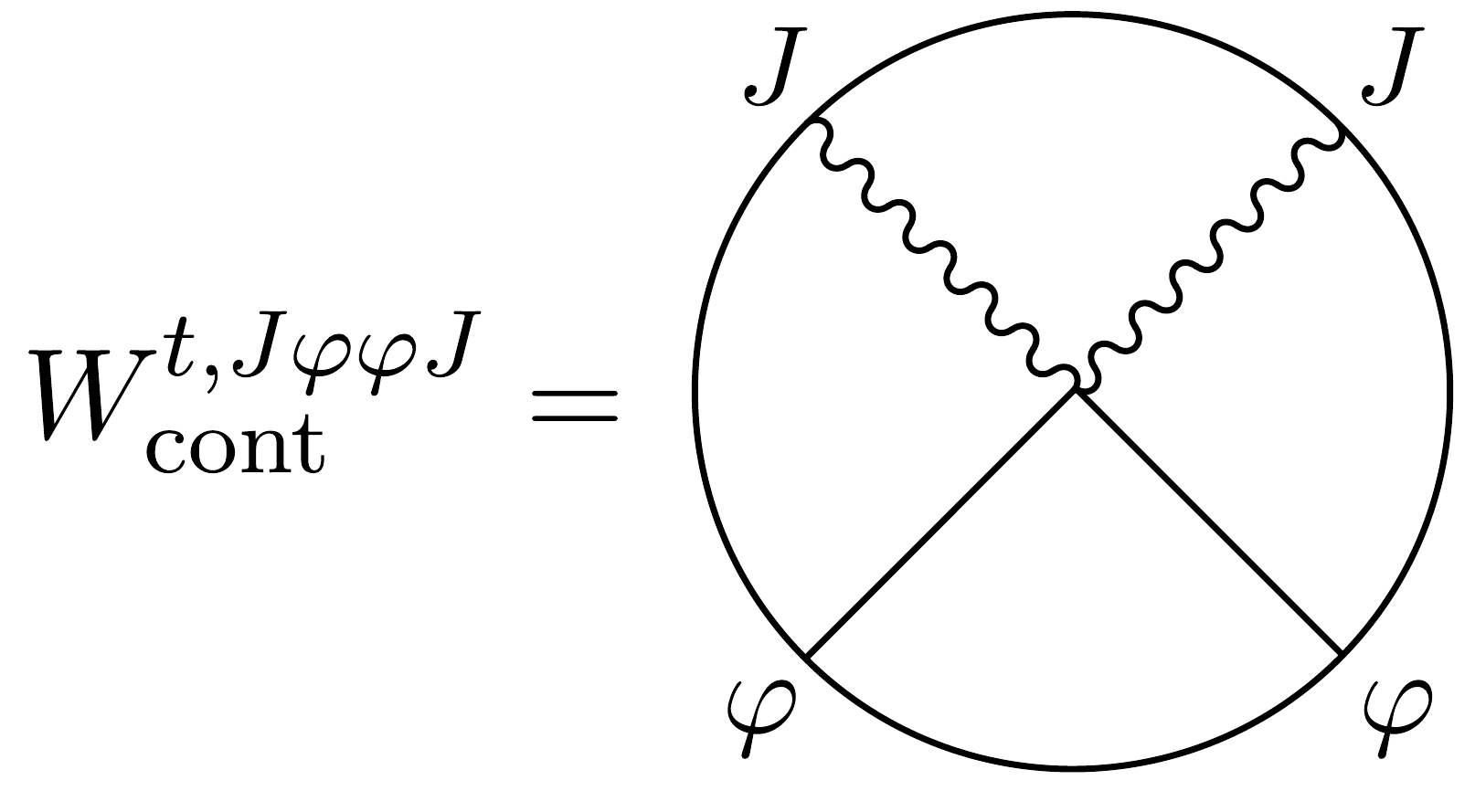}.
\end{align}
To see this, we recall that the boundary-to-bulk propagator for the Yang-Mills field in AdS$_{4}$ is,
\begin{align}
K^{\text{YM}}_{\mu}(p,z)=-i e^{-\sqrt{p^{2}}z}\epsilon_{\mu},
\end{align} 
where we assume $\epsilon\cdot p=0$. Then the quartic contact diagram for $\<J \varphi\varphi J\>$ in AdS$_{4}$ is:
\begin{align}
W^{J\varphi\varphi J}_{\text{cont}}(p_i)\bigg|_{d=3}&=\int \frac{dz}{z^{4}}z^{2}K^{\text{YM}}_{\mu}(p_1,z)K_{\varphi}(p_2,z)K_{\varphi}(p_3,z)K^{\text{YM},\mu}(p_4,z)
\nonumber \\
&=\frac{\epsilon_1\cdot \epsilon_4}{E_{\text{tot}}}.
\end{align}

The power of $z^{-4}$ in the first line comes from the AdS measure $\sqrt{-g}$ while the factor of $z^2$ comes from contracting the indices using the AdS Poincar\'e patch metric.
The combination $W^{t,J\varphi\varphi J}_{J}(p_i)-ig^{2}_{\text{YM}}W^{J\varphi\varphi J}_{\text{cont}}(p_i)$ is then a Polyakov-Regge block for $\<J\varphi \varphi J\>$ and will obey the unsubtracted dispersion formula.

\subsubsection*{Four Currents}
Finally, we will study the following $t$-channel exchange diagram for pure Yang-Mills in AdS$_{4}$,
\begin{align}
\includegraphics[scale=.33]{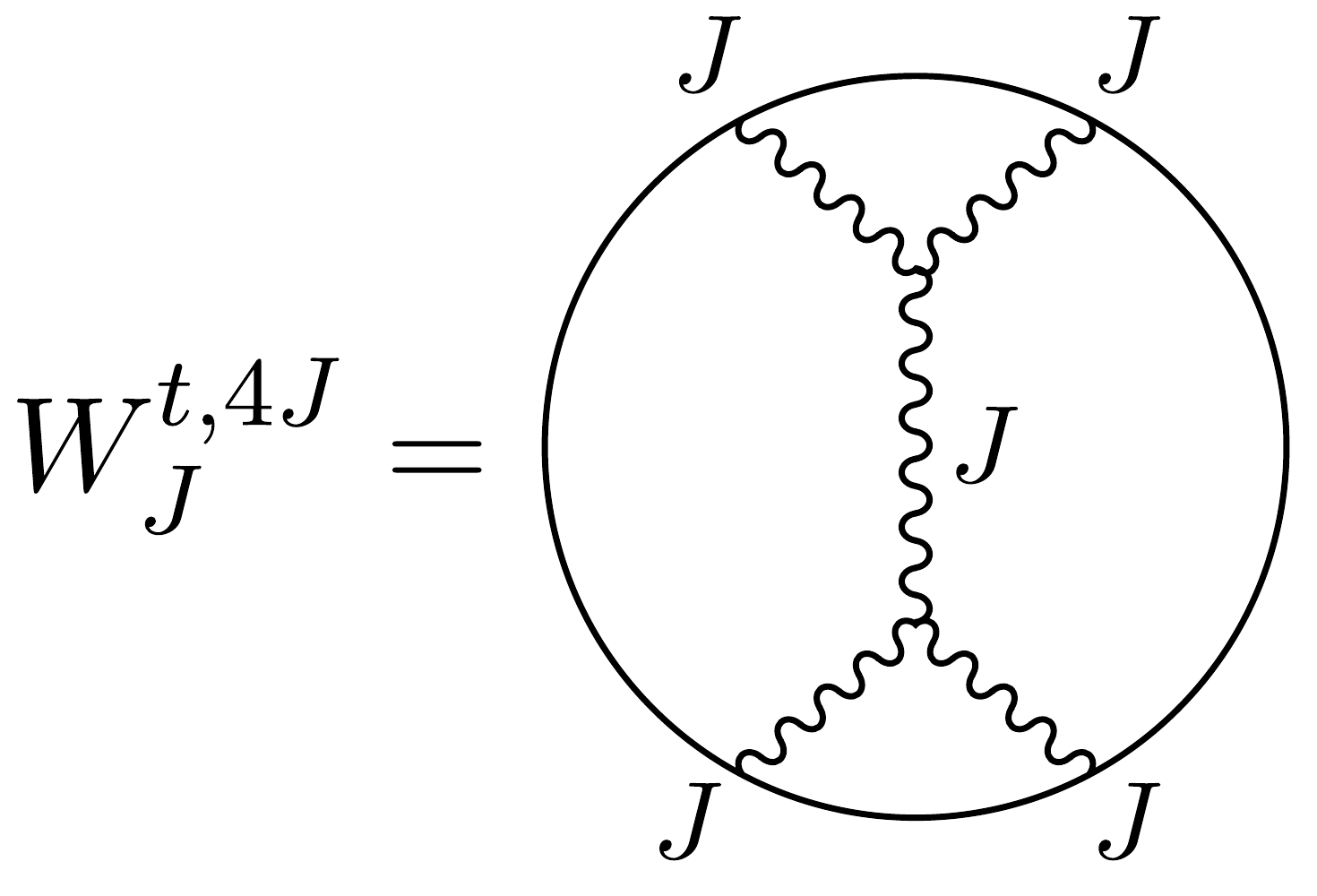},
\end{align}
which is known in closed form \cite{Albayrak:2018tam}:
\begin{align}
W^{\text{YM}}_{J}(p_i)\bigg|_{d=3}=& \ \frac{-ig^{2}}{E_{T}}\frac{V^{14\mu}(p_1,p_4,-p_{14})V^{23\nu}(p_2,p_3,p_{14})}{\left(|p_{14}|+|p_1|+|p_4|\right)\left(|p_{14}|+|p_2|+|p_3|\right)}
\nonumber 
\\
&\left(\eta_{\mu\nu}+p_{14,\mu}p_{14,\nu}\frac{|p_{14}|+E_T}{|p_{14}|(|p_1|+|p_4|)(|p_2|+|p_3|)}\right),\label{eq:ext_Currents_YM}
\end{align}
where $|p|\equiv\sqrt{p^2}$. The vertex factors are defined by,
\begin{align}
V_{\mu\nu\rho}(p_1,p_2,p_3)=\eta_{\mu\nu}(p_1-p_2)_{\rho}+\eta_{\nu\rho}(p_2-p_3)_{\mu}+\eta_{\rho\mu}(p_3-p_1)_{\nu}. 
\end{align}
We have also used the notation $V_{12,\rho}(p_i)\equiv \epsilon_{1}^{\mu}\epsilon_{2}^{\nu}V_{\mu\nu\rho}(p_i)$ where $\epsilon_i$ are the external polarizations. 

We will not write out the full form of the exchange Witten diagram, but it is not difficult to see that \eqref{eq:ext_Currents_YM} will have a branch cut starting at $t=0$ and running to $t=\infty$.
If we take the large $t$ Regge limit of \eqref{eq:ext_Currents_YM} we find,
\begin{align}
\lim\limits_{|t|\rightarrow\infty}W^{\text{YM}}_{J}(p_i)\bigg|_{d=3}=\frac{ig^{2}\epsilon_1\cdot\epsilon_4\epsilon_2\cdot\epsilon_3}{E_{\text{tot}}},\label{eq:larget_limitYM}
\end{align}
for fixed $\z_i$, $s<0$. We can note that the momentum dependence of \eqref{eq:larget_limitYM} is exactly the same as the momentum dependence of the quartic contact diagram for $\<JJJJ\>$ in AdS$_{4}$, see for example \cite{Albayrak:2018tam}. 
However, \eqref{eq:larget_limitYM} has a different polarization dependence than the contact diagram in Yang-Mills theory.
This is not relevant for the dispersion formula or for defining Polyakov-Regge blocks.
For the dispersion formula it is only important that \eqref{eq:larget_limitYM} can be expressed as some local, contact diagram in AdS$_{4}$, even if the diagram does not appear in Yang-Mills theory itself.

\section{Discussion}

In this work we studied momentum space dispersion formulas for QFT correlation functions and their applications to  the study of CFTs.
In the first part, we showed using two independent methods that the QFT dispersion formula can be rewritten in terms of a double advanced product.
In Section \ref{sec:disp_from_analyticity} we reviewed how the analyticity properties of retarded correlators can be used to derive dispersion formulas for the time-ordered correlators.
Dispersion formulas of this form were originally derived as an intermediate step in finding dispersion formulas for the on-shell S-matrix \cite{Sommer:1970mr,Bogolyubov:1990kw}.
We also showed that the double advanced product appears in the QFT dispersion formula by using the positive spectrum condition, unitarity, and operator ordering identities.
The derivation given in Section \ref{sec:disp_from_largest_time} used the properties of the time-ordered product in combination with an infinite momentum limit.
This method was originally used in \cite{tHooft:1973wag,Remiddi:1981hn} to derive dispersion formulas for Feynman diagrams in flat-space.
Both derivations require that at most one external Lorentz invariant is above threshold.
Once the correlator is computed in these kinematics, we can then analytically continue to general configurations. 
In both Section \ref{sec:disp_from_analyticity} and Section \ref{sec:disp_from_largest_time} we did not need to make any assumptions on the spectrum of the QFT.

In the second part of the paper, we studied how the QFT dispersion formula could be used to study CFT correlators.
In Section \ref{sec:equivalence} we proved that for CFT four-point functions, the QFT dispersion formula is equivalent to the CFT dispersion formula \cite{Carmi:2019cub}, modulo possible semi-local terms.
We reviewed how to generate the Polyakov-Regge expansion using the CFT dispersion formula and explained how to generate the corresponding expansion in momentum space.
We then showed that the Polyakov-Regge blocks associated to each dispersion formula are related by a Fourier transform.
In Section \ref{sec:app_to_CFT} we tested the momentum space dispersion formula on holographic four-point functions.
In order to prove that Witten diagrams obeyed boundary dispersion formulas, we used bulk dispersion formulas for AdS two-point functions.

There are many open questions to consider on CFT dispersion formulas in momentum space.
In this work we studied dispersion formulas which are valid for general QFT correlators inside their primitive domain of analyticity. 
One question to consider is: what is the role of conformal symmetry in extending or simplifying these dispersion formulas?
For example, it would be interesting to study CFT correlators in larger regions of momentum space.
For massive QFTs, one needs to study the envelope of holomorphy for the primitive domain of analyticity in order to derive crossing symmetry for the on-shell S-matrix \cite{Bros:1964iho,Bros:1965kbd,Sommer:1970mr}.
While a CFT does not have an S-matrix itself, finding the full domain of analyticity for CFT correlators in Fourier space could be a simpler problem than in generic QFTs.
Knowing the domain of analyticity for CFT correlators could also be used to derive double-dispersion formulas for CFT four-point functions and to study how anomalous thresholds appear in the flat-space limit of AdS/CFT \cite{Komatsu:2020sag}.
It would also be interesting if the double-advanced product studied here could be used to derive dispersion formulas for higher-point amplitudes \cite{AlvarezEstrada:1973yy,Chandrasekaran:2018qmx}.\footnote{At higher-points there are obstructions to single-variable dispersion formulas due to overlapping cuts and anomalous thresholds \cite{Landshoff_1,Landshoff_2}. We thank Sebastian Mizera for discussions on this point.}

It would also be useful to make a more direct connection between momentum and position space methods used in the conformal bootstrap.
For example, can we prove that the momentum space and CFT dispersion formulas are directly related by a Fourier transform?
This would be useful for higher-point correlators, where we have dispersion formulas in momentum space, but the corresponding CFT dispersion formulas are not currently known.
In addition, in this work we did not discuss how to formulate the Lorentzian inversion formula \cite{Caron-Huot:2017vep} in momentum space.
Having a Fourier representation of the inversion formula would allow us to translate results on spinning holographic correlators in momentum space \cite{Raju:2010by,Raju:2011mp,Raju:2012zr,Raju:2012zs,Isono:2018rrb,Isono:2019wex,Albayrak:2018tam,Albayrak:2019yve,Albayrak:2019asr,Albayrak:2020bso,Baumann:2020dch} to statements about the spectrum and couplings of spinning double-trace operators.\footnote{For computations of spinning double-trace anomalous dimensions using the inversion formula in position space see \cite{Sleight:2018epi,Sleight:2018ryu,Albayrak:2020rxh,Caron-Huot:2021kjy}.} 
This data is not immediately accessible from the momentum space OPE studied in Section \ref{sec:app_to_CFT} because there we studied the OPE of a momentum space discontinuity. 
At tree-level in $1/N$, taking a discontinuity effectively projects out double-trace operators.

The Lorentzian inversion formula and the conformal dispersion formula also have beautiful connections to the study of light-ray operators and analytic functionals. 
The Lorentzian inversion formula computes matrix elements of light-ray operators, $\mathbf{L}[\O](z)$ \cite{Kravchuk:2018htv}, where $z$ is a point on the celestial sphere.
Event shapes of light-ray operators are simplest to compute in momentum eigenstates and can be used to derive bounds on the CFT data.
For example, positivity of one-point energy correlators gives the well-known conformal collider bounds \cite{Hofman:2008ar} while two-point event shapes lead to superconvergent sum rules \cite{Kologlu:2019bco,Kologlu:2019mfz}.
There is also a direct connection between two-point event shapes, analytic functionals, and the CFT dispersion formula \cite{Caron-Huot:2020adz}.
Given the equivalence between the CFT and momentum space dispersion formulas, along with the simplicity of event-shapes in momentum eigenstates, it would naturally be interesting to investigate analytic functionals and the inversion formula using momentum space methods.

Momentum space dispersion formulas can also be used to derive bounds on AdS EFTs analogous to those of \cite{Adams:2006sv}.
In the past, bounds on AdS gravity have been derived by leveraging rigorous CFT consistency conditions such as unitarity and causality \cite{Camanho:2014apa,Afkhami-Jeddi:2016ntf,KPZ2017,Costa:2017twz,Meltzer:2017rtf,Afkhami-Jeddi:2018own,Belin:2019mnx,Kologlu:2019bco}.
It is also possible to generalize the recent bounds on flat space EFTs \cite{Bellazzini:2020cot,Tolley:2020gtv,Caron-Huot:2020cmc,Arkani-Hamed:2020blm,Sinha:2020win} to AdS/CFT using dispersive methods \cite{Caron-HuotWIP,Gopakumar:2021dvg}.
One motivation to rederive these bounds in momentum space is to understand how the bounds generalize to bulk theories that break a subset of the AdS isometries.
This problem is relevant for studying slow holographic renormalization group flows, which is the AdS analog of inflation \cite{Kaplan:2014dia}.
Momentum space dispersion formulas can be derived in the absence of conformal symmetry and are therefore a natural tool to study holographic RG flows and inflation \cite{Baumann:2019ghk,Grall:2021xxm,Jazayeri:2021fvk}.
Along similar lines, understanding how the EFT-hedron \cite{Arkani-Hamed:2020blm} generalizes to AdS could elucidate the role of positive geometry in holography and the conformal bootstrap \cite{Arkani-Hamed:2018ign}.

To derive dispersion formulas in momentum space, we had to start by assuming the existence of the time-ordered product.
From the time-ordered product, we are then able to define the retarded and advanced product. 
However, to our knowledge, the existence of the time-ordered product has not been derived from the Wightman axioms, but is rather an additional assumption \cite{Streater:1989vi,Bogolyubov:1990kw}.
Given that conformal correlators are more strongly constrained than generic QFT correlators, it would be interesting if the time-ordered and causal products could be derived in CFTs from the Wightman axioms.

Finally, it is useful to place the results of this work and of \cite{Meltzer:2020qbr} in a larger context.
Using the Cutkosky rules of \cite{Meltzer:2020qbr} and the dispersion formulas studied here, we can reconstruct AdS/CFT correlators from their bulk unitarity cuts.
This generalizes the original S-matrix unitarity method  \cite{Eden:1966dnq} to AdS/CFT correlators in momentum space.
Clearly, the unitarity methods discussed in these works are just the tip of the iceberg for on-shell methods in holography.
Understanding how ideas from the modern amplitudes program \cite{Elvang:2013cua} generalize to AdS can potentially reveal new symmetries and structures in holographic correlators.
Conversely, taking the flat-space limit of AdS/CFT gives us a new way to study flat-space S-matrices \cite{Susskind:1998vk,Polchinski:1999ry,Gary:2009ae,Penedones:2010ue}.
We hope the unitarity method presented here and in \cite{Meltzer:2020qbr} give a new perspective on perturbation theory in curved spacetimes and the relation between the S-matrix and conformal bootstraps.

\section*{Acknowledgements}
We thank Adam Bzowski, Simon Caron-Huot, Marc Gillioz, Savan Kharel, Sebastian Mizera, Julio Parra-Martinez, Eric Perlmutter, David Simmons-Duffin, Allic Sivaramakrishnan, Kostas Skenderis, Edward Witten, and Roman Zwicky for discussions. We would also thank Savan Kharel, Julio Parra-Martinez, and Allic Sivaramakrishnan for comments on the draft. The research of DM is supported by the Walter Burke Institute for Theoretical Physics and the Sherman Fairchild Foundation. This material is based upon work supported by the U.S. Department of Energy, Office of Science, Office of High Energy Physics, under Award Number DE-SC0011632. 

\appendix

\section{Primitive Domain of Analyticity}
\label{app:Domain_Analyticity}
In this appendix we will study sections of the primitive domain of analyticity,
\begin{align}
\mathcal{D}=\{\Im(k_{I})\neq 0, \ \Im(k_I)^2<0\} \cup \{\Im(k_I)=0, \ k_I^{2}>0\},\label{eq:APPregionAnalyticity}
\end{align} 
that were used to derive dispersion formulas for four- and five-point functions.

\subsection{Four-Points}
In this appendix we will show that \eqref{eq:APPregionAnalyticity} contains the entire cut $t$-plane for $s<0$ and $\z_i<0$. 
This section will mostly be a review of material presented in section 5 of \cite{Sommer:1970mr}.

We will choose our external momenta to be,
\begin{align}
k_1&=\left(0,-1,k_1^{\perp}\right),\label{eq:app4ptKin1}
\\
k_2&=\left(0,1,k_2^{\perp}\right),\label{eq:app4ptKin2}
\\
k_3&=\left(a+ib,a'+ib',k_3^{\perp}\right),\label{eq:app4ptKin3}
\\
k_4&=\left(-(a+ib),-(a'+ib'),k_4^{\perp}\right),\label{eq:app4ptKin4}
\end{align}
where $k_i^{\perp}$ are the momenta in the $d-2$ transverse directions.
We leave momentum conservation in the transverse directions implicit.
We will assume $d\geq3$ so that the momenta can be taken to be linearly independent, up to momentum conservation.

Given the kinematics \eqref{eq:app4ptKin1}-\eqref{eq:app4ptKin4}, we automatically have $\z_{1,2}<0$ and $s<0$. 
To also fix $\z_{3,4}<0$ we will impose the following conditions on $\{a,b,a',b'\}$:
\begin{align}
-a^2+a'^2+b^2-b'^2&=1,\label{eq:condition1aabb}
\\
ab&=a'b'.\label{eq:condition2aabb}
\end{align}
Under these assumptions we have,
\begin{align}
\z_3&=-(1+(k_{3}^{\perp})^2),
\\
\z_4&=-(1+(k_{4}^{\perp})^2).
\end{align} 
With the conditions \eqref{eq:condition1aabb} and \eqref{eq:condition2aabb} imposed, we have $\z_i<0$ for all $i$. Using the momenta \eqref{eq:app4ptKin1}-\eqref{eq:app4ptKin4} we also have $\sum\limits_{i}\z_i-s<0$. This guarantees that the $t$- and $u$-channel cuts do not overlap.

Next, we want to choose our momenta such that they stay within the primitive domain of analyticity \eqref{eq:APPregionAnalyticity}.
From \eqref{eq:app4ptKin1} and \eqref{eq:app4ptKin2} we see $p_1$, $p_2$, and $p_1+p_2$ are real and spacelike and therefore satisfy the second condition of \eqref{eq:APPregionAnalyticity}.
To ensure $p_3$ and $p_4$ also lie in the primitive domain we require
\begin{align}
|b|>|b'|.\label{eq:condition3aabb}
\end{align}
Then we automatically have that the sums $p_1+p_3$ and $p_1+p_4$ also lie in the primitive domain of analyticity.

Finally, we need to determine what values $t$ can take in the configuration \eqref{eq:app4ptKin1}-\eqref{eq:app4ptKin4} given the conditions \eqref{eq:condition1aabb}, \eqref{eq:condition2aabb}, and \eqref{eq:condition3aabb}. To impose the condition \eqref{eq:condition3aabb} we will fix $b>0$ and set,
\begin{align}
b'=Yb,\label{eq:sol_b1}
\end{align}
for $-1\leq Y\leq1$. We then solve \eqref{eq:condition1aabb} and \eqref{eq:condition2aabb} for $a$ and $b$,
\begin{align}
a&=Ya',\label{eq:sol_a}
\\
b&=\sqrt{\frac{1-a'^2 \left(1-Y^2\right)}{1-Y^2}}.\label{eq:sol_b}
\end{align}
Requiring that $b$ is real and positive gives
\begin{align}
-\sqrt{\frac{1}{1-Y^2}}< a'<\sqrt{\frac{1}{1-Y^2}}.\label{eq:conditionAY}
\end{align}
Parameterizing $t$ in terms of $Y$ and $a'$, we find:
\begin{align}
\Re t =&-(2+2a'+(k_2^{\perp}+k_3^{\perp})^2),
\\
\Im t =&-2Y\sqrt{\frac{1-a'^{2}(1-Y^{2})}{1-Y^{2}}}.
\end{align}
We are free to vary $a'$ over the entire real line, so $\Re t $ can take arbitrary values.
For $\Im t$ we first note if $-1\leq a'\leq1$, then the condition \eqref{eq:conditionAY} holds for all $Y\in [-1,1]$. In this case $\Im t$ can also take arbitrary values. As $Y\rightarrow \pm 1$ we have $\Im t\rightarrow \mp\infty$.
If $|a'|>1$ then \eqref{eq:conditionAY} implies,
\begin{align}
Y\in [-1,-\sqrt{1-a'^{-2}}) \ \cup \ (\sqrt{1-a'^{-2}},1].
\end{align}
In this case as $Y\rightarrow \pm \sqrt{1-a'^{-2}}$ we have $\Im t\rightarrow 0$. As before, when $Y\rightarrow \pm 1$, we have $\Im t\rightarrow \mp \infty$. We can therefore vary the momenta \eqref{eq:app4ptKin1}-\eqref{eq:app4ptKin4} such that $t$ runs over the complex plane, minus two cuts on the real line. 

\subsection{Five-Points}
In this appendix we will prove that the primitive domain of analyticity for the five-point correlator contains the entire cut $s_{35}$ plane. 
First we will parameterize the external momenta as,
\begin{align}
k_1&=\left(0,0,k_1^{\perp}\right),\label{eq:app5ptKin1}
\\
k_2&=\left(0,-1,k_2^{\perp}\right),\label{eq:app5ptKin2}
\\
k_3&=\left(0,1,k_3^{\perp}\right),\label{eq:app5ptKin3}
\\
k_4&=\left(a+ib,a'+ib',k_4^{\perp}\right),\label{eq:app5ptKin4}
\\
k_5&=\left(-(a+ib),-(a'+ib'),k_5^{\perp}\right).\label{eq:app5ptKin5}
\end{align}
We will assume $d\geq 4$ so that we can take the momenta to be linearly independent, up to momentum conservation. 
We also impose that the set $\{a,a',b,b'\}$ satisfies the same conditions as before, \eqref{eq:condition1aabb}, \eqref{eq:condition2aabb} and \eqref{eq:condition3aabb}. With these assumptions all the external norms $\z_i<0$ and the vectors $k_i$ lie in the primitive domain of analyticity \eqref{eq:APPregionAnalyticity}. 
We will use \eqref{eq:sol_b1}, \eqref{eq:sol_a}, and \eqref{eq:sol_b} to write the Mandelstam invariants as functions of $a'$ and $Y$. 

Given this external momenta, we find that the Mandelstams $\{s_{12},s_{13},s_{23},s_{14},s_{15},s_{45}\}$ are all negative and independent of $a'$ and $Y$. The remaining four Mandelstams $\{s_{24},s_{25},s_{34},s_{35}\}$ will depend on $a'$ and $Y$. We will take our independent variable to be $s_{35}$ and parameterize the Mandelstams $\{s_{24},s_{25},s_{34}\}$ using \eqref{eq:s5pt_relations2}-\eqref{eq:s5pt_relations4}. 
For $s_{35}$ we find,
\begin{align}
\Re s_{35}&=-\left(2 a'+(k_3^{\perp}+k_5^{\perp})^2+2\right),
\\
\Im s_{35}&=-2Y\sqrt{\frac{1-a'^{2}(1-Y^{2})}{1-Y^{2}}}.
\end{align}
Using the same arguments as before, we see $s_{35}$ can take any value in the cut complex plane.
Finally, we can choose the transverse momenta $k_i^{\perp}$ such that the cuts which run to positive infinity, \eqref{eq:branch_cut_1} and \eqref{eq:branch_cut_2}, do not overlap with the cuts running to negative infinity, \eqref{eq:branch_cut_3} and \eqref{eq:branch_cut_4}.

\section{AdS Unitarity and Dispersion}
\label{app:AdSUnitarity}
In this Appendix we will review how to derive the AdS cutting rules from the QFT unitarity condition \eqref{eq:QFT_Optical}. We will also review the diagrammatic interpretation of the non-Lorentz invariant dispersion formulas.
\subsection{AdS Cutting Rules}
To study the unitarity condition,
\begin{align}
\sum\limits_{r=0}^{n}(-1)^{r}\sum\limits_{\alpha\in\Pi_r(n)}\<\overline{T}[\f(x_{\alpha_1})\ldots \f(x_{\alpha_r})]T[\f(x_{\alpha_{r+1}})\ldots \f(x_{\alpha_n})]\>=0, \label{eq:APPQFT_Optical}
\end{align}
for AdS/CFT correlators we need the bulk Feynman rules for the partially time-ordered correlators. To compute these correlators we can use the Schwinger-Keldysh rules \cite{Schwinger:1960qe,Keldysh:1964ud}, which are equivalent to the cutting rules as derived by Veltman \cite{Veltman:1963th}. For the correlator $\<\overline{T}[\f(x_{1})\ldots \f(x_r)]T[\f(x_{r+1})\ldots\f(x_n)]\>$ we draw all possible cut Witten diagrams such that,\footnote{These cutting rules differ slightly from \cite{Meltzer:2020qbr} because we do not include an extra factor of $-1$ for each external point to the right of the cut. The source of these factors is the $(-1)^{r}$ on the right hand side of \eqref{eq:APPQFT_Optical} which in \cite{Meltzer:2020qbr} was absorbed into the definition of the cutting rules.}
\begin{enumerate}
\item The external operators in the (anti-)time-ordered symbol lie to the right and left of the cut, respectively.
\item For each cut line replace the Feynman, or time-ordered, propagator by the appropriate Wightman propagator such that in each propagator positive energy is flowing from the left to the right of the cut.
\item For each propagator to the (right) left of the cut, use the (anti-)time-ordered  propagator. 
\item For each internal vertex to the left and right of the cut, multiply by $ig$ and $-ig$ respectively.
\end{enumerate}
The condition that positive energy flows through the cut comes from the fact \\ $\<\overline{T}[\f(p_{1})\ldots \f(p_r)]T[\f(p_{r+1})\ldots\f(p_n)]\>$ is only non-zero for $p_{r+1}+\ldots+p_n\in\overline{V}_+$. In the above rules we assumed the interactions were non-derivative, so the vertex factor is a constant, but this assumption can be relaxed.
Then the unitarity condition \eqref{eq:APPQFT_Optical} says that the sum over all cut Witten diagrams, with the additional weighting of $(-1)^r$, vanishes. One can prove this by directly studying the Witten diagrams, as was shown in \cite{Meltzer:2020qbr}.
\subsection{Dispersion for Witten Diagrams}
Through the cutting rules, we can also translate the dispersion formula \eqref{eq:dispersion_Largesttime} into a statement about individual Witten diagrams. The simplest non-trivial case is a three-point function. For $n=3$, \eqref{eq:dispersion_Largesttime} becomes
\begin{align}
\<T[\f(x_1)\f(x_2)\f(x_3)]\>=& \ \<\f(x_1)T[\f(x_2)\f(x_3)]\>
\nonumber
\\ &+\theta(x_{32}^0)\left(\<\f(x_3)T[\f(x_1)\f(x_2)]\>-\<\overline{T}[\f(x_1)\f(x_3)]\f(x_2)\>\right)
\nonumber \\ &+\theta(x_{23}^0)\left(\<\f(x_2)T[\f(x_1)\f(x_3)]\>-\<\overline{T}[\f(x_1)\f(x_2)]\f(x_3)\>\right).\label{eq:APP3ptdispersion}
\end{align}
Each correlator can be computed in position space using the AdS cutting rules given in the previous section. We cannot directly use the momentum space rules for the second and third line due to the explicit $\theta$-function factors. To interpret \eqref{eq:APP3ptdispersion} diagrammatically we can think of the $\theta$-functions as giving some non-Lorentz invariant propagator \cite{tHooft:1973wag,Remiddi:1981hn}. For example, we can draw a Witten diagram for $\theta(x_{32})\<\f(x_3)T[\f(x_1)\f(x_2)]\>$ as,
\begin{align}
\includegraphics[scale=.25]{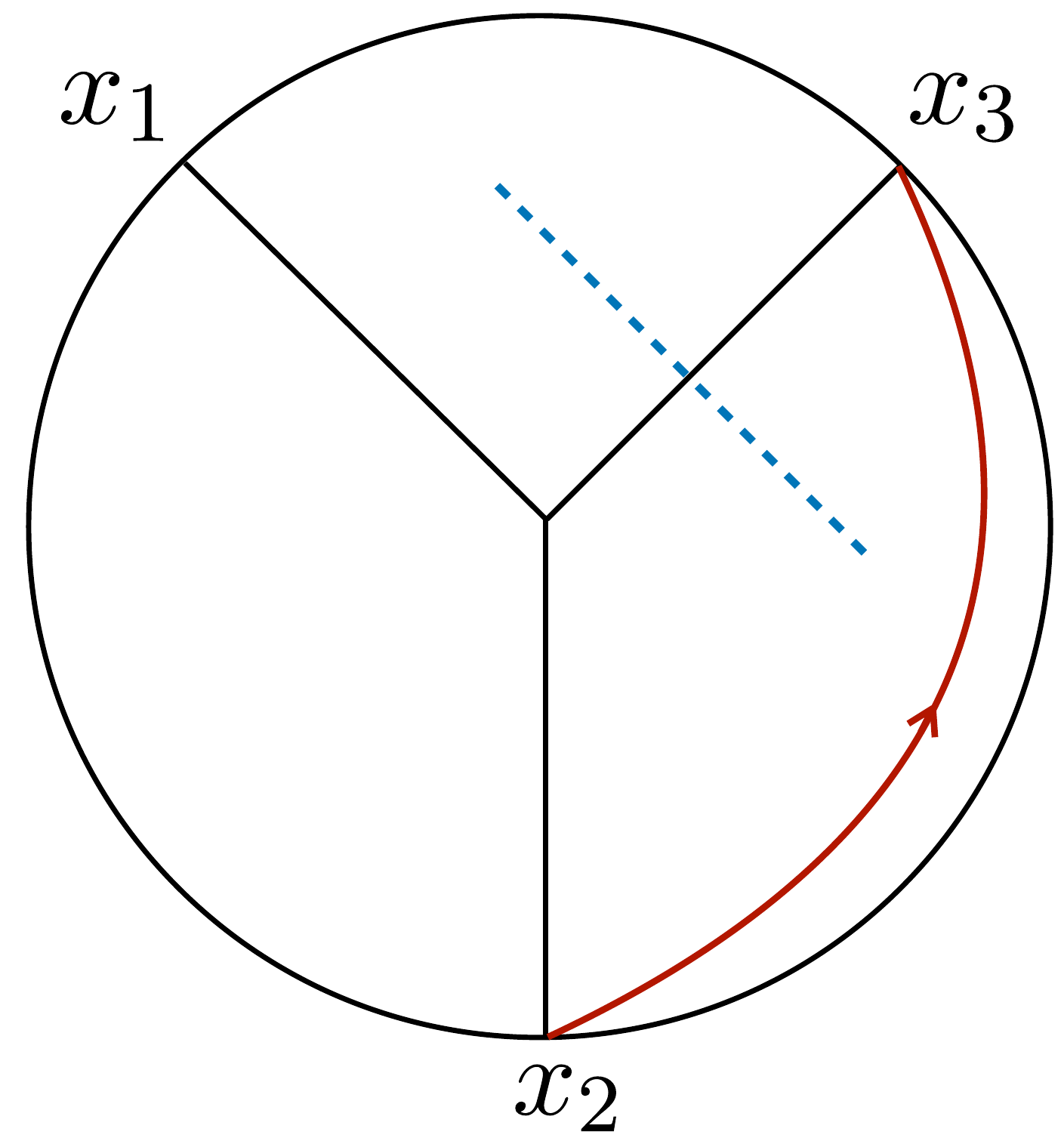}
\end{align}
To explain the notation, the blue line corresponds to the usual unitarity cut which can be used to compute $\<\f(x_3)T[\f(x_1)\f(x_2)]\>$ in position space. The red line is the ``propagator" which comes from multiplying with the $\theta$ function. This propagator is not cut. We have chosen the red line to point from $x_2$ to $x_3$ because after Fourier transforming we have,
\begin{align}
\int d^{d}x_{1}d^{d}x_{2}d^{d}x_{3}\theta(x_{32})&\<\f(x_3)T[\f(x_1)\f(x_2)]\> e^{i(p_1\cdot x_1+p_2\cdot x_2+p_3\cdot x_3)}\nonumber
\\
&=\int \frac{d^{d}k}{2\pi}\frac{i}{k^0+i\epsilon}\delta(\vec{k})\ll \f(p_3+k)T[\f(p_1)\f(p_2-k)]\rr.\label{eq:disp_three_pt}
\end{align}
Thinking of the $\theta$-function as a propagator gives a useful way to organize Witten diagram computations.
The final result in momentum space is that we have a dispersive integral of the form \eqref{eq:disp_three_pt} for our cut three-point Witten diagram.
Applying these Feynman rules to each term in \eqref{eq:APP3ptdispersion} gives a non-Lorentz invariant dispersion formula for an individual Witten diagram. To turn this into a Lorentz invariant relation, one again needs to consider an infinite momentum limit \cite{Remiddi:1981hn}.

\section{Holographic Momentum Space Blocks}
\label{app:holoblocks} 
In this appendix we will prove \eqref{eq:gblocktoWitten},
\begin{align}
\hat{g}^t_{\O}(\z_i,s,t,u)=\text{disc}_{t}W^{t}_{\O}(\z_i,s,t,u), \label{eq:APPgblocktoWitten}
\end{align}
or that the momentum space conformal blocks of $\disc_{t}\tau_4$ are equal to cut Witten diagrams.
This relation is connected to ideas in previous works \cite{Polyakov:1974gs,Caron-Huot:2017vep}.
In \cite{Polyakov:1974gs} it was shown that taking a single discontinuity in momentum space computes a causal double commutator. 
Then in \cite{Caron-Huot:2017vep} it was shown that only the primary operator $\O$ contributes to the OPE of $\dDisc_{t}W^t_\O$, or the double-commutator of an exchange diagram.
These results make it clear the relation \eqref{eq:APPgblocktoWitten} should be true and the goal of this appendix is to prove it in momentum space using the AdS Cutkosky rules \cite{Meltzer:2020qbr}.

In this appendix we will assume the external momenta are spacelike, $\z_i<0$.
This allows us to write the conformal block in terms of partially time-ordered three-point functions \eqref{eq:CBblockdisctPt2}.
Using this representation of the conformal block, the identity \eqref{eq:APPgblocktoWitten} becomes:
\begin{align}
\disc_{t}W^{t}_{\O}(\z_i,s,t,u)=& -\mathcal{P}_{\O}(p_{14},\epsilon_1,\epsilon_2)\ll\overline{T}[\f(p_2)\f(p_3)]\O(p_{14},\epsilon_1)\rr
\ll\O(p_{23},\epsilon_2)T[\f(p_1)\f(p_4)]\rr
\nonumber \\
&\hspace{2.6in}+(p_1,p_4)\leftrightarrow(p_2,p_3).
\label{eq:Wequalsdisc1}
\end{align}
That is, taking a discontinuity of a Witten diagram factorizes it into a product of three-point functions.\footnote{Factorization conditions for Witten diagrams in momentum space have also been studied in \cite{Isono:2018rrb,Isono:2019wex}. The momentum space Polyakov blocks of \cite{Isono:2018rrb,Isono:2019wex} differ from the Polyakov-Regge blocks by a sum of AdS contact diagrams.} 
For the remainder of this section we will assume $p_1+p_4\in \overline{V}_+$ so that we can drop the second term in \eqref{eq:Wequalsdisc1}. It is straightforward to generalize to the opposite configuration where $p_2+p_3\in \overline{V}_+$ and only the second term is non-zero.

To prove \eqref{eq:Wequalsdisc1}, we first need the explicit form of $t$-channel exchange Witten diagrams,
\begin{align}
W^{t}_{\O}(p_i)=\int\limits_{0}^{\infty} \frac{dz_1dz_2}{z_{1}^{d+1}z_{2}^{d+1}}&K_{\f}(p_1,z_1)K_{\f}(p_4,z_1)
(iV_{L})G_{\O}(p_{14},z_1,z_2,\xi_1,\xi_2)
\nonumber \\
&(iV_{R})K_{\f}(p_2,z_2)K_{\f}(p_3,z_2).\label{eq:discWitten_factPt0}
\end{align}
Here $K_{\f}$ is the scalar boundary-to-bulk propagator. 
$G_{\O}$ is the bulk-to-bulk propagator for the spinning AdS field $\Phi_{\Delta,J}$ dual to the CFT primary $\O_{\Delta,J}$. We are using an index-free notation for the bulk fields,
\begin{align}
\Phi_{\Delta,J}(p,z,\xi)\equiv \Phi^{A_1\ldots A_{J}}_{\Delta,J}(p,z)\xi_{A_1}\ldots \xi_{A_J}.
\end{align}
Finally, $V_{L,R}$ are the left and right vertex factors. The overall factor of $i$ multiplying each vertex is a useful convention. We have suppressed the arguments of $V_{L,R}$, but they are also differential operators in $\xi_{1,2}$ and contract the indices of the propagator with the bulk three-point vertices. 
Acting with a $t$-channel discontinuity on \eqref{eq:discWitten_factPt0} only affects the bulk-to-bulk propagator:\footnote{To commute the $t$-channel discontinuity with the $z$-integral we need to assume the $z$-integral converges. We can either cut off the $z$ integral or assume the integral converges for some range of $d$ and $\Delta_{\f}$ and then define the Witten diagram and its discontinuity by analytic continuation \cite{Bzowski:2016kni}.} 
\begin{align}
\disc_{t}W^{t}_{\O}(p_i)=\int\limits_{0}^{\infty} \frac{dz_1dz_2}{z_{1}^{d+1}z_{2}^{d+1}}&K_{\f}(p_1,z_1)K_{\f}(p_4,z_1)
(iV_{L})\disc_{t}G_{\O}(p_{14},z_1,z_2,\xi_1,\xi_2)
\nonumber \\
&(iV_{R})K_{\f}(p_2,z_2)K_{\f}(p_3,z_2).\label{eq:discWitten_factPt1}
\end{align}

To prove \eqref{eq:discWitten_factPt1} factorizes into a product of CFT three-point functions, we use that the Feynman bulk-to-bulk propagator $G_{\O}$ is by definition equal to a bulk time-ordered two-point function in free-field theory,
\begin{align}
G_{\O}(p,z_1,z_2,\xi_1,\xi_2)=\ll T[\Phi_{\Delta,J}(-p,z_1,\xi_1)\Phi_{\Delta,J}(p,z_2,\xi_2)]\rr_{\text{free}}.
\end{align} 
Then the discontinuity of the bulk-to-bulk propagator is a sum of Wightman, or on-shell, bulk-to-bulk propagators,
\begin{align}
\disc_{t}G_{\O}(p,z_1,z_2,\xi_1,\xi_2)&=G^+_{\O}(p,z_1,z_2,\xi_1,\xi_2)+(p,z_1,\xi_1)\leftrightarrow (-p,z_2,\xi_2),\label{eq:disct_Gbb}
\\
G^+_{\O}(p,z_1,z_2,\xi_1,\xi_2)&=\ll \Phi_{\Delta,J}(-p,z_2,\xi_2)\Phi_{\Delta,J}(p,z_1,\xi_1)\rr_{\text{free}}.\label{eq:posenergyprop}
\end{align}
In \eqref{eq:discWitten_factPt1} we assumed $p_1+p_4\in \overline{V}_{+}$ and as a result only the first term in \eqref{eq:disct_Gbb} will be non-zero.
We can now factorize the bulk-to-bulk Wightman two-point function by using the completeness relation \eqref{eq:completeness}:
\begin{align}
\ll \Phi_{\Delta,J}(-p,z_2,\xi_2)\Phi_{\Delta,J}(p,z_1,\xi_1)\rr_{\text{free}} &=\mathcal{P}_{\O}(p,\epsilon_1,\epsilon_2)\ll\Phi(-p,z_2,\xi_2)\O(p,\epsilon_2)\rr_{\text{free}}
\nonumber \\ &\hspace{1.05in}\times
  \ll\O(-p,\epsilon_1)\Phi(p,z_1,\xi_1)\rr_{\text{free}}
\nonumber \\[2pt]
&=\mathcal{P}_{\O}(p,\epsilon_1,\epsilon_2) K_{\O}^{+}(p,z_2,\epsilon_2,\xi_2)K_{\O}^-(-p,z_1,\epsilon_1,\xi_1).\label{eq:factorized_wightman}
\end{align}
We are working at tree-level in the bulk, so only the operator $\O_{\Delta,J}$ dual to $\Phi_{\Delta,J}$ contributes to \eqref{eq:factorized_wightman}.

In \eqref{eq:factorized_wightman} $K_{\O}^{\pm}(p,z,\epsilon,\xi)$ are the positive and negative energy Wightman boundary-to-bulk propagators. By definition they are only non-zero for $p\in\overline{V}_{\pm}$. 
In our conventions $K_{\O}^+$ corresponds to positive energy flowing from the boundary to the bulk while for $K^{-}_\O$ positive energy is flowing from the bulk to the boundary.
The on-shell, boundary-to-bulk propagators can also be computed by taking a discontinuity of the Feynman boundary-to-bulk propagator:
\begin{align}
\disc_{t}K_{\O}(p,z,\xi,\epsilon)=K^+_{\O}(p,z,\xi,\epsilon)+K^-_{\O}(p,z,\xi,\epsilon),\label{eq:discK}
\end{align}
where here $t=-p^2$. 

The identities \eqref{eq:disct_Gbb}-\eqref{eq:factorized_wightman} imply that the cut Witten diagram \eqref{eq:discWitten_factPt1} factorizes into a product of three-point Witten diagrams,
\begin{align}
\disc_{t}W^{t}_{\O}(p_i)&=\mathcal{P}_{\O}(p_{14},\epsilon_1,\epsilon_2) W^{-}_{\f\f\O}(p_1,p_4,\epsilon_1)W^{+}_{\f\f\O}(p_2,p_3,\epsilon_2),\label{eq:disctWFactorized}
\\
W^{-}_{\f\f\O}(p_1,p_4,\epsilon_1)&=\int\frac{dz}{z^{d+1}}K_{\f}(p_1,z)K_{\f}(p_4,z)(iV_{L})K^-_{\O}(p_{23},z,\epsilon_1,\xi_1),
\\
W^{+}_{\f\f\O}(p_2,p_3,\epsilon_2)&=\int\frac{dz}{z^{d+1}}K_{\f}(p_2,z)K_{\f}(p_3,z)(iV_{R})K^+_{\O}(p_{14},z,\epsilon_2,\xi_2).
\end{align} 
As a reminder, the $V_{L,R}$ are differential operators in $\xi_i$ that remove the bulk polarization dependence from the final expressions. 

Finally, to prove \eqref{eq:Wequalsdisc1} we will show,
\begin{align}
\ll\O(p_{23},\epsilon_1)T[\f(p_1)\f(p_4)]\rr&=W^{-}_{\f\f\O}(p_1,p_4,\epsilon_1),\label{eq:cutWitten3ptPropto1}
\\[3pt]
\ll\overline{T}[\f(p_2)\f(p_3)]\O(p_{14},\epsilon_2)\rr&=-W^{+}_{\f\f\O}(p_2,p_3,\epsilon_2).\label{eq:cutWitten3ptPropto2}
\end{align}
To prove these relations, we can use the AdS cutting rules \cite{Meltzer:2020qbr}, which we reviewed in Appendix \ref{app:AdSUnitarity}. 
The first equality \eqref{eq:cutWitten3ptPropto1} follows trivially from the cutting rules.
For the second equality, the cutting rules imply:
\begin{align}
\ll\overline{T}[\f(p_2)\f(p_3)]\O(p_{14},\epsilon_2)\rr=\int\frac{dz}{z^{d+1}}K^*_{\f}(p_2,z)K^*_{\f}(p_3,z)(-iV_{R})K^+_{\O}(p_{14},z,\epsilon_2,\xi_2).\label{eq:cutWitten3ptPropto2pt2}
\end{align} 
To derive the momentum space dispersion formula, we assumed all the external momenta $p_i$ are spacelike.
When $p$ is spacelike we have, 
\begin{align}
K^*_{\f}(p)= -K_{\f}(p),\label{eq:bulkbdycomplex}
\end{align}
and we see \eqref{eq:cutWitten3ptPropto2pt2} implies \eqref{eq:cutWitten3ptPropto2}. The relation \eqref{eq:bulkbdycomplex} follows from the explicit form of the boundary to bulk propagator given in \eqref{eq:scalarBbdy}.\footnote{Alternatively, one can use that $K_{\f}(p)$ comes from a bulk time-ordered correlator in free-field theory. Therefore, the boundary to bulk propagator will also obey the coincidence relations \eqref{eq:coinc_vev} for spacelike momenta.}

Finally, we have shown
\begin{align}
\disc_{t}W^{t}_{\O}(p_i)=-\mathcal{P}_{\O}(p_{14},\epsilon_1,\epsilon_2)&\ll\overline{T}[\f(p_2)\f(p_3)]\O(p_{14},\epsilon_1)\rr
\nonumber \\
&\ll\O(p_{23},\epsilon_2)T[\f(p_1)\f(p_4)]\rr,
\end{align}
when $p_1+p_4\in \overline{V}_{+}$.
This proves \eqref{eq:Wequalsdisc1} and therefore the original relation \eqref{eq:APPgblocktoWitten}.

So far we have studied the conformal blocks of $\disc_{t}\tau_{4}(\z_i,s,t,u)$ when $\z_i$, $s<0$.
For this choice of momenta, $\disc_{t}\tau_{4}(\z_i,s,t,u)$ is equal to two different correlation functions,
\begin{align}
-2\disc_{t}\<T[\f(p_1)\ldots \f(p_4)]\>=& \  \<A[\f(p_2);\f(p_3)]A[\f(p_1);\f(p_4)]\>+(p_2,p_3)\leftrightarrow (p_1,p_4) \nonumber
\\[3pt]
=& \ \<\overline{T}[\f(p_2)\f(p_3)]T[\f(p_1)\f(p_4)]\>+(p_2,p_3)\leftrightarrow (p_1,p_4), \label{eq:TTbequiv}
\end{align}
but this relation does not hold for generic, external momenta.
It is then natural to ask: For generic momenta are the conformal blocks $\hat{g}_{\O}(\z_i,s,t,u)$ associated to the causal double-commutator or to the partially time-ordered correlator in the first and second lines of \eqref{eq:TTbequiv}, respectively?
The correct answer is that $\hat{g}^{t}_{\O}(\z_i,s,t,u)$ are the conformal blocks for the causal double-commutator.
To prove this, we use that causal restrictions in position space imply analyticity properties in momentum space.
Once we know the conformal block for the causal double-commutator in the configuration $\z_i$, $s<0$, we can analytically continue to general momenta.
For the details of this argument, see Appendix B of \cite{Meltzer:2020qbr}.

There are also holographic representations for conformal blocks in position space.
For example, the conformal blocks of Euclidean four-point functions $\<\f(x_1)\ldots\f(x_4)\>_{\text{E}}$ are equal to geodesic, exchange Witten diagrams \cite{Hijano:2015zsa}. 
To compute the geodesic exchange diagrams, we integrate the two internal vertices over geodesics that connect the two boundary points, as opposed to integrating over all of AdS. 
By contrast, for the cut Witten diagram, $\disc_{t}W^{t}_{\O}$, we integrate the vertices over all of AdS, but we put the internal propagator on shell. 
It would be interesting to better understand the connection between these holographic representations of conformal blocks.

\section{Regge Limit of Momentum Space Blocks}
\label{app:Regge_Blocks}
In this appendix we will study the large $t$ Regge limit of the conformal blocks $\hat{g}^{t}_{\Delta,J}$ and Polyakov-Regge blocks $\widehat{P}^{t|s}_{\Delta,J}$.
First, we will prove \eqref{eq:cond_conv_PH}, which gives the large $t$ scaling of $\hat{g}^{t}_{\Delta,J}$ as a function of $\Delta_{\f}$.
From this result, we can prove \eqref{eq:spinPH1}-\eqref{eq:spinPH3}, which gives the effective spins, $J_0$ and $\tilde{J}_0$, of the Polyakov-Regge block $\widehat{P}^{t|s}_{\Delta,J}$ as a function of $\Delta_{\f}$

In this appendix we will also continue to assume that the external momenta are spacelike, $\z_i<0$.
Under this assumption, the conformal block $\hat{g}^{t}_{\O}$ can be written in terms of partially time-ordered three-point functions, see \eqref{eq:CBblockdisctPt2}.
We can then use the results of \cite{Gillioz:2020wgw} to write down the conformal blocks explicitly:\footnote{In comparison to \cite{Gillioz:2020wgw}, we will use the mostly plus metric and will be considering the $t$-channel conformal blocks.}
\begin{align}
\hat{g}^{t}_{\Delta,J}(\z_i,s,t,u)=-t^{\frac{1}{2}(4\Delta_{\f}-3d)}\sum\limits_{m=0}^{J}C_{\Delta,J,m}\mathcal{C}^{\frac{d-3}{2}}_{m}(\cos(\theta))V^{\overline{T}}_{\Delta,J,m}(w_2,w_3)V^{T}_{\Delta,J,m}(w_1,w_4)
\nonumber
\\
\hfill + \ (p_2,p_3)\leftrightarrow (p_1,p_4).\label{eq:momentum_block_explicit}
\end{align}
Here $C_{\Delta,J,m}$ are a set of positive constants defined in equation 2.40 of \cite{Gillioz:2020wgw}, $C_{m}^{\alpha}(x)$ are the Gegenbauer polynomials and
\begin{align}
w_i=\frac{\z_i}{t}.
\end{align}
The angle $\theta$ is defined by,
\begin{align}
\cos(\theta)=\frac{t(u-s)+(\z_1-\z_4)(\z_2-\z_3)}{\sqrt{(t-\z_1-\z_4)^{2}-4\z_1\z_4}\sqrt{(t-\z_2-\z_3)^{2}-4\z_2\z_3}}.
\end{align}

The highest spin component, $m=J$, of the time-ordered vertex factor is,
\begin{align}
V^{T}_{\Delta,J,J}(w_1,w_4)=& \frac{i^{J-1}(4\pi)^{d+1}2^{-(2\Delta_{\f}+\Delta+J+1)}(\Delta-1)_J}{\Gamma^{2}\left(\frac{\Delta+J}{2}\right)\Gamma\left(\frac{2\Delta_{\f}-\Delta+J}{2}\right)\Gamma\left(\frac{2\Delta_{\f}-\tilde{\Delta}+J}{2}\right)}\left[(1-w_1-w_4)^2-4w_1w_4\right]^{J/2}
\nonumber
\\
\bigg[&f_{\Delta_{\f}\Delta_{\f};\Delta,J}(-w_1)^{\Delta_\f-d/2}(-w_4)^{\Delta_\f-d/2}F_{\Delta_{\f}\Delta_{\f};\Delta,J}(w_1,w_4)
\nonumber
\\
&+f_{\Delta_{\f}\tilde{\Delta}_{\f};\Delta,J}(-w_1)^{\Delta_\f-d/2}F_{\Delta_{\f}\tilde{\Delta}_{\f};\Delta,J}(w_1,w_4)
\nonumber
\\
&+f_{\tilde{\Delta}_{\f}\Delta_{\f};\Delta,J}(-w_4)^{\Delta_\f-d/2}F_{\tilde{\Delta}_{\f}\Delta_{\f};\Delta,J}(w_1,w_4)
\nonumber
\\
&+f_{\tilde{\Delta}_{\f}\tilde{\Delta}_{\f};\Delta,J}F_{\tilde{\Delta}_{\f}\tilde{\Delta}_{\f};\Delta,J}(w_1,w_4)\bigg],
\label{eq:highest_spin_vertex}
\end{align}
where $\tilde{\Delta}=d-\Delta$. The constant $f_{\Delta_{1}\Delta_{2};\Delta,J}$ is a product of $\Gamma$-functions,
\begin{align}
f_{\Delta_{1}\Delta_{2};\Delta,J}=\frac{\Gamma(d/2-\Delta_1)\Gamma(d/2-\Delta_2)\Gamma(\frac{\Delta_1+\Delta_2-\tilde{\Delta}+J}{2})}{\Gamma(1-\frac{\Delta_1+\Delta_2-\Delta+\ell}{2})}.
\end{align}
Finally $F_{\Delta_1,\Delta_2;\Delta,J}(w_1,w_4)$ is the Appell $F_4$ function,
\begin{small}
\begin{align}
F_{\Delta_1,\Delta_2;\Delta,J}(w_1,w_4)=F_{4}\left(\frac{\Delta_1+\Delta_2-\Delta+J}{2},\frac{\Delta_1+\Delta_2+J-\tilde{\Delta}}{2};\Delta_1-\frac{d}{2}+1,\Delta_2-\frac{d}{2}+1;w_1,w_4\right).
\label{eq:AppellF4}
\end{align}
\end{small}

The lower-spin vertex factors, $V^{T}_{\Delta,J,m}(w_1,w_4)$ with $0\leq m<J$, are determined by a recursion relation \cite{Gillioz:2020wgw}:
\begin{align}
V^{T}_{\Delta,J,m}(w_1,w_4)=&\frac{m+1}{(J-m)(\Delta+m-1)}\bigg[\frac{1}{\sqrt{(1-w_1-w_4)^2-4w_1w_4}}\widehat{D}_0V^{T}_{\Delta,J,m+1}(w_1,w_4)
\nonumber
\\[5pt]
&+\frac{(m+2)(\Delta-m-d)(d-1+J+m)}{(d-1+2m)(d+1+2m)}V^{T}_{\Delta,J,m+2}(w_1,w_4)
\bigg],
\label{eq:recursionAPP}
\\[20pt]
\widehat{D}_{0}=&\ 2[(w_1-w_4-1)w_1\partial_{w_1}+(w_1-w_4+1)w_4\partial_{w_4}
\nonumber
\\
&\ +(d-\Delta_{\f}-1)(w_1-w_4)].
\end{align}
When solving this recursion relation, we set $V^{T}_{\Delta,J,J+1}=0$.
For $\widehat{D}_{0}$ we took equation 3.13 in \cite{Gillioz:2020wgw} and assumed the external scalars were identical.
Finally, the vertex factor for the anti-time-ordered product is defined by 
\begin{align}
V^{\overline{T}}_{\Delta,J,m}(w_2,w_3)=(V^{T}_{\Delta,J,m}(w_2,w_3))^*.
\end{align}

We can now take the large $t$ Regge limit of \eqref{eq:momentum_block_explicit}.
Since we are taking $t$ large with $\z_i<0$ and fixed, the $w_i$ variables scale like:
\begin{align}
w_i\rightarrow 0 \quad \text{such that } \ \frac{w_i}{w_j}=\frac{\z_i}{\z_j} \ \text{ is fixed}.
\end{align}
In this limit we can use the defining sum for the Appell $F_4$ function,
\begin{align}
F_{4}(a_1,a_2;b_1,b_2;z_1,z_2)=\sum\limits_{m,n=0}^{\infty}\frac{(a_1)_{m+n}(a_2)_{m+n}}{(b_1)_{m}(b_2)_{n}m!n!}z_1^mz_2^n,
\end{align}
to expand $F_4$ in integer powers of $t^{-1}$.
Similarly, if we take $t$ large with $\z_i$, $s<0$ we have $\cos(\theta)\rightarrow -1$ and the Gegenbauers also have a Taylor series expansion in $t^{-1}$.
Moreover, both expansions will be analytic in the remaining momenta.

Next we can consider the vertex factors.
For simplicity, we can start with the highest-spin vertex factor \eqref{eq:highest_spin_vertex}.
In the limit $w_i\rightarrow 0$ it scales like,
\[
    \lim\limits_{w_i\rightarrow 0}V^{T}_{\Delta,J,J}(w_1,w_4)\sim 
\begin{cases}
    (-w_1)^{\Delta_{\f}-d/2}(-w_4)^{\Delta_{\f}-d/2},& \text{ if } \Delta_{\f}<d/2,\\
    1,&  \text{ if } \Delta_{\f}>d/2.\numberthis \label{eq:scaling_vertex}
\end{cases}
\]

For the lower-spin vertex factors, $V^{T}_{\Delta,J,m}$ with $m<J$, we can use the recurrence relation \eqref{eq:recursionAPP}.
From this recurrence relation, we see that the lower-spin vertex factors have the same analytic structure as the highest spin vertex \eqref{eq:highest_spin_vertex}.
In addition, the lower-spin vertex factors have either the same scaling as \eqref{eq:highest_spin_vertex}, or are further suppressed, when we take the $w_i$ small at the same rate.
Since the lower-spin factors do not introduce any new non-analyticities or change the scaling of the block in the Regge limit, we will restrict our attention to the highest-spin factor \eqref{eq:highest_spin_vertex} in the remainder of this appendix.

The previous results show that the large $t$ scaling of $\hat{g}^{t}_{\O}$ is fixed by the overall power of $t$ in \eqref{eq:momentum_block_explicit} and the powers of $(-w_i)^{\Delta_{\f}-d/2}$ in \eqref{eq:highest_spin_vertex}.
If $\Delta_{\f}<d/2$, we find the following scaling,
\begin{align}
\lim\limits_{t\rightarrow \infty} \hat{g}^{t}_{\Delta,J}(\z_i,s,t,u)\sim  t^{\frac{1}{2}(d-4\Delta_{\f})}\prod\limits_{i=1}^{4}(-\z_i)^{\Delta_{\f}-d/2},\label{eq:APPscal1}
\end{align}
which comes from using the first term in the brackets of \eqref{eq:highest_spin_vertex} for both the left and right vertex factors.
When $\Delta_{\f}>d/2$ we instead have:
\begin{align}
\lim\limits_{t\rightarrow \infty}\hat{g}^{t}_{\Delta,J}(\z_i,s,t,u)\sim t^{\frac{1}{2}(4\Delta_{\f}-3d)},\label{eq:APPscal2}
\end{align}
which comes from using the last term in the brackets of \eqref{eq:highest_spin_vertex} for both vertex factors.
The scalings \eqref{eq:APPscal1} and \eqref{eq:APPscal2} correspond to the first and second line of \eqref{eq:cond_conv_PH}.

From these results, it is then not difficult to prove \eqref{eq:spinPH1}-\eqref{eq:spinPH3}.
We can start by assuming $\Delta_{\f}\leq d/4$.
In this case \eqref{eq:APPscal1} implies that the block $\hat{g}^t_{\Delta,J}$ is constant or grows at large $t$.
In interacting, unitary CFTs, we additionally have the lower bound $\Delta_{\f}>\frac{1}{2}(d-2)$.
The two bounds are compatible if $d<4$ and the scalar $\f$ does not have a large anomalous dimension, e.g. it can be the $\s$ operator in the $3d$ Ising model \cite{Poland:2018epd}.

To define the momentum space Polyakov-Regge block when $\Delta_{\f}\leq d/4$, we can add and subtract a finite number of terms that grow at large $t$,
\begin{align}
\hat{g}^{t}_{\Delta,J}(\z_i,s,t,u)= \left(\hat{g}^{t}_{\Delta,J}(\z_i,s,t,u)-\sum\limits_{a}t^{a}\mathcal{S}_{a}(\z_i,s,u)\right)+\sum\limits_{a}t^{a}\mathcal{S}_{a}(\z_i,s,u).\label{eq:sub_g}
\end{align}
The values of the exponents $``a"$ and the functions $\mathcal{S}_a$ are chosen such that the term in the parentheses is superbounded in momentum space.
The Polyakov-Regge block is then defined by:
\begin{align}
\widehat{P}^{t|s}_{\Delta,J}(\z_i,s,t,u)&= \widehat{P}^{(1),t|s}_{\Delta,J}(\z_i,s,t,u)+\widehat{P}^{(2),t|s}_{\Delta,J}(\z_i,s,t,u),
 \label{eq:decomp_PV0} \\[2pt]
 \widehat{P}^{(1),t|s}_{\Delta,J}(\z_i,s,t,u)&=\frac{1}{2\pi i}\int\limits_{0}^{\infty}\frac{dt'}{t'-t-i\epsilon}\left(\hat{g}^{t}_{\Delta,J}(\z_i,s,t',u)-\sum\limits_{a}t'^{a}\mathcal{S}_{a}(\z_i,s,u)\right), \label{eq:app_superbounded}
 \\
\widehat{P}^{(2),t|s}_{\Delta,J}(\z_i,s,t,u)&=\frac{1}{2\pi i}\sum\limits_{a} \mathcal{S}_{a}(\z_i,s,u)\int\limits_{0}^{\infty}\frac{dt'}{t'-t-i\epsilon}t'^{a},
\nonumber
\\ &=\frac{i}{2\sin(a\pi)}\sum\limits_{a} \mathcal{S}_{a}(\z_i,s,u)(-t)^a.
\label{eq:non_analytic}
\end{align}
By definition, the dispersion integral in \eqref{eq:app_superbounded} converges and $\widehat{P}^{(1),t|s}_{\Delta,J}$ is superbounded in momentum space.
The dispersion integrals in \eqref{eq:non_analytic} diverge, but we have a finite sum of pure powers and can define the integrals by analytic continuation.
The resulting expression, $\widehat{P}^{(2),t|s}_{\Delta,J}$, manifestly grows at large $t$.
From \eqref{eq:APPscal1} we see the leading term we need to subtract is non-analytic in all of the momenta and therefore $\widehat{P}^{(2),t|s}_{\Delta,J}$ will also contain non-analytic terms that grow at large $t$.
This proves that if $\Delta_{\f}\leq d/4$, then $\widehat{P}^{t|s}_{\Delta,J}$ is not superbounded in position space, or that $J_0\geq0$.
The effective spin in the momentum space Regge limit, $\tilde{J}_0$ is bounded from below by $J_0$, see \eqref{eq:Jtilde_max}, so we also have $\tilde{J}_0\geq0$. 
This proves \eqref{eq:spinPH1}.

The second case, \eqref{eq:spinPH2}, is trivial to prove. 
If $\frac{d}{4}<\Delta_{\f}<\frac{3d}{4}$, then the defining integral for $\widehat{P}^{t|s}_{\Delta,J}$ converges and we have $\tilde{J}_0<0$.
From \eqref{eq:Jtilde_max} this automatically implies that $J_0<0$.

To prove the final case, \eqref{eq:spinPH3}, we can follow the same strategy used to prove \eqref{eq:spinPH1}.
First we note that for $\Delta_{\f}\geq \frac{3d}{4}$, the leading behavior given by \eqref{eq:APPscal2} is semi-local.
It is not difficult to go further and show that when $\Delta_{\f}\geq \frac{3d}{4}$, the only terms that grow with $t$ are semi-local.
To construct a term which is non-analytic in all of the momenta, we need at least one additional power of $(-w_i)^{\Delta_{\f}-d/2}$ from both the left and right vertex factors in \eqref{eq:momentum_block_explicit}.
This gives the overall scaling,
\begin{align}
\hat{g}^{t}_{\Delta,J}(\z_i,s,t,u)\bigg|_{\text{non-analytic}}\sim t^{-d/2}, \quad \text{ if } \ \Delta_{\f}\geq \frac{3d}{4}.
\end{align}
This proves that if $\Delta_{\f}\geq \frac{3d}{4}$, then all the subtraction terms are semi-local.

To define $\widehat{P}^{t|s}_{\Delta,J}$ for $\Delta_{\f}\geq \frac{3d}{4}$, we can follow the same procedure as before.
By adding and subtracting the finite number of semi-local terms that grow at large $t$, we find,
\begin{align}
\widehat{P}^{t|s}_{\Delta,J}(\z_i,s,t,u)&= \widehat{P}^{(3),t|s}_{\Delta,J}(\z_i,s,t,u)+\widehat{P}^{(4),t|s}_{\Delta,J}(\z_i,s,t,u),
 \label{eq:decomp_P} \\[2pt]
 \widehat{P}^{(3),t|s}_{\Delta,J}(\z_i,s,t,u)&=\frac{1}{2\pi i}\int\limits_{0}^{\infty}\frac{dt'}{t'-t-i\epsilon}\left(\hat{g}^{t}_{\Delta,J}(\z_i,s,t',u)-\sum\limits_{a}t'^{a}\mathcal{Q}_{a}(\z_i,s,u)\right), \label{eq:non_analytic_3}
 \\
\widehat{P}^{(4),t|s}_{\Delta,J}(\z_i,s,t,u)&=\frac{1}{2\pi i}\sum\limits_{a} \mathcal{Q}_{a}(\z_i,s,u)\int\limits_{0}^{\infty}\frac{dt'}{t'-t-i\epsilon}t'^{a},
\nonumber
\\ &=\frac{i}{2\sin(a\pi)}\sum\limits_{a} \mathcal{Q}_{a}(\z_i,s,u)(-t)^a.
\label{eq:analytic_4}
\end{align}
The exponents $``a"$ and the functions $\mathcal{Q}_a$ are chosen such that the integral in \eqref{eq:non_analytic_3} converges.
By definition, $\widehat{P}^{(3),t|s}_{\Delta,J}$ will be superbounded in momentum space.
The dispersion integrals in \eqref{eq:analytic_4} are defined by analytic continuation and the resulting term, $\widehat{P}^{(4),t|s}_{\Delta,J}$, is explicitly semi-local.
This proves that if $\Delta_{\f}\geq \frac{3d}{4}$, then $\widehat{P}^{t|s}_{\Delta,J}$ is superbounded in the position space Regge limit, $J_0<0$, but is not superbounded in the momentum space limit, $\tilde{J}_0\geq0$.
This proves the final case, \eqref{eq:spinPH3}.

\section{Details of Largest Time Derivation}
\label{app:details}
In this appendix we will fill in the details of Section \ref{sec:disp_from_largest_time} on how to derive Lorentz invariant dispersion formulas from the infinite momentum limit. 
For convenience, we reproduce the kinematics \eqref{eq:dispersion_n_pointV2},
\begin{align}
p_1&=(P,P,p_{1}^{\perp})  \label{eq:APPnpt_kinematics1}
\\
p_2&=(-P,-P,p_{2}^{\perp}) 
\\
p_i&=(0,0,p_i^{\perp}) \quad \text{for} \quad i=3,\ldots,n  \label{eq:APPnpt_kinematics3}
\end{align}
and the dispersion formula which follows from the largest time equation,
\begin{align}
\hspace{-1cm}\<T[\f(p_1)\ldots \f(p_n)]\>=&\sum\limits_{r=0}^{n-2}(-1)^{r+1}\sum\limits_{\alpha\in \Pi_{r}(n-2)}\int\frac{d^{d}k}{2\pi}\frac{-i}{k^0+i\epsilon}\delta(\vec{k})\bigg[
\nonumber
\\
&
\<\overline{T}[\f(p_n+k)\f(p_{\alpha_1})...\f(p_{\alpha_{r}})]T[\f(p_{\alpha_{r+1}})...\f(p_{\alpha_{n-2}})\f(p_{n-1}-k)]\>
\nonumber 
\\
+&\<\overline{T}[\f(p_{n-1}+k)\f(p_{\alpha_1})...\f(p_{\alpha_{r}})]T[\f(p_{\alpha_{r+1}})...\f(p_{\alpha_{n-2}})\f(p_n-k)]\>\bigg]. \label{eq:APPdispersion_n_pointV1}
\end{align}

In practice there are 8 different classes of correlators we need to study depending on how $\{\f(p_1),\f(p_2)\}$ and $\{\f(p_{n-1}\pm k),\f(p_n\mp k)\}$ are grouped. In the expressions below we will suppress the $n-4$ operators with momenta purely in the $d-2$ transverse directions. We have the following four classes of correlators where $\f(p_{n-1}-k)$ is in the time-ordered product:
\begin{eqnarray}
& \hspace{.1in}   \<\overline{T}[\f(p_{n}+k)\ldots]T[\f(p_1)\f(p_2)\f(p_{n-1}-k)\ldots]\> ,
\\[3pt]  
& \hspace{.1in} \<\overline{T}[\f(p_{n}+k)\f(p_1)\f(p_2)\ldots]T[\f(p_{n-1}-k)\ldots]\>,
\\[3pt]
& \hspace{.1in}   \<\overline{T}[\f(p_{n}+k)\f(p_1)\ldots]T[\f(p_2)\f(p_{n-1}-k)\ldots]\>,
\\[3pt]
& \hspace{.1in} \<\overline{T}[\f(p_{n}+k)\f(p_2)\ldots]T[\f(p_1)\f(p_{n-1}-k)\ldots]\>,
\end{eqnarray}
and an additional four correlators with $p_{n}\leftrightarrow p_{n-1}$. For the remainder of this section we will set $n=4$ to simplify notation and consider the correlators:
\begin{eqnarray}
& \hspace{.1in}   \<\f(p_{4}+k)T[\f(p_1)\f(p_2)\f(p_{3}-k)]\> , \label{eq:checkscaling1}
\\[3pt]
& \hspace{.1in} \<\overline{T}[\f(p_{4}+k)\f(p_1)\f(p_2)]\f(p_{3}-k)\>,\label{eq:checkscaling2}
\\[3pt]
& \hspace{.1in}   \<\overline{T}[\f(p_{4}+k)\f(p_1)]T[\f(p_2)\f(p_{3}-k)]\>,\label{eq:checkscaling3}
\\[3pt]
& \hspace{.1in} \<\overline{T}[\f(p_{4}+k)\f(p_2)]T[\f(p_1)\f(p_{3}-k)]\>.\label{eq:checkscaling4}
\end{eqnarray}
The $n>4$ result will follow immediately as we will discuss at the end of this appendix.

The correlator \eqref{eq:checkscaling1} only has support for $p_4+k\in \overline{V}_-$.
To solve this $\theta$-function constraint, we make the following change of variables,
\begin{align}
k^0=-\sqrt{\z_4'-\z_4}. \label{eq:appCOV4pt1}
\end{align}
As a reminder, the unprimed variables refer to Lorentz invariants constructed from the external momenta while the primed variables are constructed from the internal momenta shifted by $k$.

Since $k^0$ in \eqref{eq:appCOV4pt1} is independent of $P$, the measure terms in \eqref{eq:APPdispersion_n_pointV1} remain independent of $P$ after making this change of variables. We then only need to check how the internal invariants scale with $P$. We find:
\begin{align}
t'&\sim -2\sqrt{\z_4'-\z_4}P,
\\
u'&\sim 2\sqrt{\z_4'-\z_4}P,
\end{align}
with all the other invariants approaching a constant. We can drop \eqref{eq:checkscaling1} in the dispersion integral if the corresponding unshifted correlator, $\ll\f(p_{4})T[\f(p_1)\f(p_2)\f(p_{3})]\rr$ vanishes in the following limit:
\begin{align}
\lim\limits_{t\rightarrow \infty}\ll\f(p_{4})T[\f(p_1)\f(p_2)\f(p_{3})]\rr=0,
\end{align}
with $\zeta_i$, $s<0$ held fixed.

For the second correlator \eqref{eq:checkscaling2} we find identical results up to relabeling. This correlator only has support for $p_3-k\in\overline{V}_+$, so we set,
\begin{align}
k^0=-\sqrt{\z_3'-\z_3}. \label{eq:appCOV4pt2}
\end{align}
We see $k^0$ in \eqref{eq:appCOV4pt2} is again independent of $P$, which implies the measure terms remain independent of $P$. We also find the same scaling for the internal invariants as before,
\begin{align}
t'&\sim -2\sqrt{\z_3'-\z_3}P,
\\
u'&\sim 2\sqrt{\z_3'-\z_3}P.
\end{align}
If we assume the unshifted correlator $\<\overline{T}[\f(p_1)\f(p_2)\f(p_{4})]\f(p_{3})\>$ decays in the large $t$ Regge limit, then we can also drop the second correlator  \eqref{eq:checkscaling2} from the dispersion integral.

The third correlator \eqref{eq:checkscaling3} introduces the scaling \eqref{eq:4ptFunnyScaling1}-\eqref{eq:4ptFunnyScaling3}. This correlator only has support for $p_2+p_3-k\in\overline{V}_{+}$, so we set,
\begin{align}
k^0=-P-\sqrt{P^{2}+t'-t}. \label{eq:COV4pt3}
\end{align}
If we use \eqref{eq:COV4pt3} and take the large $P$ limit, we find the measure terms scale like,
\begin{align}
\frac{dk^{0}}{k^0+i\epsilon}\sim \frac{dt'}{4P^{2}}.
\end{align}
In addition, the internal invariants of $\<\overline{T}[\f(p_1)\f(p_{4}+k)]T[\f(p_2)\f(p_{3}-k)]\>$ scale like,
\begin{align}
u'\sim 8P^{2}, \label{eq:APP4ptFunnyScaling1}
\\
\z_3'\sim 4P^{2},
\\
\z_4'\sim 4P^{2}. \label{eq:APP4ptFunnyScaling3}
\end{align}
We can drop this term if the unshifted correlator $\<\overline{T}[\f(p_1)\f(p_{4})]T[\f(p_2)\f(p_{3})]\>$ grows slower than $P^{2}$ in the above scaling limit.

Lastly, we have the correlator \eqref{eq:checkscaling4}. As we already showed in Section \ref{sec:FourPtTimeOrdering}, in the large $P$ limit this gives a finite contribution corresponding to the $u$-channel piece of the final result \eqref{eq:dispersion_4pt_LT}.
One can then repeat an identical analysis for the remaining four correlators, which amounts to making the replacement $p_3\leftrightarrow p_4$.   
The conditions given in this section are sufficient to derive a Lorentz invariant relation from the infinite momentum limit, but we have not proven that they are necessary. 
It is possible that there are cancellations between different correlators in the large $P$ limit, in which case some of our assumptions can be relaxed.

To generalize to $n>4$, we can note the only non-zero contributions to the dispersion formula come from terms where $\f(p_1)$ is in the time-ordered product and $\f(p_2)$ is in the anti-time-ordered product. We also need $\f(p_{n-1})$ and $\f(p_n)$ to sit in different ordering products. The additional $n-4$ operators are effectively spectators which do not participate in the $k$ integrals or the large $P$ limit. Then to arrive at \eqref{eq:dispersion_n_pointV2}, we take our four point result and sum over all possible partitions of the remaining $n-4$ operators.

\bibliographystyle{utphys}
\bibliography{biblio_dispersion}
\end{document}